\documentclass[twocolumn]{aastex63} 
\usepackage{amsmath}
\usepackage{natbib}
\usepackage{float}

\newcommand\ldl{$\lambda/\Delta\lambda$}
\newcommand\lhalbol{$\log_{10}L_{H\alpha}/L_{bol}$}

\newcommand\teff{$T_{\rm eff}$}
\newcommand\logg{$\log{g}$}
\newcommand\vsini{$v\sin{i}$}
\newcommand\kms{km s$^{-1}$}

\graphicspath{{./}}

\submitjournal{\apjs}

\begin{document}

\title{The Brown Dwarf Kinematics Project (BDKP). VI. Ultracool Dwarf Radial and Rotational Velocities from SDSS/APOGEE High-resolution Spectroscopy}

\shorttitle{APOGEE DR17 Ultracool Dwarfs}
\shortauthors{Hsu et. al}


\correspondingauthor{Chih-Chun Hsu}
\email{chsu@northwestern.edu}

\author[0000-0002-5370-7494]{Chih-Chun Hsu}
\affiliation{Center for Interdisciplinary Exploration and Research in Astrophysics (CIERA), Northwestern University,
1800 Sherman, Evanston, IL 60201, USA}
\affiliation{Center for Astrophysics and Space Science, Department of Physics, University of California San Diego, La Jolla, CA 92093, USA}

\author[0000-0002-6523-9536]{Adam J.\ Burgasser} 
\affiliation{Department of Astronomy \& Astrophysics, University of California San Diego, La Jolla, CA 92093, USA}

\author[0000-0002-9807-5435]{Christopher A. Theissen} 
\affiliation{Department of Astronomy \& Astrophysics, University of California San Diego, La Jolla, CA 92093, USA}

\author[0000-0002-7961-6881]{Jessica L. Birky} 
\affiliation{Department of Astronomy, University of Washington, Seattle, WA 98195, USA}

\author[0000-0003-2094-9128]{Christian Aganze}
\affiliation{Center for Astrophysics and Space Science, Department of Physics, University of California San Diego, La Jolla, CA 92093, USA}
\affiliation{Department of Physics, Stanford University, Stanford, CA 94305, USA}

\author[0000-0003-0398-639X]{Roman Gerasimov}
\affiliation{Center for Astrophysics and Space Science, University of California San Diego, La Jolla, CA 92093, USA}

\author[0000-0002-7224-7702]{Sarah J. Schmidt}
\affiliation{Leibniz-Institute for Astrophysics Potsdam (AIP), An der Sternwarte 16, D-14482, Potsdam, Germany}

\author[0000-0002-6096-1749]{Cullen H. Blake}
\affiliation{University of Pennsylvania Department of Physics and Astronomy, 209 S 33rd St, Philadelphia, PA 19104, USA}

\author[0000-0001-6914-7797]{Kevin R. Covey}
\affiliation{Department of Physics \& Astronomy, Western Washington University, Bellingham WA 98225-9164, USA}

\author[0000-0002-6906-2379]{Elizabeth Moreno-Hilario}
\affiliation{Instituto de Astronom\'ia, Universidad Nacional Aut\'onoma de M\'exico, Ciudad Universitaria, CDMX, C.P. 04510, Mexico}

\author[0000-0001-5072-4574]{Christopher R.\ Gelino}
\affiliation{NASA Exoplanet Science Institute, Mail Code 100-22, California Institute of Technology, 770 South Wilson Avenue, Pasadena, CA 91125, USA}
\affiliation{Infrared Processing and Analysis Center, Mail Code 100-22, California Institute of Technology, 1200 E. California Boulevard, Pasadena, CA 91125, USA}

\author[0000-0001-7351-6540]{Javier Serna}
\affiliation{Instituto de Astronom\'{i}a, Universidad Aut\'{o}noma de M\'{e}xico \\
Ensenada, B.C, M\'{e}xico}

\author[0000-0002-8725-1069]{Joel R. Brownstein}
\affiliation{Department of Physics and Astronomy, University of Utah, 115 S. 1400 E., Salt Lake City, UT 84112, USA}

\author[0000-0001-6476-0576]{Katia Cunha}
\affiliation{Steward Observatory, University of Arizona, 933 North Cherry Avenue, Tucson, AZ 85721-0065, USA}
\affiliation{Observat\'orio Nacional/MCTIC, R. Gen. Jos\'e Cristino, 77, 20921-400, Rio de Janeiro, Brazil}

\begin{abstract}
We present precise measurements of radial (RV) and projected rotational ({\vsini}) velocities of a sample of 258 M6 to L2 dwarfs with multi-epoch, high-resolution ({\ldl} = 22500), near-infrared (1.514--1.696~$\mu$m) spectroscopic observations reported in the Apache Point Observatory Galactic Evolution Experiment (APOGEE) Data Release 17. 
The spectra were modeled using a Markov Chain Monte Carlo forward-modeling method which 
achieved median precisions of $\sigma_\text{RV}$ = 0.4~{\kms} and $\sigma_{v\sin{i}}$ = 1.1~{\kms}.
One-half of our sample (138 sources) are previously known
members of nearby young clusters and moving groups, and we identified three new kinematic members of the Argus or Carina Near moving groups, 2MASS~J05402570+2448090, 2MASS~J14093200+4138080, and 2MASS~J21272531+5553150.
Excluding these sources, we find that the 
majority of our sample has kinematics consistent with the Galactic thin disk, and eleven sources are associated with the intermediate thin/thick disk. 
The field sample has a velocity dispersion of 38.2$\pm$0.3~{\kms}, equivalent to an age of 3.30$\pm$0.19 Gyr based on empirical age-velocity dispersion relations;
and a median {\vsini} of 17~{\kms}. 
For 172 sources with multi-epoch observations, we identified 37 as having significant radial velocity variations, and determined preliminary orbit parameters for 26 sources with four or more epochs, nine of which are short-period binary candidates. 
For 40 sources with photometric variability periods from the literature less than 5 days and {\vsini} $>$ 20~{\kms}, we find a decline in projected radii $R\sin{i}$ with age congruent with evolutionary models.
Finally, we also present multi-epoch RV and {\vsini} measurements for additional 444 candidate ultracool dwarfs.
\end{abstract}


\section{Introduction} \label{sec:intro}



Ultracool dwarfs (UCDs) are stellar and substellar objects with effective temperatures below 3000~K, 
masses below 0.1~M$_{\odot}$, and spectral types spanning the late-M, L, T, and Y dwarf classes \citep{Kirkpatrick:2005aa,Burgasser:2006aa,Cushing:2011aa}.
Over the past two decades, tens of thousands of UCDs have been discovered through large sky surveys such as the Two Micron All Sky Survey (2MASS; \citealp{Skrutskie:2006aa}), the Sloan Digital Sky Survey (SDSS; \citealp{York:2000aa}), the \textit{Wide-field Infrared Survey Explorer} (\textit{WISE}; \citealp{Cutri:2012aa}), and \textit{Gaia} \citep{Gaia-Collaboration:2018aa, Gaia-Collaboration:2021ac}. Given such a large sample of UCDs, statistically studies of their physical properties becomes possible, particularly through the use of high-resolution spectroscopy.

High-resolution spectroscopy provides several unique opportunities for characterizing the physical properties of stars. 
These data provide precise radial velocities (RVs) which combined with proper motions and distances yield 3D kinematics that can be used to determine membership in Galactic populations (thin/thick disk, halo; \citealt{Bensby:2003aa}) and young clusters and moving groups \citep{Gagne:2018ab}. 
The dispersion of spatial velocities can also
be used to statistically infer population ages
\citep{Wielen:1977aa, Aumer:2009aa}.
Multi-epoch precise RVs enable the identification of single-line or double-line low-mass binaries (e.g. \citealp{Blake:2010aa, Burgasser:2016aa, Triaud:2020aa, Hsu:2023aa}).
Projected rotational velocities trace angular momentum evolution as a function of age and magnetic activity \citep{Zapatero-Osorio:2006aa, Herbst:2007aa, Irwin:2011aa, Bouvier:2014aa} and enable assessment of spin/orbit alignment in binary and exoplanet systems (Rossiter–McLaughlin effect; \citealp{Rossiter:1924aa, McLaughlin:1924aa, Triaud:2010aa}).
High-resolution spectroscopy can also be used to measure elemental abundances, including rare species and isotopes (e.g., $^{13}$CO; \citealp{Tsuji:2016aa, Souto:2017aa, Crossfield:2019aa}); and detailed line modeling can be used to measure magnetic fields through Zeeman splitting, wavelength-dependent limb darkening, and map surface structure through Doppler imaging (e.g., \citealp{Shulyak:2010aa, Crossfield:2014aa}).

Despite these opportunities, the current sample of reported high-resolution spectroscopy for ultracool dwarfs is limited to a few hundred systems due to their intrinsic faintness and red/infrared spectral energy distributions 
(SEDs; \citealt{Hsu:2021aa}).
For comparison, the Apache Point Observatory Galactic Evolution Experiment (APOGEE; \citealt{Majewski:2017aa, Wilson:2019aa}), part of the Sloan Digital Sky Survey (SDSS; \citealp{Blanton:2017aa}), provides high-resolution ({\ldl} $\sim$22,500) near-infrared spectra 
(1.514--1.696~\micron)
of more than 730,000 stars \citep{Abdurrouf:2022aa}.
The single-epoch spectra of APOGEE have provided the chemical abundances for planet hosts \citep{Souto:2017aa, Souto:2018aa, Wilson:2018aa, Souto:2020aa}, Galactic chemical populations \citep{Hayden:2015aa, Cunha:2015aa, Bovy:2016ab, Donor:2020aa}, and cluster kinematics \citep{Ness:2016ab, Da-Rio:2017aa}.
The acquisition of multi-epoch spectra has enabled discovery of thousands of binary systems, including $\sim$4,000 single-lined spectroscopic binary (SB1) companions to giants \citep{Price-Whelan:2018aa} and $>$7,000 double-lined spectroscopic binaries (SB2) among Main Sequence stars \citep{Skinner:2018aa,Kounkel:2021aa}.
However, APOGEE does not provide robust RVs and {\vsini}s for UCDs, as 
both the APOGEE Stellar Parameter and Chemical Abundances Pipeline (ASPCAP \citealt{Garcia-Perez:2016aa}), and a more recent radial velocity pipeline \texttt{Doppler} \citep{Nidever:2021aa}, have difficulties fitting molecule-rich UCD spectra below {\teff} $=$ 3,500~K.

Attempts have been made to model APOGEE spectra through other means.
Metallicities and abundances of early-to-mid M dwarfs (M0-M5) have been extensively explored in \cite{Schmidt:2016aa, Souto:2017aa, Souto:2018aa, Rajpurohit:2018aa, Souto:2020aa, Souto:2021aa, Souto:2022aa, Wanderley:2023aa}.
APGOEE UCDs were relatively less explored in the literature.
For example, \citep{Birky:2020aa} measured the {\teff}s, metallicities, and spectral types of 5875 M dwarfs across M0 to M9 using a data-driven approach. 
\cite{Skinner:2018aa} identified 44 M dwarf double-lined spectral binaries among a sample of 1350 M dwarfs\footnote{Their sample is composed of all early-to-mid M dwarfs except for one M7 SB2 2MASS J03122509+0021585 (a.k.a. LSPM J0312+0021), but the APOGEE data have low quality and were flagged as bad data in DR17.}.
\citet{Deshpande:2013aa} reported RV and {\vsini} measurements for 253 M dwarfs
with APOGEE data using a forward-modeling approach.
However, the sample is dominated by early- to mid-M dwarfs, with only 27 sources falling in the UCD regime\footnote{These were derived from the $V - J$ colors of their sample, and only 10 of which have independent spectral types of M6 and later.}.
As discussed below, there are in fact hundreds of UCDs with APOGEE data.
Additionally, the RVs reported by \citet{Deshpande:2013aa} were determined by $\chi^2$ minimization using BT-Settl models \citep{Allard:1997aa}. As shown in this paper, these older atmosphere models are unable to provide robust RVs and {\vsini}s in the UCD regime due to the missing FeH opacities.
\cite{Gilhool:2018aa} measured {\vsini} for 714 M dwarfs, with 17 M6--M8 in their sample.
More recently, \cite{Sarmento:2021aa} analyzed 313 M dwarfs to measure RV and {\vsini} using high signal-to-ratio (S/N $>$ 200) spectra (but included 37 lower S/N benchmark sources to calibrate and explore their measurement limits). Due to their selection criteria for high S/N spectra, only one UCD, M8e 2MASS J05392474$+$4038437, was selected in their sample.
The availability of several more UCD APOGEE spectra in Data Release 17 (DR17; \citealt{Abdurrouf:2022aa}) and improvements in stellar modeling motivate a re-evaluation of these data, with the aim of building 
a larger and more precise sample of accurate RV and {\vsini} for UCDs.


In this article, we present multi-epoch measurements of precise radial and projected rotational velocities for 258 spectroscopically-classified and 444 candidate ultracool dwarfs in the APOGEE sample, using a Markov Chain Monte Carlo forward-modeling method. This approach has been shown to provide precise and robust RV and {\vsini} measurements for high-resolution near-infrared spectra of late-M, L, and T dwarfs \citep{Blake:2010aa, Burgasser:2016aa, Hsu:2021aa}, and here we apply the method to APOGEE observations of UCDs.
In Section~\ref{sec:sample}, we define our sample, including low-resolution optical spectra of twelve sources without prior classifications in the literature.
In Section~\ref{sec:methods}, we describe our forward-modeling approach in detail.
In Section~\ref{sec:stellar_parameters}, we discuss our RV and {\vsini} measurements, and inferred effective temperatures and surface gravities, based on our fits.
In Section~\ref{sec:analysis}, we combine our RV measurements with {\em Gaia} astrometry to determine 3D space motions, and use these to characterize Galactic populations and orbits, cluster memberships, and kinematic ages for the sample.
We identify potential spectroscopic binary candidates among sources with multi-epoch observations, and make preliminary orbit fits for 26 sources with four or more epochs of data.
We also examine rotational velocity statistics, including trends with spectral type and distributions of projected radii and rotation axis inclinations for sources with published photometric variability periods. 
We summarize our key results in Section~\ref{sec:sum}.

\section{Sample and Observations} \label{sec:sample}

\subsection{Sample Construction} \label{sec:sample_construct}

Our ultracool dwarf sample was curated from APOGEE DR17 \citep{Abdurrouf:2022aa} of SDSS-IV \citep{Blanton:2017aa} based on observations obtained with the 2.5-meter Sloan Foundation Telescope at the Apache Point Observatory \citep{Gunn:2006aa} and 2.5-meter du Pont Telescope at the Las Campanas Observatory \citep{Bowen:1973aa}.
The majority of our targets were proposed in two SDSS Ancillary Science Programs, SDSS-III Project 176 (PIs Suvrath Mahadevan and Cullen Blake; ``A Radial Velocity Survey of Bright M Dwarfs with APOGEE: Companions, {\vsini}, Fe/H'') and SDSS-IV Project 288 (PI: Adam Burgasser; ``APOGEE-2 and eBOSS Observations of the Lowest-Mass Stars and Brown Dwarfs in the Solar Neighborhood'')
\footnote{These programs are `APOGEE\_ANCILLARY', `APOGEE\_MDWARF', `APOGEE2\_ANCILLARY', `APOGEE2\_APOKASC', `APOGEE2\_CALIB\_CLUSTER', `APOGEE2\_CIS', `APOGEE2\_CNTAC', `APOGEE2\_GAIA\_OVERLAP', `APOGEE2\_GAIA\_OVERLAP', `APOGEE2\_K2', `APOGEE2\_MANGA\_LED', `APOGEE2\_MDWARF', `APOGEE2\_NORMAL\_SAMPLE', `APOGEE2\_ONEBIN\_GT\_0\_3', `APOGEE2\_SFD\_DERED', `APOGEE2\_SHORT', `APOGEE2\_ULTRACOOL', and `APOGEE2\_YOUNG\_CLUSTER'.}.
These programs are well-aligned with one of the main stellar science goals of the APOGEE survey, which is the identification of close binary systems and substellar companions through radial velocity variables \citep{Blanton:2017aa}.
We drew all of our sources from the DR17 allStar catalog,\footnote{Available at \url{https://data.sdss.org/sas/dr17/apogee/spectro/aspcap/dr17/synspec/allStar-dr17-synspec.fits}.}
and constructed two UCD samples for our analysis dubbed the ``gold'' and ``full'' samples. The analysis in this paper focuses on the gold sample. We present the construction and measurements of the full sample in the Appendix~\ref{appendix:full_sample}. 

\begin{figure}
\centering
\includegraphics[width=\linewidth, trim=20 10 10 0]{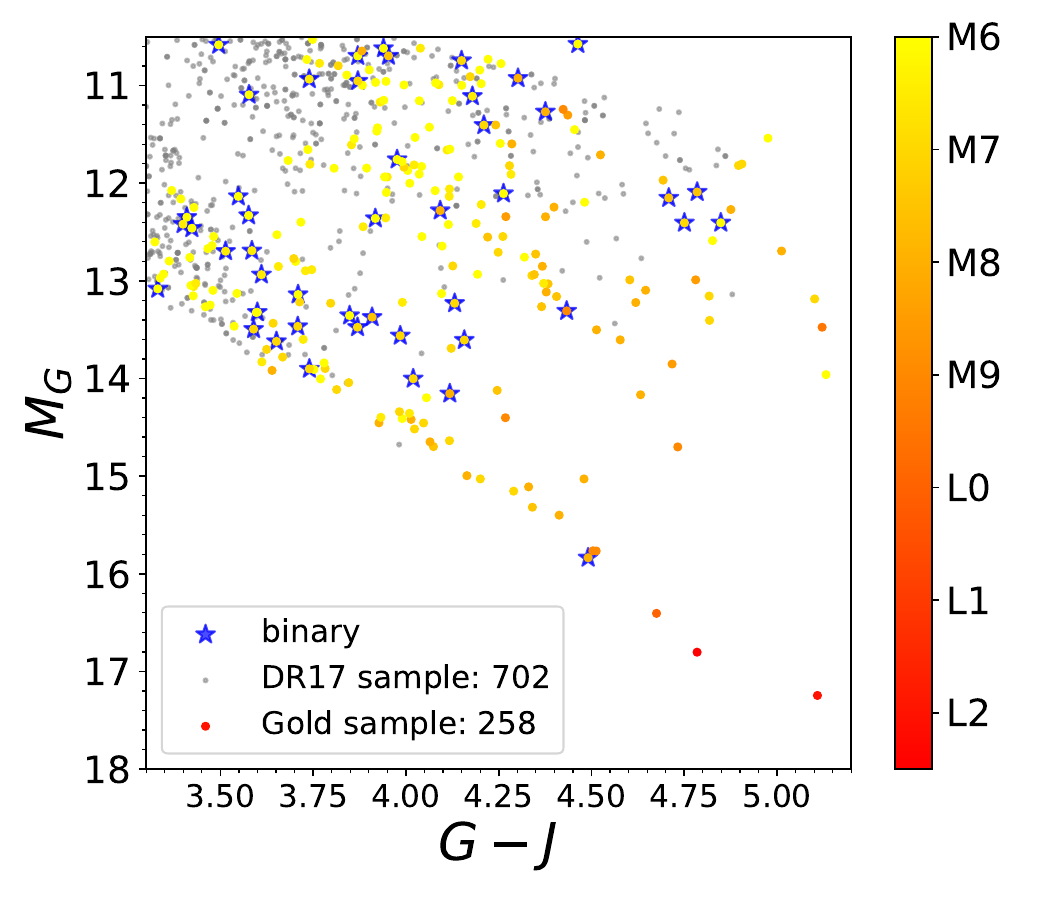}
\includegraphics[width=\linewidth, trim=20 10 10 0]{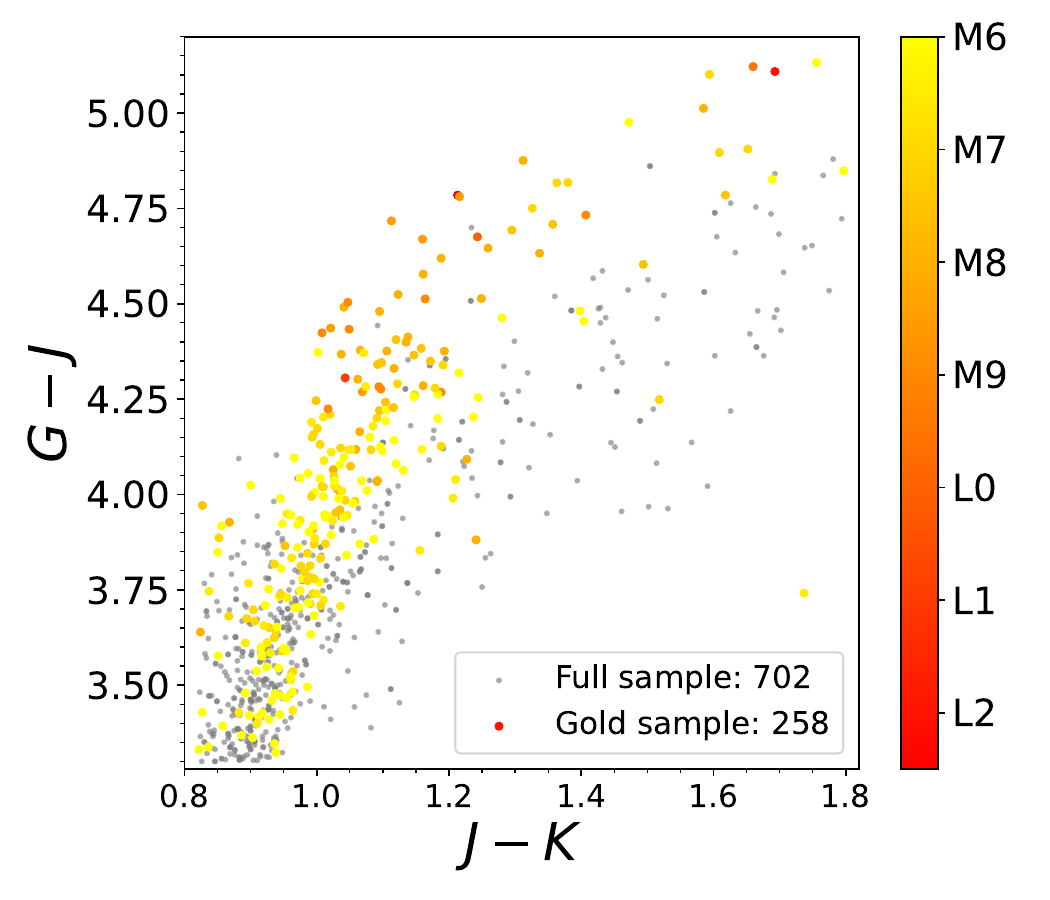}
\caption{Color-magnitude properties of the UCD APOGEE sample.
The full sample is indicated by grey dots, the gold sample by dots color-coded with spectral type. Blue stars highlight spectroscopic binary candidates.
\textit{\replaced{top right}{Top}}: M$_{G}$ versus $G-J$.
\textit{\replaced{bottom}{Bottom}}: $G-J$ versus $J-K$.
}
\label{fig:JK_MJ}
\end{figure}

Our ``gold'' sample was based on sources with classifications of M6 and later as reported in SIMBAD \citep{Wenger:2000aa}, 
the Late-Type Extension to MoVeRS (LaTE-MoVeRS; \citealp{Theissen:2016aa,Theissen:2017aa}), \cite{Reyle:2018aa}, and \cite{Best:2021aa}, or reported in this work.
We use the most recent spectroscopic classification as our adopted value.
Two sources where rejected from the sample based classifications earlier than M6 from additional follow-up spectroscopy discussed below.
We validated the remaining sources by visually inspecting 
optical (SDSS) and near-infrared images (2MASS), and confirmed correct placement on color-magnitude and color-color diagrams combining 2MASS and \textit{Gaia} EDR3 \citep{Gaia-Collaboration:2021ac} photometry and astrometry (Figure~\ref{fig:JK_MJ}).
We further imposed a limit on the median spectral signal-to-noise ratio (SNR) of the APOGEE data to be $>$10.
After removal of spectral type earlier than M6 using our optical spectra (Section~\ref{sec:classify}) and color-type relation from \cite{Kiman:2019aa},
these criteria resulted in a sample of 931 spectra of 258 sources summarized in Table~\ref{table:sample_gold}, which is the gold sample.

\begin{figure*}[!htpb]
\centering
\includegraphics[width=0.45\linewidth]{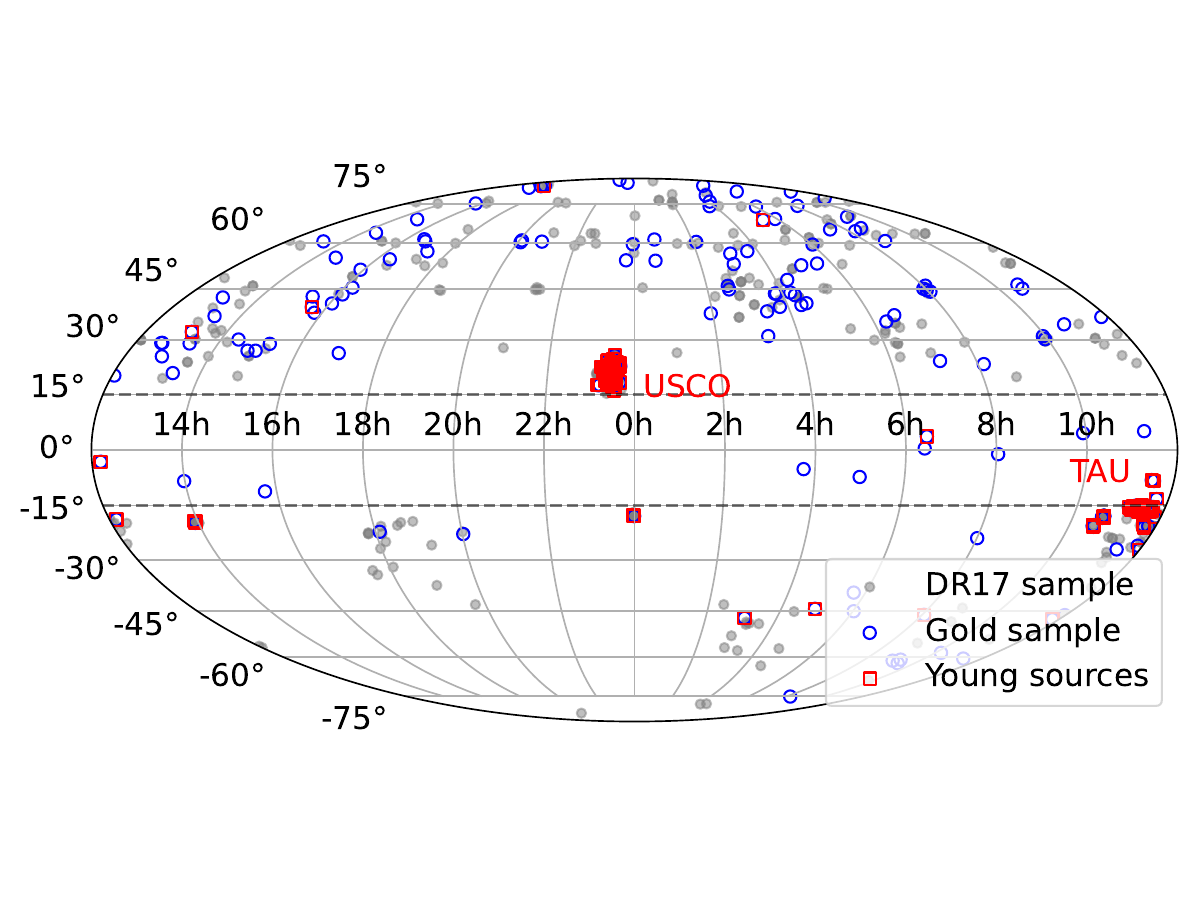}
\includegraphics[width=0.45\linewidth]{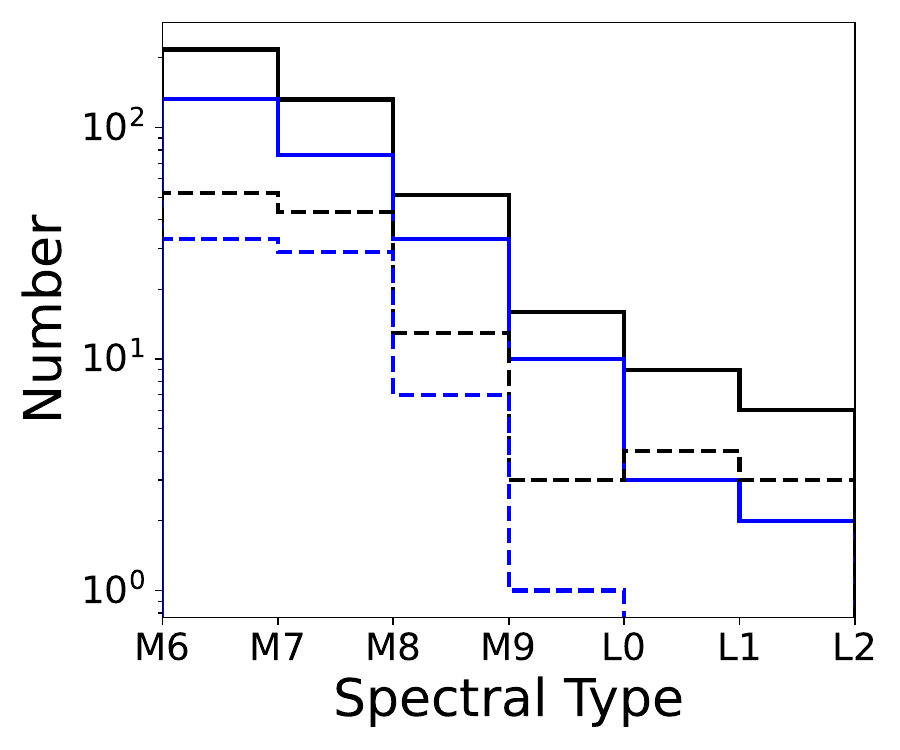} \\
\includegraphics[width=0.45\linewidth]{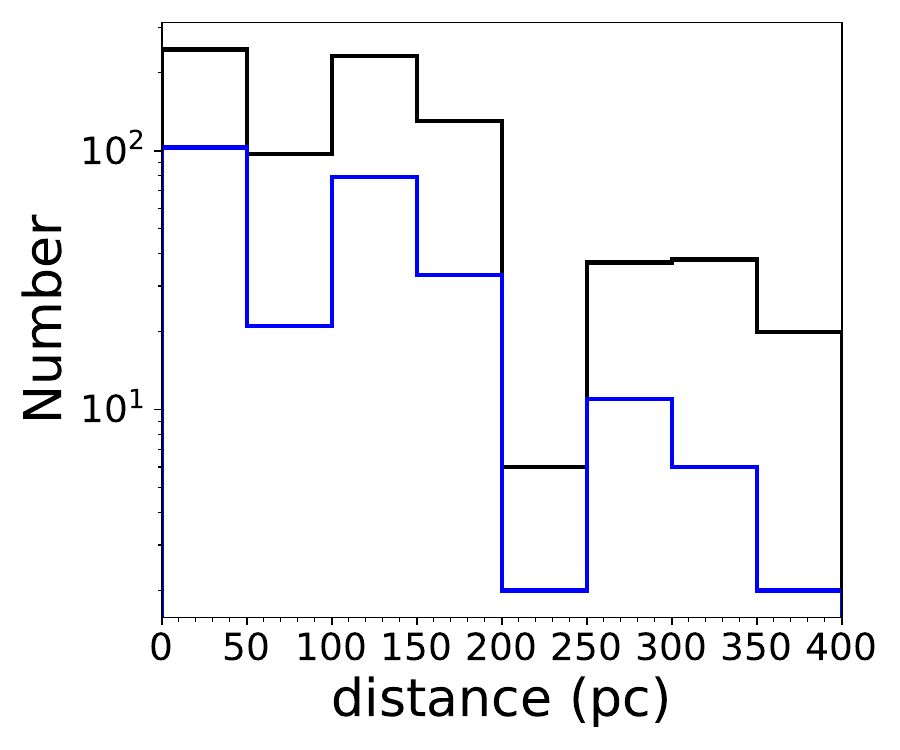}
\includegraphics[width=0.45\linewidth]{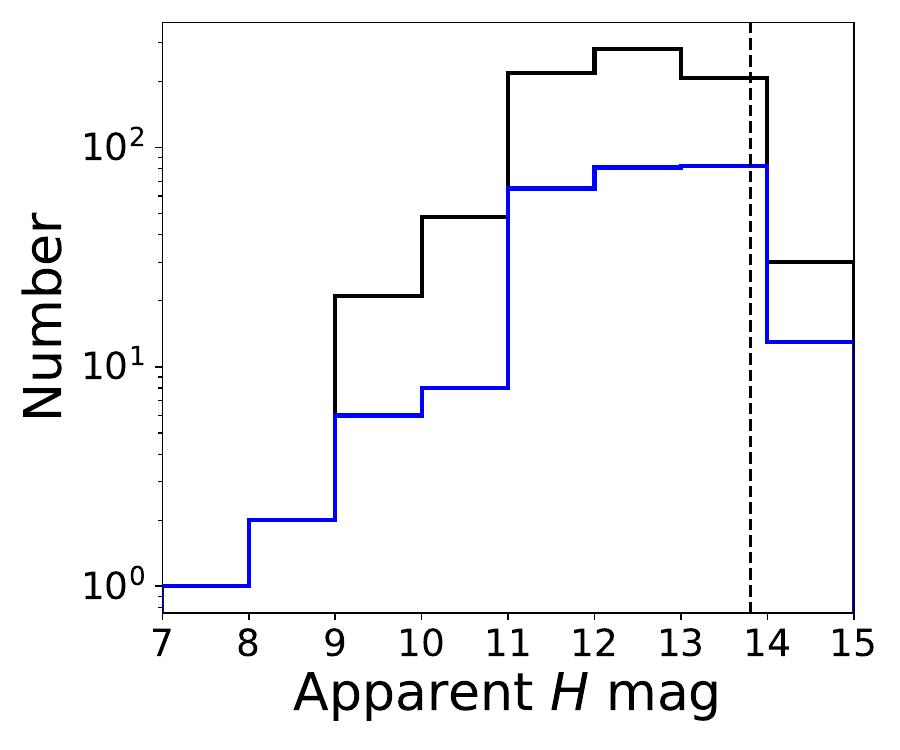}
\caption{Distributions of observables for the APOGEE UCD sample.
Gold sample sources are indicated by blue lines and symbols, 
full sample sources are indicated by black lines and symbols, 
\textit{Top left}: Sky distribution of our samples in Galactic coordinates. 
Known young sources in our gold sample are highlighted by red squares, and the major two clusters Upper Scorpius (USCO) and Taurus (TAU) are labeled in red texts near the cluster locations.
The Galactic latitudes at $+/-$15$^{\circ}$ are shown in grey dashed lines.
\textit{Top right}: Spectral type distributions for our samples, and gold sample spectral types drawn from the literature (see Table~\ref{table:sample_gold} for references) or observations presented here, and full sample spectral types estimated from the \citet{Kiman:2019aa} \textit{Gaia} $G-G_\mathrm{RP}$/spectral type relation. 
Dashed line histograms indicate the distributions of sources with four or more observation epochs. 
\textit{Bottom left}: {\it Gaia} EDR3 parallax measurement distributions for our samples. 
Note that the maximum distance value is 400~pc.
The single source within 10~pc ($\pi > 100$~mas) is the M6.5Ve G 51$-$15 (2MASS J08294949+2646348).
\textit{Bottom right}: Apparent 2MASS $H$-band magnitude distributions for our samples. The SNR limit = 100 ($H$ $\sim$ 13.8~mag) is labeled in the vertical black dashed line.
}
\label{fig:sample}
\end{figure*}

Figure~\ref{fig:sample} displays the observable properties of our gold and full samples. 
The distribution of targets across the sky is highly dependent on APOGEE's survey pointings. Notably, half of the gold sample includes known members of nearby young moving groups (the Upper Scorpius and Taurus young associations),
targeted as part of the programs described in \cite{Zasowski:2013aa, Zasowski:2017aa, Beaton:2021aa} and \citet{Santana:2021aa}.
The spectral types for our gold sample range from M6 to L2, with 252 late-M dwarfs and 6 L dwarfs.
For the full sample that is not in the gold sample, the \textit{Gaia} $G-G_\mathrm{RP}$/spectral type relation of \cite{Kiman:2019aa} indicates spectral types of M4 to L4, with 184 late-M dwarfs and 17 L dwarfs.
The majority of our gold sample (80\%) and full sample (90\%) have distances larger than 30~pc, with the M6.5Ve G 51$-$15 (2MASS J08294949+2646348) being the closest source at a \textit{Gaia} distance of 3.5810$\pm$0.0008~pc. 
The apparent $H$-band magnitudes of our targets extend to 14.65, which is slightly fainter than the APOGEE magnitude limit of 13.8 for SNR = 100.

The majority of both samples (66\% and 51\%) have APOGEE spectral observations taken over multiple epochs, enabling more precise determinations of radial velocities and the possibility of measuring radial velocity variations. We defined multi-epoch subsamples in our gold and full samples as those with at least four observations satisfying SNR $>$ 10, each separated by at least 0.85 day. A total of 71 sources in the gold sample and 115 sources in the full sample satisfy these criteria, and the radial velocity variables among the subsample are discussed in Section~\ref{sec:binaries}.

Finally, we note that the requirement for a \textit{Gaia} detection likely resulted in the rejection of true UCDs in the full sample due to sensitivity limitations \citep{Theissen:2018aa},
and APOGEE targeting was not intended to be uniform across the sky \citep{Zasowski:2013aa}. 
Therefore, our APOGEE UCD sample is not expected to be magnitude- or volume-complete, but rather representative of the broader UCD population. 

\begin{longrotatetable}
\begin{deluxetable*}{cllllrrrrrrrr}
\tablecaption{APOGEE DR17 Sample\label{table:sample_gold}}
\tablewidth{700pt}
\tabletypesize{\scriptsize}
\tablehead{
\colhead{} & \colhead{} & \colhead{} & \colhead{} & \colhead{} & \colhead{} & \colhead{} & \colhead{} & \colhead{} &
\colhead{} & \colhead{} & \colhead{} & \colhead{} \\
\colhead{APOGEE ID} & 
\colhead{RA} & \colhead{Dec} & 
\colhead{$Gaia$ eDR3 Source ID} & 
\colhead{SpT} & \colhead{SpT Ref} & \colhead{$N_\text{obs}$} & \colhead{2MASS $H$} & 
\colhead{$\mu_{\alpha}$} & \colhead{$\mu_{\delta}$} & \colhead{$\mu$ Ref} & \colhead{$\pi$} &  \colhead{$\pi$ Ref} \\ 
\colhead{} & \colhead{(deg)} & \colhead{(deg)} & \colhead{} & \colhead{} & \colhead{} & \colhead{} & \colhead{(mag)} & \colhead{(mas yr$^{-1}$)} & \colhead{(mas yr$^{-1}$)} &
\colhead{} & \colhead{(mas)} & \colhead{}
} 
\startdata
2MASS J00034394$+$8606422 & 0.933088 & $+$86.111732 & 574059045248283008 & M6.0 & (93) & 10 & $11.738 \pm 0.031$ & $175.94 \pm 0.04$ & $-25.83 \pm 0.04$ & (99) & $30.81 \pm 0.03$ & (99) \\ 
2MASS J00312793$+$6139333 & 7.866401 & $+$61.659252 & 430215470915266560 & M7 & (79, 97) & 6 & $12.483 \pm 0.028$ & $369.52 \pm 0.07$ & $158.0 \pm 0.07$ & (99) & $27.86 \pm 0.07$ & (99) \\ 
2MASS J00381273$+$3850323 & 9.553082 & $+$38.842308 & 368514898441866368 & M6 & (91) & 6 & $12.459 \pm 0.033$ & $163.84 \pm 0.05$ & $12.5 \pm 0.05$ & (99) & $17.21 \pm 0.06$ & (99) \\ 
2MASS J00452143$+$1634446 & 11.339304 & $+$16.579082 & 2781513733917711616 & L2beta & (44) & 3 & $12.059 \pm 0.035$ & $359.07 \pm 0.2$ & $-47.91 \pm 0.14$ & (99) & $65.41 \pm 0.18$ & (99) \\ 
2MASS J00514593$-$1221458 & 12.941377 & $-$12.362734 & 2376328546438257536 & M6.1 & (93) & 2 & $11.89 \pm 0.022$ & $-32.76 \pm 0.06$ & $-38.85 \pm 0.06$ & (99) & $26.36 \pm 0.04$ & (99) \\ 
2MASS J01154176$+$0059317 & 18.924037 & $+$0.992157 & 2535273088355573120 & M6 & (100) & 2 & $12.651 \pm 0.027$ & $163.82 \pm 0.15$ & $24.1 \pm 0.13$ & (99) & $15.77 \pm 0.13$ & (99) \\ 
2MASS J01215816$+$0101007 & 20.492347 & $+$1.016877 & 2535133209860543616 & M6.5 & (100) & 2 & $12.429 \pm 0.026$ & $161.94 \pm 0.16$ & $-78.8 \pm 0.09$ & (99) & $26.49 \pm 0.12$ & (99) \\ 
2MASS J01243124$-$0027556 & 21.130173 & $-$0.465459 & 2533946665015570304 & M7 & (51) & 2 & $11.506 \pm 0.029$ & $66.72 \pm 0.32$ & $-170.47 \pm 0.18$ & (99) & $29.17 \pm 0.25$ & (99) \\
\enddata
\end{deluxetable*}
\end{longrotatetable}

\subsection{Additional Spectral Observations} \label{sec:classify}

\begin{figure}[!htpb]
\centering
\includegraphics[width=\linewidth]{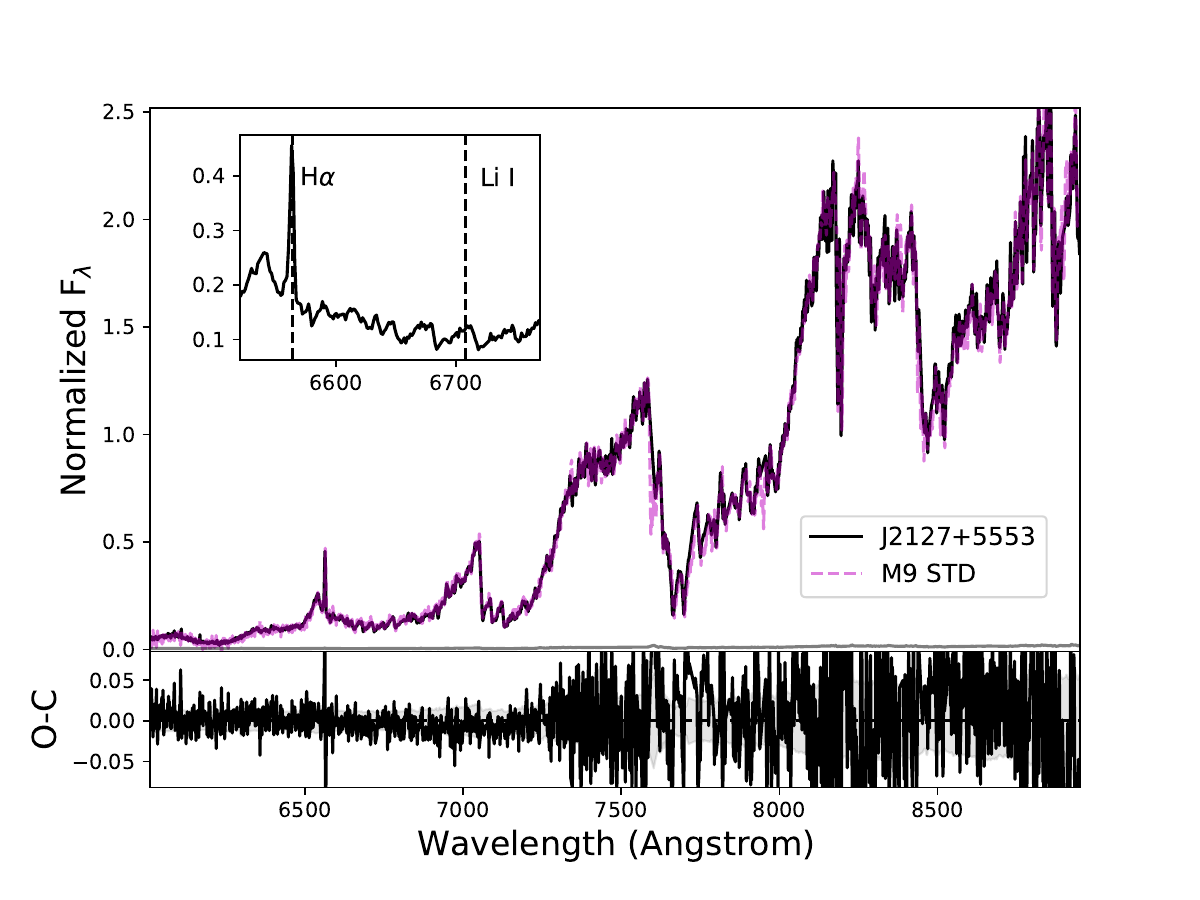}
\caption{Normalized Shane/Kast spectrum of 2MASS J21272531+5553150 (black line) compared to the best-match M9 spectral template from \citet[magenta line]{Bochanski:2007ab}.
The lower panel compares the difference between these spectra to the measurement uncertainty (grey band). The inset box highlights the region around H$\alpha$ emission at 6563~{\AA} and Li~I absorption at 6708~{\AA}.
}
\label{fig:kast}
\end{figure}

The majority of sources in our full sample lack published spectroscopic classifications.\footnote{High-resolution spectra are not ideal for spectral classifications; see Section~\ref{sec:teff_logg}.}
To bolster this sample, we obtained additional low-resolution optical spectra with the Kast Double Spectrograph \citep{Miller:1994aa} on the Shane 3-m Telescope at the Lick Observatory over multiple nights between 2018 January 21 and 2022 March 11. We used the 600/7500 grating and 2$\arcsec$ slit to obtain 6000-9000~{\AA} spectra at an average resolution of $\lambda/\Delta\lambda \approx 1800$. Data acquisition included observations of flat-field and arc lamps for pixel response and wavelength calibration, nightly observations of a spectral flux standard from \citet{Hamuy:1992aa,Hamuy:1994aa} for relative flux calibration, and observations of a nearby G2~V or A0~V star at similar airmass for telluric absorption and continuum correction. All data were reduced using the \texttt{kastredux} package\footnote{\url{https://github.com/aburgasser/kastredux}.} using default settings.
An example spectrum of the M9 dwarf 2MASS J21272531+5553150 (aka LSPM J2127+5553) is shown in Figure~\ref{fig:kast}.
Table~\ref{table:kast} summarizes the observations and corresponding measurements. 
Spectral classifications were determined by the closest match to SDSS dwarf spectral templates from \citet{Bochanski:2007ab,Schmidt:2014aa}, and \citet{Kesseli:2017aa}, and indicate types ranging from M5 to M9.
Metallicity index ($\zeta$)
measurements \citep{Lepine:2007ab} are uniformly greater than 0.9 and indicate near-solar metallicities for all sources.
We detected H$\alpha$ emission in all of the sources, and measured relative H$\alpha$ emission luminosity ({\lhalbol}) from equivalent widths and the $\chi$ factor relations of \citet{Douglas:2014aa} and \citet{Schmidt:2014aa}; these values range over $-4.9 <$ {\lhalbol} $< -3.8$. 


\begin{deluxetable*}{llccclcc}
\tablecaption{Shane/Kast Observations of APOGEE Targets\label{table:kast}}
\tablecolumns{4}
\tablehead{
\colhead{2MASS Source ID} & 
\colhead{Obs.\ Date} & 
\colhead{Airmass} & 
\colhead{Exp.\ Time} & 
\colhead{S/N\tablenotemark{a}} & 
\colhead{SpT\tablenotemark{b}} & 
\colhead{$\zeta$\tablenotemark{c}} & 
\colhead{\lhalbol\tablenotemark{d}} \\
\colhead{} & 
\colhead{(UT)} & 
\colhead{} &
\colhead{(s)} & 
\colhead{} & 
\colhead{} & 
\colhead{} & 
\colhead{} \\
}
\startdata
2MASS J14554964+0321420 & 2021 May 15 &     1.21 & 2400 &   73 & M5.0 & 1.195$\pm$0.004 & $-$4.81$\pm$0.12 \\
2MASS J15042797+0942464 & 2021 May 15 &     1.42 & 3000 &   47 & M5.0 & 1.093$\pm$0.007 & $-$3.75$\pm$0.08 \\
2MASS J07552256+2755318 & 2021 Nov 27 &     1.02 & 3000 &  130 & M6.0 & 1.118$\pm$0.002 & $-$4.01$\pm$0.09 \\
2MASS J11210854+2126274 & 2021 May 15 &     1.05 & 2400 &   97 & M6.0 & 1.048$\pm$0.002 & $-$4.11$\pm$0.09 \\
2MASS J08080189+3157054 & 2021 Jan 17 &     1.11 & 3000 &   55 & M7.0 & 1.007$\pm$0.003 & $-$4.22$\pm$0.16 \\
2MASS J12215013+4632447 & 2018 Jan 21 &     1.02 & 2400 &  156 & M7.0 & 1.008$\pm$0.001 & $-$4.17$\pm$0.16 \\
2MASS J13342918+3303043 & 2020 Feb 05 &     1.02 & 3600 &   84 & M7.0 & 1.031$\pm$0.002 & $-$4.91$\pm$0.12 \\
2MASS J16572919+2448509 & 2022 Mar 11 &     1.06 & 3000 &  113 & M7.0 & 0.919$\pm$0.002 & $-$4.05$\pm$0.13 \\
2MASS J21381698+5257188 & 2020 Dec 14 &     1.22 & 3000 &   85 & M7.0 & 1.005$\pm$0.002 & $-$4.17$\pm$0.14 \\
2MASS J12493960+5255340 & 2021 May 16 &     1.06 & 3000 &   16 & M9.0 & 1.506$\pm$0.043 & $-$4.93$\pm$0.14 \\
2MASS J19544358+1801581 & 2020 Aug 14 &     1.07 & 3600 &   98 & M9.0 & 1.207$\pm$0.003 & $-$4.70$\pm$0.11 \\
2MASS J21272531+5553150 & 2020 Aug 14 &     1.06 & 3600 &   96 & M9.0 & 1.080$\pm$0.003 & $-$4.78$\pm$0.10 \\
\enddata
\tablenotetext{a}{Median signal-to-noise ratio in the 7200--7400~{\AA} region.}
\tablenotetext{b}{Closest match to the SDSS dwarf spectral templates defined in \citet{Bochanski:2007ab,Schmidt:2014aa}, and \citet{Kesseli:2017aa}.}
\tablenotetext{c}{Metallicity index defined in \citet{Lepine:2007ab}, where $\zeta > 0.875$ indicates a dwarf (solar) metallicity classification.}
\tablenotetext{d}{Relative luminosity in H$\alpha$ emission based on the measured H$\alpha$ equivalent width and $\chi$ correction factors compiled by \citet{Douglas:2014aa} and \citet{Schmidt:2014aa}.}
\end{deluxetable*}

\section{APOGEE Spectral Analysis} \label{sec:methods}

\subsection{Spectral Data} \label{sec:apogee-data}

Spectra were drawn from the APOGEE single epoch, individual visit spectra (apVisit) files \citep{Abdurrouf:2022aa}, covering chips a (1.657--1.696 \micron), b (1.585--1.644 \micron), and c (1.514--1.581 \micron).
APOGEE data reduction is described in detail in \cite{Nidever:2015aa, Holtzman:2018aa, Jonsson:2020aa}; and \citet{Abdurrouf:2022aa}.
Each apVisit spectrum underwent dark, flat-field, cosmic ray, flux, sky, and telluric corrections, as well as wavelength calibrations.
In particular, the wavelength solutions were derived from a combination of sky lines and ThArNe and UNe hollow-cathode lamps \citep{Nidever:2015aa}.
Bad pixels were masked out using the bit mask for each chip\footnote{We specifically masked data with integer bit mask values of ``0'' (BADPIX; bad pixel mask or pixels from strong persistence jump), ``1'' (CRPIX; cosmic ray contaminated pixel), ``2'' (SATPIX; saturated pixel), ``3'' (UNFIXABLE; unfixable pixel), ``4'' in (BADDARK; bad pixels from dark frames), ``5'' (BADFLAT bad pixels from flat-field lamp frames), ``6'' (BADERR; pixels with high error), ``12'' (SIG\_SKYLINE; pixels near the large flux from sky lines), and ``14'' (NOT\_ENOUGH\_PSF; less than 50\% of point-spread function in good pixels).}.
In order to simultaneously calibrate the wavelength solution imprinted in the spectra, the telluric absorption profile has been retained for each reduced apVisit spectrum (HDU7).

\subsection{Forward Modeling} \label{sec:modeling}

To infer the physical properties of our sources---effective temperature ({\teff}), surface gravity ({\logg}), radial velocity (RV), and rotational velocity ({\vsini})---we employed a Markov Chain Monte Carlo (MCMC) forward-modeling technique that simultaneously models the telluric and stellar absorption present in each APOGEE apVisit spectrum. We used the \texttt{Spectral Modeling Analysis and RV Tool} (\texttt{SMART}; \citealp{Hsu:2021ab}), which follows methods previously described in \citet{Blake:2010aa, Burgasser:2016aa, Hsu:2021aa}; and \citet{Theissen:2022aa}.

Each APOGEE spectrum is forward-modeled using the following equation:
\begin{multline}
\begin{aligned}
D[p] = \Bigg( C[p] \times \Bigg[ \bigg(M \Big[p^* \big(\lambda \big[ 1 + \frac{RV^*}{c}\big] \big) , T_\text{eff}, \log{g} \Big] \\
\ast \kappa_D (\nu_\text{micro}) \ast \kappa_R (v\sin{i} ) \bigg) \times T \big[ p^*(\lambda, \mathrm{AM}, \mathrm{PWV}) \big] \Bigg] \\ \ast \kappa_G (\Delta \nu_\text{inst}) \Bigg)  + C_\text{flux} \, ,
\end{aligned}
\end{multline}
Here, $D[p]$ is the data model as a function of pixel $p$; 
$C[p]$ is a fifth-order polynomial representing a continuum correction;
$M[p]$ is the stellar solar-metallicity atmosphere model parameterized by {\teff} and {\logg}; 
$p^*(\lambda)$ is a wavelength-to-pixel conversion function, initially provided by the APOGEE reduction pipeline with an additional constant offset parameter $C_{\Delta\lambda}$ to adjust for chip-to-chip variations;
$RV^* = RV + v_\text{bary}$ is the radial velocity of the source plus barycentric motion of the Earth at the observed epoch; 
$c$ is the speed of light; 
$\kappa_D$ is a Gaussian 
convolution kernel that applies a microturbulence velocity broadening $\nu_\text{micro}$, modeled as 2.478 $-$ 0.325 $\times \log{g}$ {\kms} \citep{Zamora:2015aa} 
; $\kappa_R$ is a rotational line broadening convolution kernel for projected rotational velocity {\vsini}, with a linear limb-darkening coefficient $\epsilon$ = 0.6 \citep{Gray:1992aa};
$T[p]$ is the telluric absorption model based on the model grid of \citet{Moehler:2014aa} and parameterized by airmass (AM) and precipitable water vapor (PWV);
$\kappa_G$ is a Gauss-Hermite convolution kernel used to account for the instrumental line spread function (LSF), with width $\nu_\text{inst}$ obtained from the APOGEE pipeline \citep{Nidever:2015aa,Bovy:2016aa}; 
and $C_\text{flux}$ is a constant additive flux offset. 

The log-likelihood function of our model fit was defined as
\begin{equation}
\ln \, \mathcal{L} = -0.5 \times \left[ \sum \chi^2/C_\text{noise}^2 + \sum{\ln ( 2 \pi (C_\text{noise} \sigma)^2 ) } \right],
\end{equation}
where the statistic
\begin{equation}
\chi^2 = \sum_{i=1}^{N}\frac{(S[p]-{\alpha}D[p])^2}{\sigma[p]^2}
\end{equation}
compares the observed spectrum $S[p]$ and uncertainty $\sigma[p]$ to the scaled forward model $D[p]$, with the scale factor $\alpha$ determined to minimize $\chi^2$. We include the constant scaling factor $C_{\mathrm{noise}}$ to account for under estimation of observational noise, as well as systematic errors such as missing line features.

Cool dwarfs have abundant molecular absorption lines in their infrared spectra, and hence careful consideration must be made for the choice of stellar model.
We explored 
synthetic model grids from 
\citet[BT-Settl CIFIST models]{Baraffe:2015aa}, 
\citet[ACES models]{Husser:2013aa}, 
\citet[MARCS models]{Gustafsson:2008aa}, and  
\citet[Sonora models]{Marley:2018aa}.
For the lowest-temperature dwarfs (L dwarfs), 
we found that the Sonora models outperform other model sets, particular redward of 1.58~$\micron$ where FeH absorption is an important source of opacity \citep{Cushing:2003aa,Souto:2017aa}, although there are still some missing features in chip c.

For each source and model set, we fit all three chips simultaneously, with the nuisance parameters $C_\text{flux}$, $C_{\Delta \lambda}$, and continuum parameters modeled separately for each chip. 
The 18 continuum parameters, LSF broadening, $\nu_\text{micro}$, and $v_\text{bary}$ were fit and applied at the end of the MCMC loop, leaving 13 parameters to be fit by the forward-modeling routine, summarized in Table~\ref{table:mcmc_parameter}.
Priors for these parameters were assumed to be uniformly distributed between bounds based on expected values for late-M and L dwarfs or model parameter limits. 
We used the \texttt{emcee} code \citep{Foreman-Mackey:2013aa} to run the MCMC with kernel-density estimator (KDE)\footnote{\url{https://docs.scipy.org/doc/scipy/reference/generated/scipy.stats.gaussian_kde.html}} moves, which allows efficient convergence of best-fit parameters. 
We deployed 100 chains of 1000 steps each, with the first 800 steps removed for ``burn-in''. 
At step 600, we used a 3-sigma-clipping mask of data minus model residuals to remove outlying flux values, which were mostly unmasked bad pixels and discrepant opacities.
Chains were visually inspected to ensure convergence, and the typical integrated auto-correlation range was $\sim$17 steps.
We report best-fit parameters as the median of the tails of the chains, with uncertainties derived from the 16\% to 84\% quantile range.

We note that one-half of our sample are young sources, and hence magnetic emission as traced by the Bracket hydrogen emission series (e.g., 
Br 16$\rightarrow$4, $\lambda$ = 1.556~{\micron}; 
Br 15$\rightarrow$4, $\lambda$ = 1.571~{\micron}; 
Br 13$\rightarrow$4, $\lambda$ = 1.611~{\micron}; 
Br 11$\rightarrow$4, $\lambda$ = 1.681~{\micron}; 
see \citealt{Sullivan:2019aa}) could be present in some sources. None of the models used include hydrogen emission; however, 
any such emission was much weaker compared to other absorption lines, as verified by visual inspection of each best-fit spectrum\footnote{The residual sky lines from the APOGEE pipeline have stronger fluxes compared to possible Bracket hydrogen emission series, but were largely masked out in our sigma clipping process.}. 

\begin{deluxetable}{lccc}
\tablecaption{Forward Modeling Parameters \label{table:mcmc_parameter}}
\tablecolumns{4}
\tablehead{
\colhead{Description} &  \colhead{Symbol (unit)}  & \colhead{Priors\tablenotemark{a}} & \colhead{Bounds}
}
\startdata
Temperature & $T_{\mathrm{eff}}$ (K) & (1800, 4000)\tablenotemark{b} & (1200, 7000)\tablenotemark{b} \\
\nodata & \nodata & (1500, 2400)\tablenotemark{c} & (200, 2400)\tablenotemark{c} \\
Surface Grav. & $\log{g}$ (cm s$^{-2}$) & (3.5, 5.5) & (3.5, 5.5) \\
Rot. Vel. & {\vsini} ({\kms}) & (0, 50) & (0, 100) \\
Radial Vel. & RV ({\kms}) & ($-$100, +100) & ($-$200, +200) \\
Flux Offset\tablenotemark{d} & $C_{\mathrm{flux}}$ & ($-$0.01, +0.01) & ($-$10$^4$, +10$^4$) \\
Wave Offset\tablenotemark{d} & $C_{\Delta \lambda}$ ({\AA}) &  ($-$0.1, +0.1) & ($-$0.5, +0.5) \\
Airmass & $AM$ & (1.0, 3.0) & (1.0, 3.0) \\
Water Vapor & $PWV$ (mm) & (0.5, 20.0) & (0.5, 20.0) \\
Noise Factor & $C_{\mathrm{noise}}$ & (1.0, 5.0) & (1.0, 10.0)
\enddata
\tablenotetext{a}{All priors assume a uniform distribution over range specified}
\tablenotetext{b}{{\teff} range for BT-Settl models}
\tablenotetext{c}{{\teff} range for Sonora models}
\tablenotetext{d}{The three APOGEE chips were fit individually.}
\end{deluxetable}

\begin{figure*}[h]
\centering
\includegraphics[width=\textwidth]{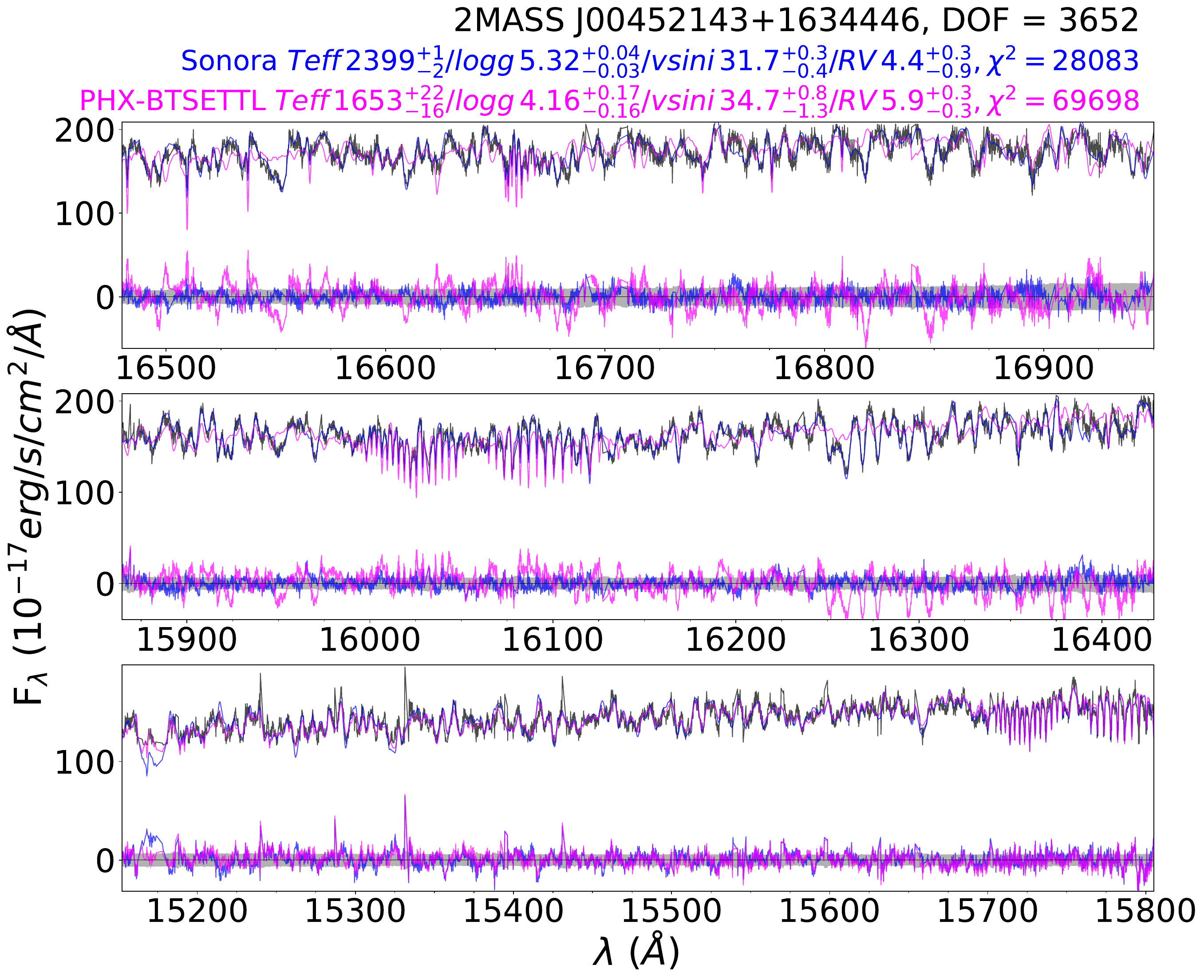}
\caption{Spectrum and best-fit forward models for the APOGEE spectrum of the L2$\beta$ 2MASS J00452143+1634446 observed on JD of 2456587.736. 
The APOGEE data are labeled in black, and the best-fit forward models with Sonora and BT-Settl models are labeled in blue and magenta, respectively. The noise and residual (data$-$model) are depicted in grey-shaded regions and colored lines corresponding to the models, respectively. 
Parameters values and their uncertainties are listed at the top of the panel.
}
\label{fig:J0045_model_compare}
\end{figure*}

\begin{figure*}[h]
\centering
\includegraphics[width=\textwidth]{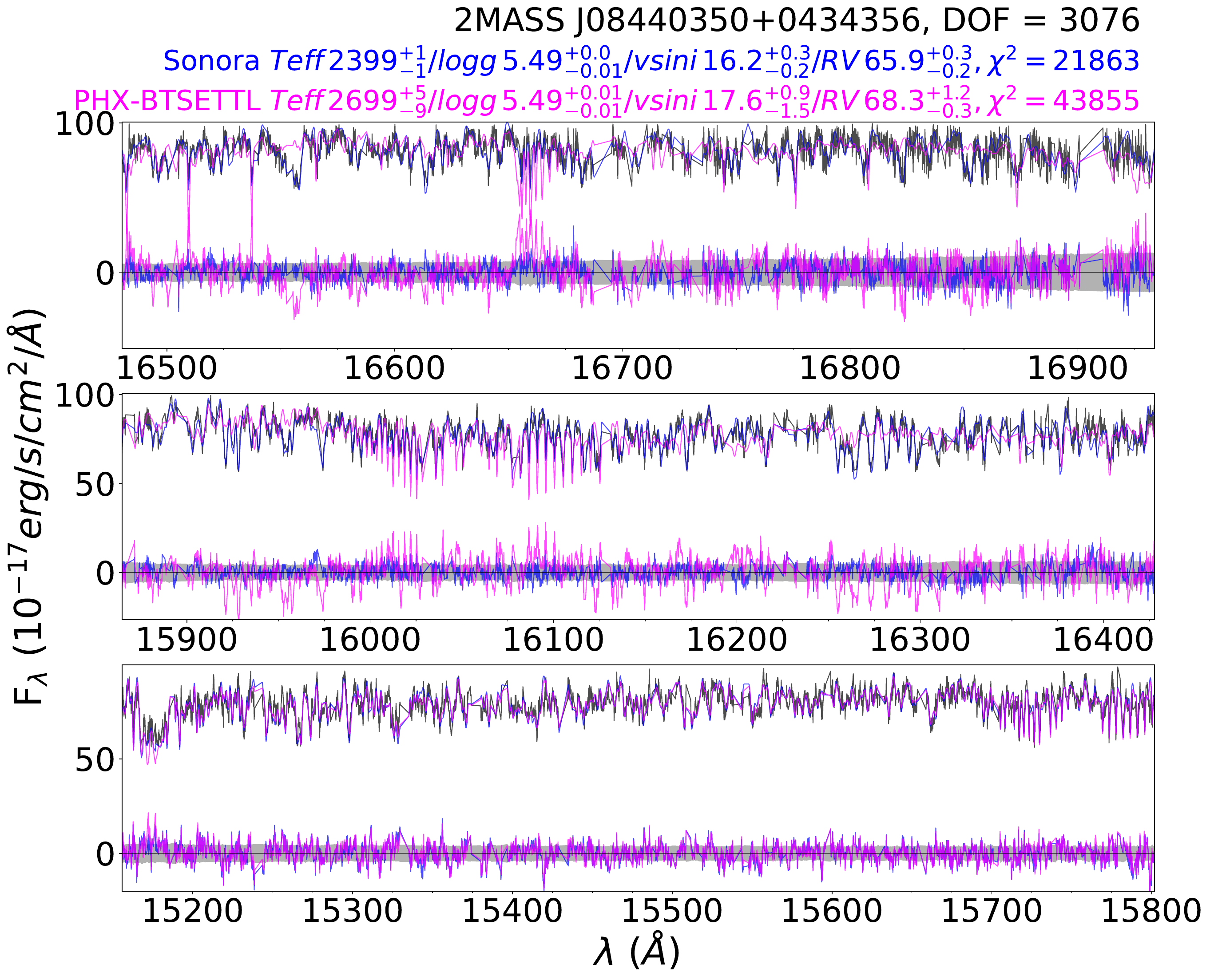}
\caption{Same as Figure~\ref{fig:J0045_model_compare} for the APOGEE spectrum of the M9 2MASS J08440350+0434356 observed on JD of 2458198.659.
}
\label{fig:J0844_model_compare}
\end{figure*}

\begin{figure*}[h]
\centering
\includegraphics[width=\textwidth]{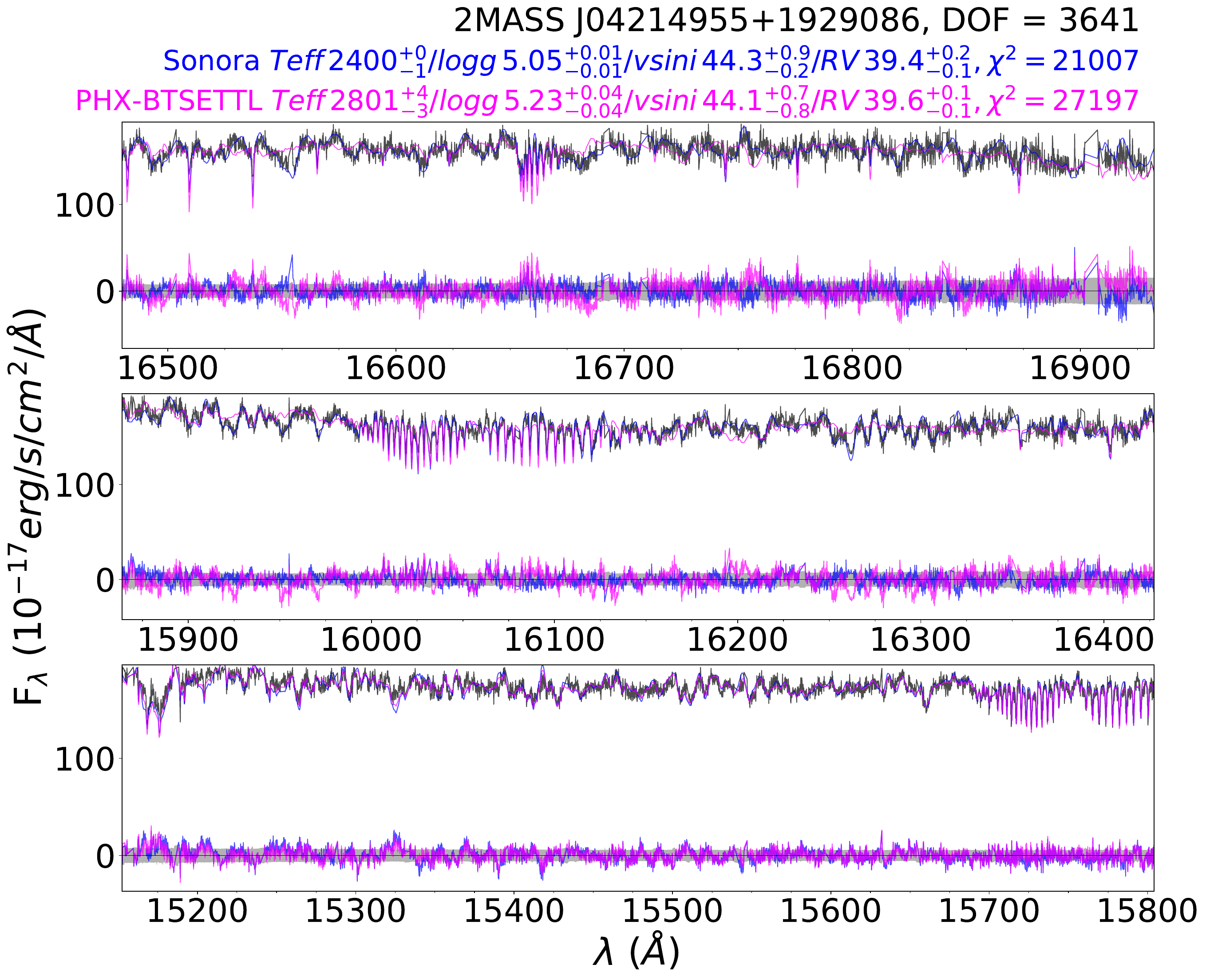}
\caption{Same as Figure~\ref{fig:J0045_model_compare} for the APOGEE spectrum of the M7+M9.5 2MASS J04214955+1929086 observed on JD of 2458820.725. 
}
\label{fig:J0421_model_compare}
\end{figure*}

To assess the quality of these fits, Figures~\ref{fig:J0045_model_compare}, \ref{fig:J0844_model_compare} and \ref{fig:J0421_model_compare} show the best-fit models using Sonora and BT-Settl models for the L2$\beta$ 2MASS J00452143+1634446, the M9 2MASS J08440350+0434356, and the M7+M9.5 binary 2MASS J04214955+1929086, respectively.
The Sonora models clearly outperform the other models based on the residuals and $\chi^2$ fit values, even though the best-fit $T_\mathrm{eff}$ reaches the $T_\mathrm{eff}$ limits of the Sonora model grid.
The best-fit models of the BT-Settl, ACES, and MARCS models give similar results, with
significant residuals redward of 1.58~{\micron} driving up $\chi^2$ values.
The Sonora models incorporate the \citet{Hargreaves:2010aa} \ion{FeH}{0} $E^4\Pi-A^4\Pi$ transitions (1.58263~$\micron$, 1.59188~$\micron$, 1.62457~$\micron$ bandhead covering from 1.58 to 1.75~$\micron$; \citealp{Wallace:2001aa, Cushing:2005aa}), and we thus attribute residuals in the other models to outdated \ion{FeH}{0} opacities.
The other models also yield lower $\log{g}$ and higher $v\sin{i}$ values than the Sonora models, which we interpret as compensation for these missing opacities; a similar effect for methane from \cite{Yurchenko:2014aa} has been noted in $J$-band spectra of T dwarfs \citep{Hsu:2021aa}.

For warmer sources (late-M dwarfs) the BT-Settl models provided superior fits to the Sonora and other models, although telluric absorption strengths are higher for these model fits, again suggesting compensation for missing opacity. Based on these initial fits, we focused our analysis on these two model sets, with an ``optimal model'' transition, as verified by visual inspection of each fit, around 2700~K $\lesssim$ {\teff} $\lesssim$ 3000~K.
At and below these transition temperatures, the Sonora models yield better fits
even at its {\teff} = 2400~K parameter limit.

\section{Results} \label{sec:stellar_parameters}

In this section, we review our RV, {\vsini}, $T_\text{eff}$, and $\log{g}$ measurements, all of which are compiled in Table~\ref{table:indmeasurement}\footnote{While not analyzed for the rest of the manuscript, we also provided measurements for an additional 444 candidate UCDs with 1543 epochs in Table~\ref{table:indmeasurement_full}.}.

\subsection{Radial Velocities} \label{sec:rv}

Our RV measurements span the range $-$76.08~{\kms} to $+$66.68~{\kms} with a median value of $-$0.78~{\kms} and a median measurement uncertainty of 0.32~{\kms}. 
To assess sources of systematic uncertainty, we evaluated the scatter in measurements inferred from 13 epochs of observations of the M2 2MASS J16495034+4745402, one of the RV standards in the \citet{Deshpande:2013aa} APOGEE sample.
We compared fits based on MARCS, BT-Settl, and ACES models, which had internal per-epoch scatter of 0.18~{\kms}, 0.19~{\kms}, and 0.16~{\kms}, respectively;
and an overall scatter between models of 0.18~{\kms}.
We therefore conservatively assume an overall systematic RV uncertainty of 0.19~{\kms}, which has been added in quadrature to the reported RV measurements for all sources, resulting in a 
final median RV precision of 0.37~{\kms}.

We explored whether this precision could be improved by modeling restricted wavelength regions where strong telluric absorption can be used to improve the wavelength calibration. 
This experiment was motivated by higher fitting scatter in regions that had by slight mismatches between the observed and modeled spectra, possibly due to offsets in the wavelength calibrations derived from arc lamp lines.
We fit regions of strong telluric absorption at
16560--16700~{\AA} on chip a, 15980--16160~{\AA} on chip b, and 15100--15500~{\AA} and 15500--16800~{\AA} on chip c (see Figure~\ref{fig:J0045_model_compare}), and included an additional pixel-to-wavelength zero-point offset term in our model.
The resulting variance in RV measurements between these regions, 0.23--0.56~{\kms}, was worse than fitting all three chips simultaneously, likely due to the lack of strong stellar lines. We conjecture that slight improvements in the APOGEE pipeline wavelength calibration could be realized by combining arc lines, sky emission lines, and telluric absorption lines to compute the overall wavelength calibration. 

\begin{figure*}[!htbp]
\gridline{\fig{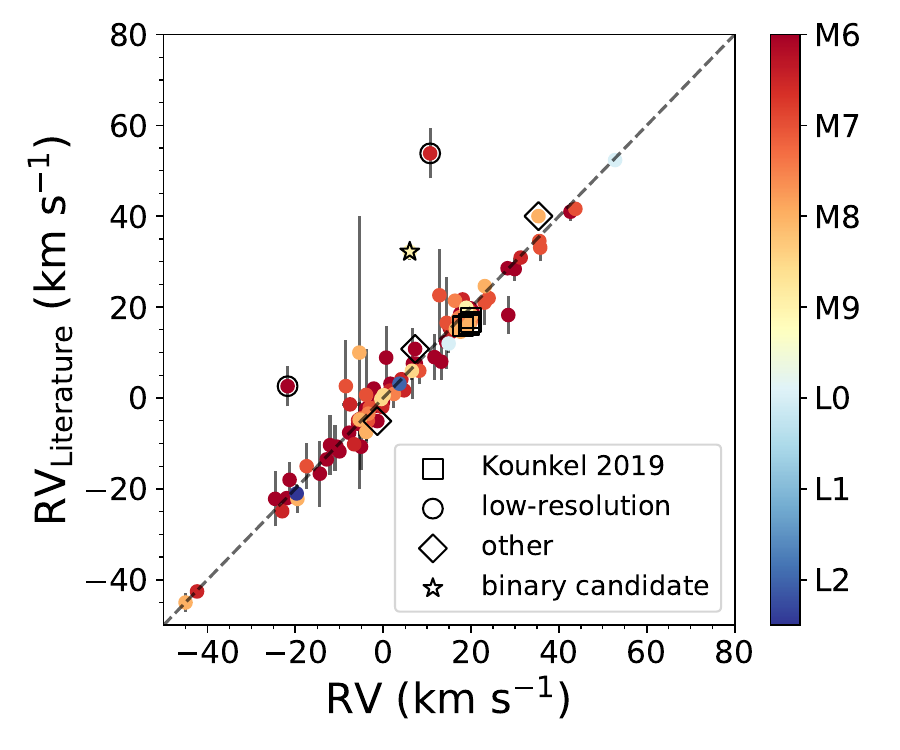}{0.5\textwidth}{}
\fig{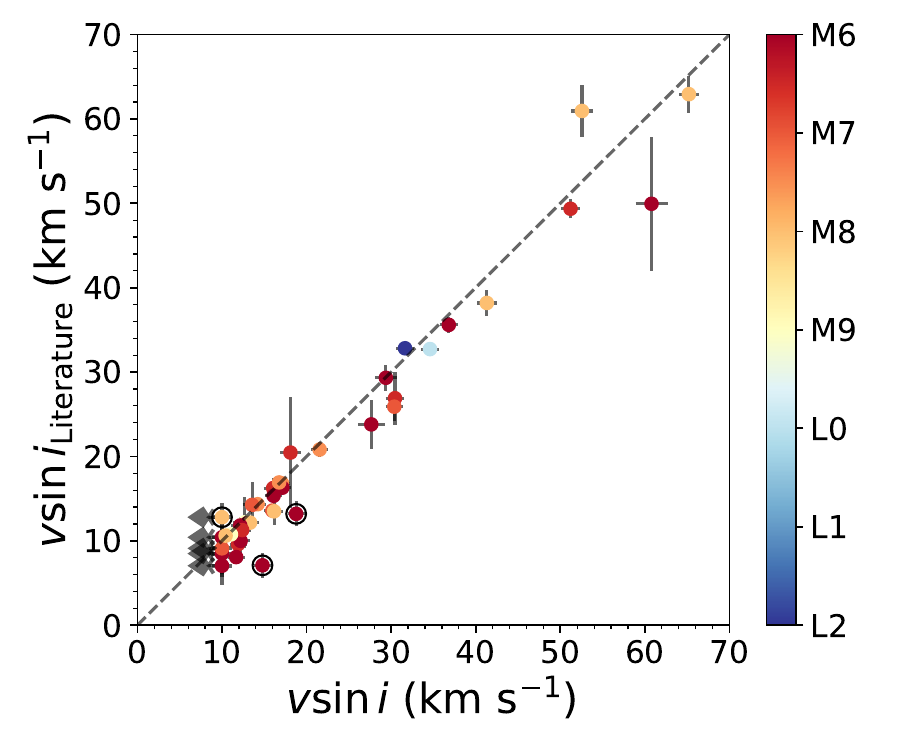}{0.5\textwidth}{}}
\caption{Comparison of RV (left) and {\vsini} (right) measurements from our APOGEE data to previous values reported in the literature (see Table \ref{table:litmearurement}). The black dashed line delineates perfect agreement. Sources are color-coded by spectral types. 
RV outliers are labeled with large symbols indicating 
binary candidates (star symbol), young sources with systematic RV offsets (\citealt{Kounkel:2019aa}; square symbol), 
measurements based on medium-resolution SpeX spectra (R$\sim$2000; \citealt{Terrien:2015aa}, circles), and other issues (diamonds; see Section~\ref{sec:rv} for details).
{\vsini} outliers are also highlighted by larger symbols, and are largely attributed to the systematic differences between Sonora and BT-Settl models.
} \label{fig:rv_vsini_pub}
\end{figure*}

We quantified the validity of our RV measurements by comparing our measured RVs with those reported in the literature, summarized in Table~\ref{table:litmearurement}.
Excluding published RVs from APOGEE DR16 \citep{Jonsson:2020aa}, A total of 67 sources in our sample have reported RVs in the literature with uncertainties. Figure~\ref{fig:rv_vsini_pub} compares these values.
The vast majority (84\%) have consistent RVs to within 3$\sigma$ deviation, while 11 sources are significant outliers\replaced{.}{:} 

\begin{itemize}
    \item Two of these outliers, 
    2MASS J08294949+2646348 ($\Delta$RV = 5.5~{\kms}), 
    2MASS J16311879+4051516 ($\Delta$RV = 4.4~{\kms}), 
    are based on measurements made with lower-resolution data ({\ldl} $\approx$ 2000) reported in \cite{Terrien:2015aa}, and may reflect underestimated uncertainties.
    \item Five other outliers, 
    2MASS J04201611+2821325,
    2MASS J04262939+2624137, 
    2MASS J04294568+2630468,
    2MASS J04330945+2246487, and
    2MASS J04363893+2258119 are all members of the Taurus Complex star-forming region 
    and have prior APOGEE measurements reported by \cite{Kounkel:2019aa}
    that are 2--3~{\kms} lower than our measurements.
    \citet{Cottaar:2014aa}; \citet{Cook:2014aa}; and \cite{Kounkel:2019aa} all report a systematic red-shift in RV measurements among the lowest temperature sources in their cluster samples ($T_\mathrm{eff} \, \leq 3400$~K), and the last study proposes a 
    systematic correction of $\Delta \mathrm{RV} = 12.84 - 0.0038 \times T_\mathrm{eff}$. 
    Accounting for this offset brings our measurements fully in line with RVs reported in \cite{Kounkel:2019aa}, but we did not report these corrected values in this work.
    \item  Another young source, the M6 2MASS J16093019$-$2059536, a reported member of the Upper Scorpius Association \citep{Slesnick:2006aa}, also has a significantly different RV from our measurements (RV$=-1.3\pm0.6$~{\kms}) compared to that reported in \citet[][RV$=-5.1\pm0.6$~{\kms}]{Dahm:2012aa}. The latter is based on optical high-resolution spectra and cross-correlation with the M8 standard VB 10. On the other hand, our measurement is fully consistent with that reported in \citet[$-$0.98 $\pm$ 0.09~{\kms}]{Jonsson:2020aa}, using the cross-correlation method with DR16 APOGEE spectra. The variance between these measurements could again be due to the RV offset found among young low-temperature sources, or variability induced by a binary (only one epoch of APOGEE data was available for this source).
    \item 2MASS J03505737+1818069 (LP 413$-$53) appears to be the currently known shortest orbital period UCD binary based on the high scatter of individual epoch RV measurements, analyzed in detailed in \cite{Hsu:2023aa}.
    \item The remaining outliers, 
    2MASS J00034394+8606422 and 2MASS J07140394+3702459, appear to be poorly fit by BT-Settl models due to a lack of FeH opacities, whereas the Sonora models provide a much better fit, resulting in a shift of 2--4~{\kms} (depending on the epoch) and bringing our measurement in line with that from the APOGEE pipeline. As the literature measurement from \cite{Deshpande:2013aa} utilized BT-Settl models, we attribute this difference to modeling systematics.
\end{itemize}



For completeness, we also compared our measured RVs with those provided by the APOGEE DR17 pipeline \texttt{Doppler} \citep{Nidever:2021aa}, which uses cross-correlation with the \texttt{The Cannon} models \citep{Ness:2015aa} trained from Synspec \citep{Hubeny:2017aa,Hubeny:2021aa}.
While performing well for warmer stars, the pipeline is known to have systematic issues with M dwarfs with {\teff} $\lesssim$ 3500~K \citep{Abdurrouf:2022aa}.
Indeed, roughly half of the sources in our sample with independent literature measurements show a $>$3$\sigma$ discrepancy with APOGEE pipeline RVs, and 15 sources in our sample have pipeline RVs $>$ 250~{\kms}. We thus consider APOGEE pipeline RVs to be unreliable for these low-temperature objects. 

In summary, all of the significant outliers between our APOGEE and literature RV measurements can be explained by methodological or astrophysical causes, and we conclude that our measurements are robust to a median precision of 0.37~{\kms}.

\subsection{Projected Rotational Velocities} \label{sec:vsini}
\begin{figure}
\centering
\includegraphics[width=\linewidth, trim=20 10 10 0]{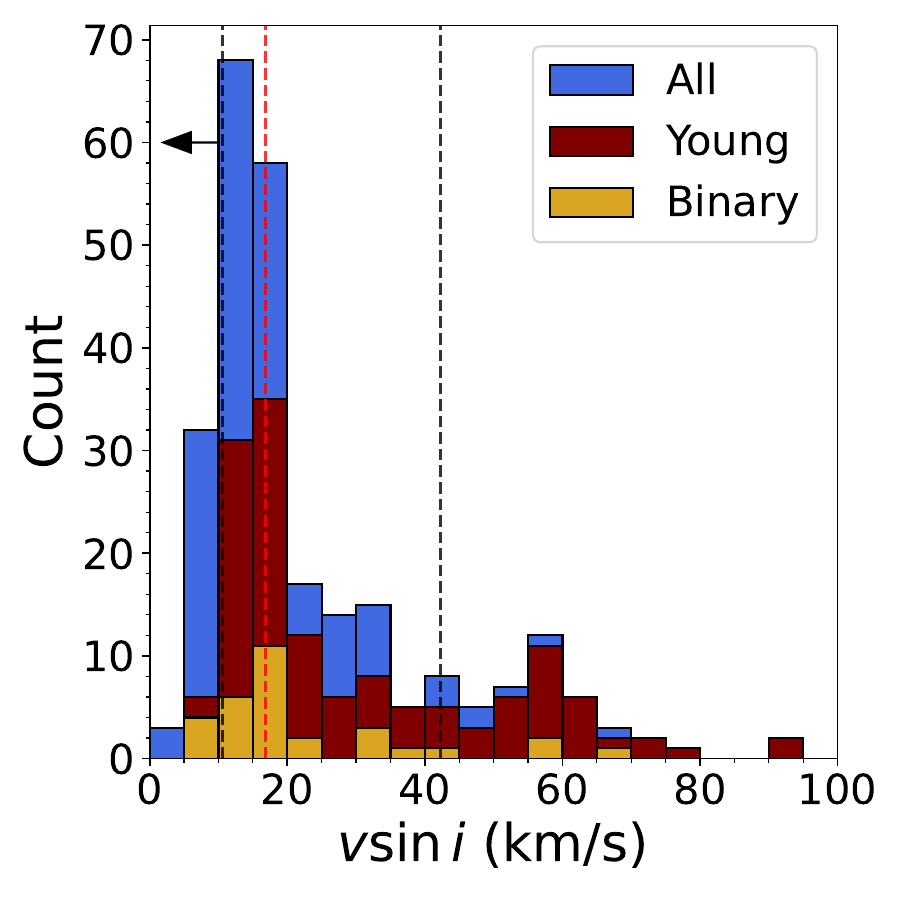}
\caption{Histogram of {\vsini} measurements.
The median and 16\%/84\% quantiles are indicated by vertical dashed lines.
Binaries, young cluster members, and overall sample are indicated by stacked yellow, red and blue histograms, respectively.
Our minimum $v\sin{i}$ detection floor at 10~{\kms} is indicated in the black arrow.
}
\label{fig:vsini_dist}
\end{figure}

Measured {\vsini} values for our sample range from 0.4~{\kms} to 92.8~{\kms}, with a median value of 17~{\kms} (Table~\ref{table:indmeasurement}). 
The distribution of our measurements is shown in Figure~{\ref{fig:vsini_dist}}.
To determine our {\vsini} detection limit, we compared this distribution for
both the Sonora and BT-Settl models and found a
sharp transition at 10~{\kms}, which we adopt as
our vsini detection limit.
This limit is more conservative than the 8~{\kms} limit reported in \cite{Gilhool:2018aa} and the 5~{\kms} limit reported in \citealp{Deshpande:2013aa}).
We adopt their final $\langle v\sin{i} \rangle$ as 10~{\kms} for sources with $\langle v\sin{i} \rangle$ $<$ 10~{\kms}.
Our median precision of sources with higher {\vsini} ($>$10~{\kms}) values is 0.5~{\kms}.
In addition to theses statistical uncertainties, we compared multi-epoch measurements (N$_\mathrm{obs} \geq 3$) for all of our non-RV varying sources, and found that the median standard deviation of {\vsini} values per source to be 0.95~{\kms}. We adopt this as an estimate of our systematic {\vsini} uncertainty and conservatively add it in quadrature
to our statistical uncertainties, resulting in a median {\vsini} uncertainty of 1.1~{\kms}.


We assessed the reliability of our {\vsini} measurements by again comparing to literature values (Figure~\ref{fig:rv_vsini_pub}). 
There are 41 sources with literature measurements, three of which are identified as $>$3$\sigma$ outliers: 
\begin{itemize}
    \item 2MASS J00034394+8606422 ($<10$~{\kms}; best-fit {\vsini} = 18.8 $\pm$ 1.0~{\kms} vs. 13.2 $\pm$ 1.5~{\kms} in \citealt{Deshpande:2013aa}),
    \item 2MASS J07140394+3702459 ($<$10~{\kms}; best-fit {\vsini} = 7.3 $\pm$ 1.1~{\kms} vs. 12.8 $\pm$ 0.5~{\kms} in \citealt{Deshpande:2013aa}), and 
    \item 2MASS J16311879+4051516 (14.8 $\pm$ 1.6~{\kms} vs. 7.1 $\pm$ 1.5~{\kms} in \citealt{Reiners:2018aa}). 
\end{itemize}
2MASS J00034394+8606422 (LP 2-291) is a known eclipsing binary ($P$=13.9182$\pm$0.0004~day; \citealp{Prsa:2022aa}), which {\vsini} varies at different epochs could be due to the (unresolved) secondary flux.
In these three cases, we find that Sonora models provide a much better fit to the APOGEE data, whereas the literature values are based on comparisons to BT-Settl models. 
We therefore attribute these deviations to systematics associated with model choice, and adopt our measured {\vsini} values for this analysis.

\begin{figure}
\centering
\includegraphics[width=\linewidth, trim=20 10 10 10]{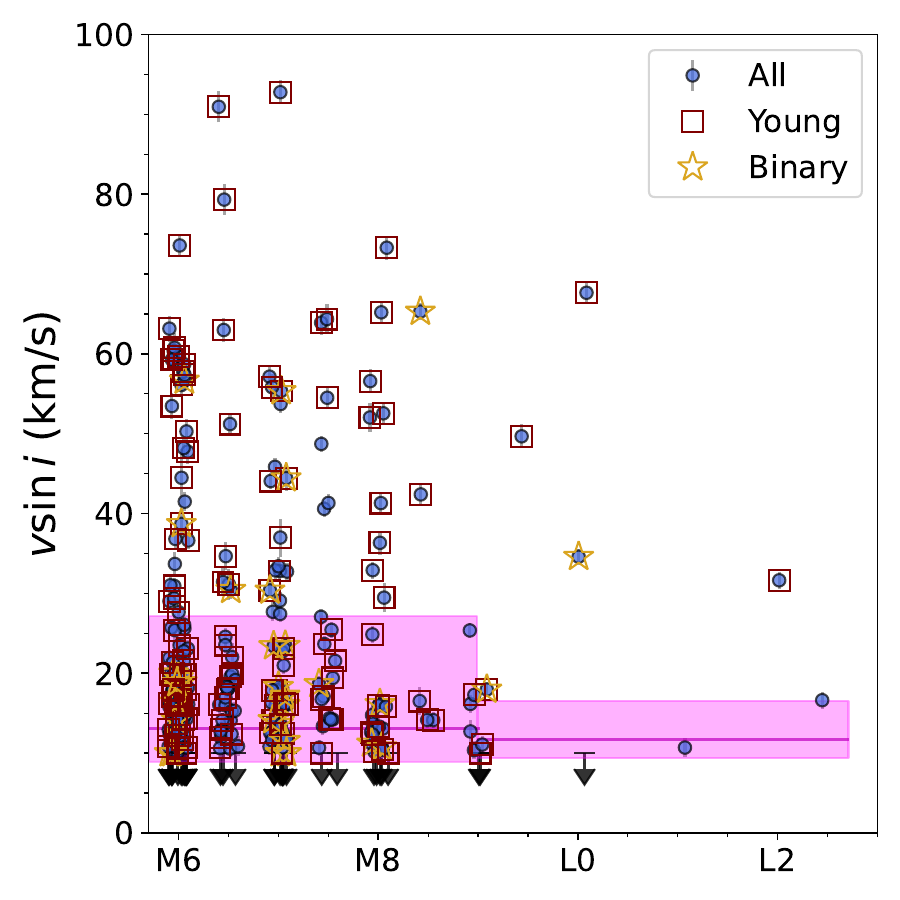}
\caption{Distribution of {\vsini} measurements as a function of spectral type.
Binary candidates are indicated by yellow stars, and young sources are indicated by red squares. 
Sources with {\vsini} $\leq$ 10~{\kms} (below our measurement limit) are indicated by downward black arrows.
The median and the 84$^\mathrm{th}$/16$^\mathrm{th}$ percentiles are shown with solid and magenta lines and shaded regions for M6--M8 and M9--L2 subtypes, respectively.
}
\label{fig:vsini_spt}
\end{figure}

The distribution of {\vsini} measurements as a function of spectral type is shown in Figure~\ref{fig:vsini_spt}. Median values are approximately constant over the M6--L2 range of $17^{+25}_{-6}$~{\kms}, with uncertainties computed from the 84$^\mathrm{th}$ and 16$^\mathrm{th}$ percentiles. This values is consistent with trends previously reported in the literature \citep{Crossfield:2014aa,Tannock:2021aa}; e.g., \citet{Hsu:2021aa} report a  median {\vsini} of 12.1~{\kms} for M4--M9 dwarfs and  
16.2~{\kms} for M9--L2 dwarfs.

\subsection{Fast Rotators}
While the majority of our sources have {\vsini} $<$ 40 {\kms}, there are fourteen sources with {\vsini} $>$ 60~{\kms}. 

Thirteen are young brown dwarfs: 
\begin{itemize}
    \item M8 2MASS J04311907+2335047 in the 1-2~Myr Taurus molecular clouds \citep{Slesnick:2006ab};
    \item L0 2MASS J05350162$-$0521489 (V$^\ast$ V2113~Ori) in the 1--2~Myr Orion Nebular Cluster \citep{Meeus:2005aa}; and 
    \item M7 2MASS J15560497$-$2106461, M6 2MASS J16003023$-$2334457, M7.5 2M16025214$-$2121296,
    M6.5 2MASS J16044303$-$2318258, 
    M8 2MASS J16045199$-$2224108, 
    M6.5 2MASS J16045581$-$2307438,
    M6 2MASS J16053077$-$2246200,
    M7.5 2MASS J16111711$-$2217173,
    M6 2MASS J16124692$-$2338408,
    M6.5 2MASS J16131600$-$2251511, and
    M6 2MASS J16200757$-$2359150,
    in the 10~Myr Upper Scorpius moving group \citep{Pecaut:2016aa}.
\end{itemize}
These sources are in an age range in which they have largely contracted and spun up, but have not had 
sufficient time to lose angular momentum through magnetized winds \citep{Kawaler:1988aa,Barnes:2003aa,Matt:2015aa}.

The other fast rotator is likely a binary. 
2MASS J15010818+2250020 (aka TVLM 513-46546) is a source with known periodic radio variability \citep{Hallinan:2006aa} and a potential giant planet companion identified by radio astrometry \citep{Curiel:2020ab}, which we also identify as an RV variable in our sample (\textit{Gaia} RUWE = 1.661; $\Delta$RV$_\mathrm{max} \, \sim$ 2~{\kms}; Section~\ref{sec:binaries}). 
While TVLM 513-46546 exhibits H${\alpha}$ emission and low surface gravity, 
it does not show \ion{Li}{1} absorption \citep{Burgasser:2015ac} and has been ruled out as a member of the Argus moving group (Section~\ref{sec:cluster_member}).


\subsection{Effective Temperatures and Surface Gravities} \label{sec:teff_logg}

\begin{figure*}
\centering
\includegraphics[width=\linewidth, trim=0 0 60 0]{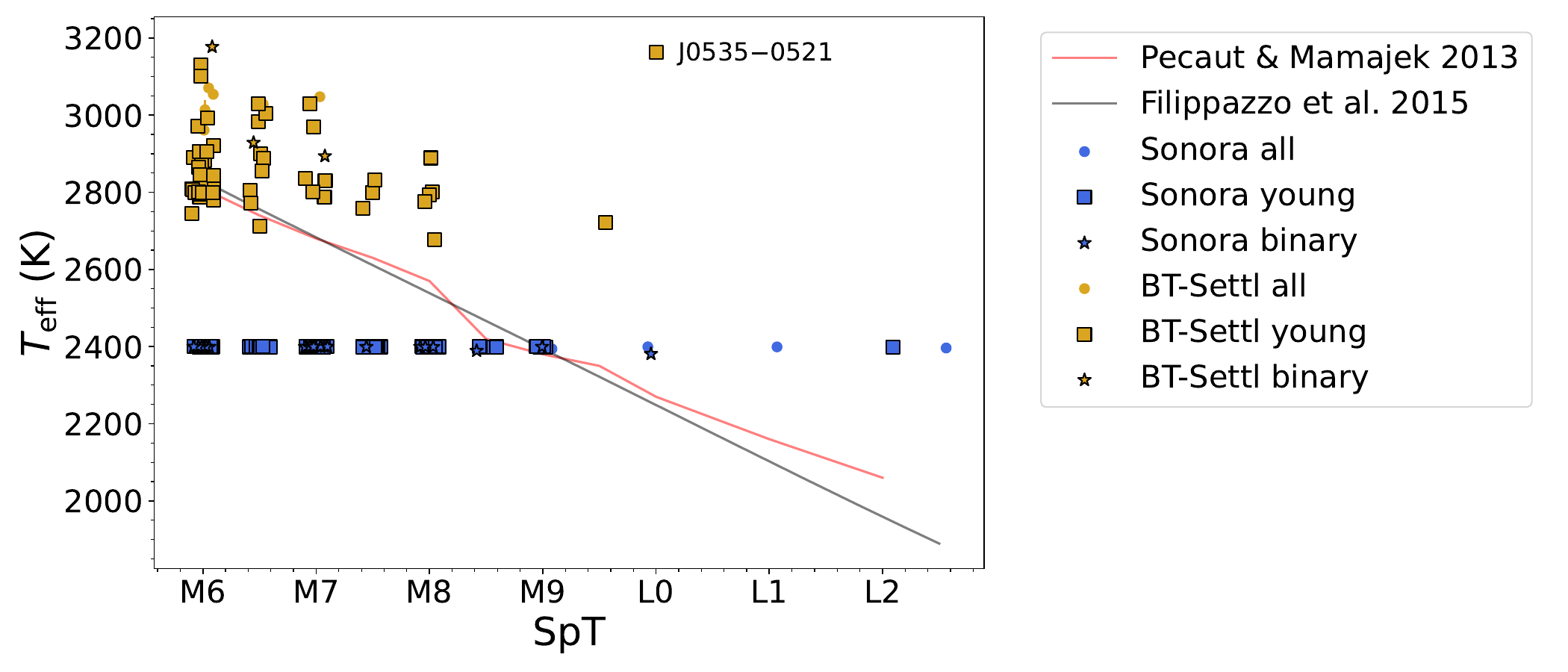}
\caption{Comparison of our best-fit {\teff}s as a function of spectral type between the Sonora (blue) and BT-Settl (yellow) models. 
The young sources and binaries are depicted in square boxes and stars, respectively.
Over-plotted is empirical {\teff}-spectral type relations from \citet[red line]{Pecaut:2013aa} and \citet[black line]{Filippazzo:2015aa}.
The young L0 2MASS J05350162$-$0521489 in the Orion Nebula Cluster is labeled.
These model fit parameters are not used in further analysis; see Section~\ref{sec:teff_logg} for details.
}
\label{fig:teff_compare}
\end{figure*}

Our fits also provided best estimates of effective temperature and surface gravity based on the particular model used. 
Of the 258 spectra modeled, 196 were best-fit by the Sonora models while 62 were best-fit by the BT-Settl models.
As noted above, the Sorona grid has a $T_\mathrm{eff}$ ceiling of 2400~K, which corresponds to a spectral type of approximately M9 \citep{Filippazzo:2015aa}. As this encompasses the majority of our sample, all of the inferred temperatures from these model fits were close to the model limit, making any inference of $T_\mathrm{eff}$ trends impossible.
Furthermore, the Sonora model grid is cloudless, with its inferred $T_\mathrm{eff}$ typically hotter by $\sim$300--500~K compared to the BT-Settl model grid \citep{Hsu:2021aa}, making the inferred $T_\mathrm{eff}$ = 2400~K for L2$\beta$ 2MASS J00452143$+$1634446 reach (expected $T_\mathrm{eff}$ $\sim$ 2060~K from \citealp{Filippazzo:2015aa}).
On the other hand, $T_\mathrm{eff}$s inferred from BT-Settl models ranged between 2676~K and 3177~K, with a median of 2860~K.
Figure~\ref{fig:teff_compare} shows the {\teff} trend as a function of spectral type. 
Ignoring the Sonora fit values, the best-fit {\teff}s from the BT-Settl models show a general decreasing trend toward later spectral types from M6 to M9.5. 
The outlier in this trend is the young L0 2MASS J05350162$-$0521489 ({\teff} = 3162$\pm$13~K) \citep{Meeus:2005aa}.
Comparing to empirical spectral type to {\teff} relations from \cite{Pecaut:2013aa} and \cite{Filippazzo:2015aa}, our best-fit BT-Settl {\teff}s are systematically higher than the empirical {\teff}s, which we attribute to the model discrepancies.

\begin{figure*}
\centering
\includegraphics[width=\linewidth, trim=0 0 60 0]{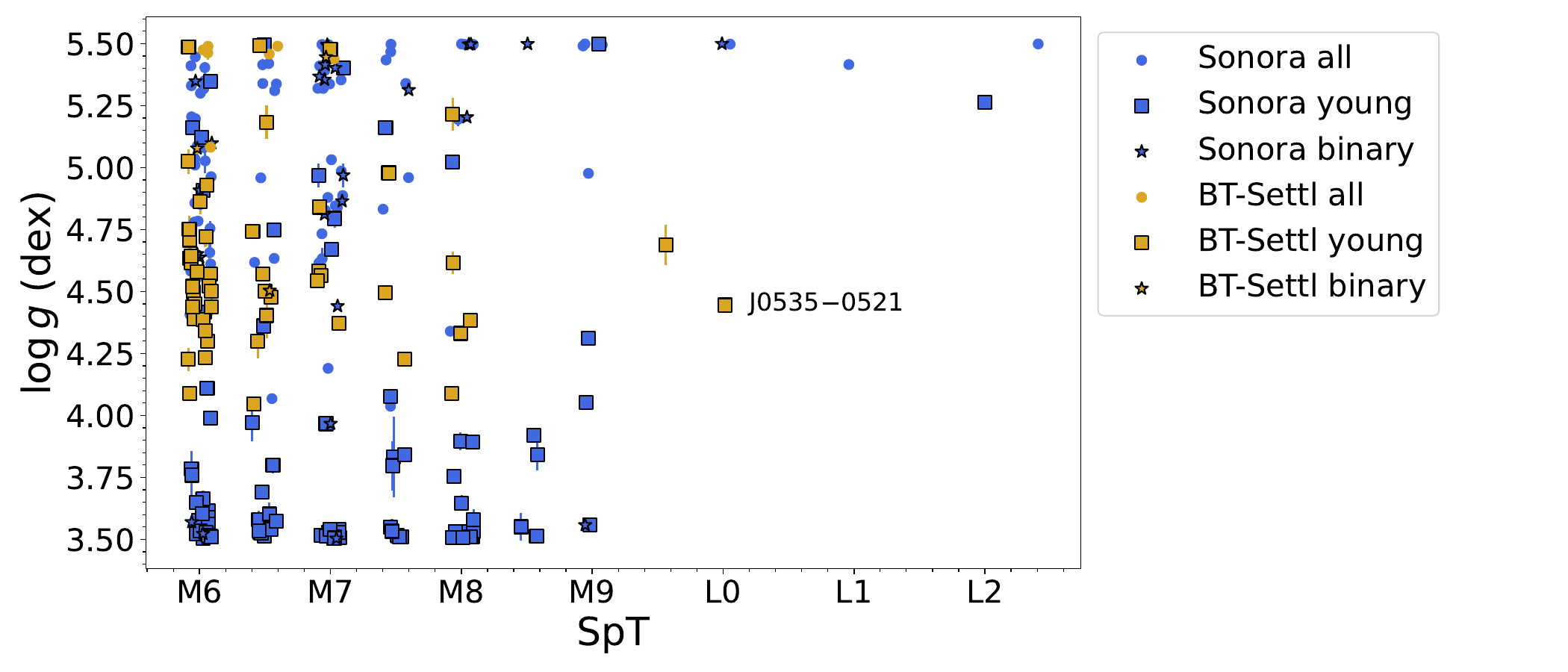}
\caption{Comparison of our best-fit {\logg}s as a function of spectral type between the Sonora (blue) and BT-Settl (yellow) models. 
The young sources and binaries are depicted in square boxes and stars, respectively.
These model fit parameters are not used in further analysis; see Section~\ref{sec:teff_logg} for details.
}
\label{fig:logg_compare}
\end{figure*}

Figure~\ref{fig:logg_compare} illustrates fit {\logg} trends as a function of spectral type.
Surface gravities from the Sonora model fits scatter across the full model parameter range of 3.5 $\leq$ {\logg} $\leq$ 5.5, with members of young clusters typically (but not consistently) having $\log{g}$ values close to the minimum.
Surface gravities from the BT-Settl model fits have a narrower range of 4.0 $\leq$ {\logg} $\leq$ 5.5, with the young cluster members again having values in the bottom half of this range.

Given the limited temperature fit range for the Sonora models, and large scatter in inferred surface gravities for both models, we do not regard these values as realistic estimates, and do not further investigate their trends. However, we verified that variations in {\teff} and {\logg} about the optimal values had minimal influence on derived RV and {\vsini} values (see also \citealt{Theissen:2022aa}).
Comparing fits between the BT-Settl and Sonora model sets yields equivalent RVs and {\vsini}s on average, with standard deviations of 0.9~{\kms} and 1.5~{\kms}, respectively.

\begin{longrotatetable}
\begin{deluxetable*}{clllrrrrrrrrrrrr}
\tablecaption{Spectral Model Fit Parameters\label{table:indmeasurement}}
\tablewidth{650pt}
\tabletypesize{\scriptsize}
\tablehead{
\colhead{} & \colhead{} & \colhead{} & \colhead{} & \colhead{} & \colhead{} & \colhead{} & \colhead{} & \colhead{} &
\colhead{} & \colhead{} & \colhead{} & \colhead{} & \colhead{} & \colhead{} & \colhead{} \\
\colhead{APOGEE ID} & 
\colhead{Plate} & \colhead{Loc.} & 
\colhead{Fiber} & 
\colhead{JD} & \colhead{bary.\tablenotemark{a}} & \colhead{S/N} & \colhead{RV\tablenotemark{b, f}} & 
\colhead{$\langle$RV$\rangle$\tablenotemark{b, c, f}} & \colhead{$v\sin i$\tablenotemark{b}}& \colhead{$\langle$$v\sin i$$\rangle$\tablenotemark{b, c, d}} & \colhead{$T_\mathrm{eff}$\tablenotemark{g}} & \colhead{$\langle$$T_\mathrm{eff}$$\rangle$\tablenotemark{c, g}} &  \colhead{$\log{g}$\tablenotemark{g}} & \colhead{$\langle$$\log{g}$$\rangle$\tablenotemark{c, g}} & \colhead{Mdl\tablenotemark{e}} \\ 
\colhead{} & \colhead{} & \colhead{ID} & \colhead{ID} & \colhead{(day)} & \colhead{(km s$^{-1}$)} & \colhead{} & \colhead{(km s$^{-1}$)} & \colhead{(km s$^{-1}$)} & \colhead{(km s$^{-1}$)} &
\colhead{(km s$^{-1}$)} & \colhead{(K)} & \colhead{(K)} & \colhead{(cm s$^{-2}$)} & \colhead{(cm s$^{-2}$)} & \colhead{}
} 
\startdata
2M0003$+$8606 & 5221 & 4217 & 275 & 2455815.921 & $11.92$ & 33 & $7.5^{+0.4}_{-0.8}$ & $7.3^{+0.3}_{-0.4}$ & $18.0^{+1.0}_{-1.0}$ & $18.8^{+1.0}_{-1.0}$ & $2399.5^{+0.4}_{-0.6}$ & $2399.6^{+0.1}_{-0.2}$ & $4.63^{+0.03}_{-0.03}$ & $4.65^{+0.01}_{-0.01}$ & S \\ 
\nodata & 5221 & 4217 & 227 & 2455822.934 & $11.89$ & 18 & $6.7^{+0.6}_{-0.4}$ & \nodata & $24.0^{+3.3}_{-1.1}$ & \nodata & $2399.3^{+0.5}_{-0.6}$ & \nodata & $4.14^{+0.04}_{-0.06}$ & \nodata & S \\ 
\nodata & 5221 & 4217 & 275 & 2455841.765 & $10.97$ & 32 & $7.1^{+0.4}_{-0.3}$ & \nodata & $18.1^{+1.9}_{-1.0}$ & \nodata & $2399.4^{+0.4}_{-0.7}$ & \nodata & $4.61^{+0.03}_{-0.03}$ & \nodata & S \\ 
\nodata & 5221 & 4217 & 275 & 2455866.685 & $7.98$ & 42 & $7.5^{+0.3}_{-0.5}$ & \nodata & $17.9^{+1.0}_{-1.0}$ & \nodata & $2399.6^{+0.3}_{-0.5}$ & \nodata & $4.67^{+0.03}_{-0.02}$ & \nodata & S \\ 
\nodata & 5221 & 4217 & 227 & 2455876.687 & $6.30$ & 39 & $7.6^{+0.3}_{-0.5}$ & \nodata & $19.8^{+1.0}_{-1.0}$ & \nodata & $2399.5^{+0.3}_{-0.7}$ & \nodata & $4.86^{+0.02}_{-0.03}$ & \nodata & S \\ 
\nodata & 5221 & 4217 & 224 & 2456653.557 & $-3.17$ & 42 & $7.1^{+0.3}_{-0.3}$ & \nodata & $17.9^{+1.0}_{-1.0}$ & \nodata & $2399.6^{+0.3}_{-0.8}$ & \nodata & $4.67^{+0.04}_{-0.03}$ & \nodata & S \\ 
\nodata & 5221 & 4217 & 224 & 2456654.555 & $-3.37$ & 40 & $7.1^{+0.3}_{-0.2}$ & \nodata & $17.9^{+1.0}_{-1.0}$ & \nodata & $2399.7^{+0.3}_{-0.5}$ & \nodata & $4.63^{+0.02}_{-0.02}$ & \nodata & S \\ 
\nodata & 5221 & 4217 & 224 & 2456655.555 & $-3.58$ & 33 & $6.9^{+0.3}_{-0.3}$ & \nodata & $18.0^{+1.0}_{-1.0}$ & \nodata & $2399.6^{+0.3}_{-0.6}$ & \nodata & $4.57^{+0.03}_{-0.02}$ & \nodata & S \\
\enddata
\tablenotetext{a}{Barycentric correction.}
\tablenotetext{b}{Systematic uncertainties of 0.19~{\kms} and 0.95~{\kms} are added to RV and {\vsini} measurements, respectively. See Sections~\ref{sec:rv} and ~\ref{sec:vsini} for details.}
\tablenotetext{c}{Weighted average over all epochs.}
\tablenotetext{d}{Averaged {\vsini} $<$ 10~km~s$^{-1}$ is below our {\vsini} detection limit; see Section~{\ref{sec:vsini}} for details.}
\tablenotetext{e}{Models used: S = Sonora 2018 \citep{Marley:2018aa}; B = BT-Settl \citep{Baraffe:2015aa}.}
\tablenotetext{f}{We did not apply the effective temperature dependent correction to RV from \cite{Kounkel:2019aa} for 2MASS J04201611+2821325, 2MASS J04262939+2624137, 2MASS J04294568+2630468, 2MASS J04330945+2246487, and 2MASS J04363893+2258119. See Section~\ref{sec:rv} for details.}
\tablenotetext{g}{Readers should not take our {\teff} and {\logg} measurements as accurate. We report these values only for the reproducibility of our work. See Section~\ref{sec:teff_logg} for details.}
\tablecomments{Measurements from individual spectra over individual or multiple epochs are combined using inverse uncertainty weighting (weight = $1/(\sigma_\text{upper}^2 + \sigma_\text{lower}^2)$); upper and lower uncertainties are also combined using inverse uncertainty-squared weighting. In cases where individual spectra have S/N $<$ 10, spectral data are combined first, then modeled.}
\end{deluxetable*}
\end{longrotatetable}

\startlongtable
\begin{deluxetable*}{crrrrrrr}
\tablecaption{RV and $v\sin{i}$ Measurements with ASPCAP and Literature Comparison \label{table:litmearurement}}
\tablewidth{700pt}
\tabletypesize{\scriptsize}
\tablehead{
\colhead{} & \colhead{} & \colhead{} & \colhead{} & \colhead{} & \colhead{} & \colhead{} & \colhead{}  \\
\colhead{APOGEE ID} & \colhead{$\langle$RV$\rangle$\tablenotemark{a, c}} & ASPCAP RV\tablenotemark{b} & Lit. RV & Lit. RV Ref. & \colhead{$\langle v\sin{i} \rangle$\tablenotemark{a}} & Lit. $v\sin{i}$ & Lit. $v\sin{i}$ Ref. \\ 
\colhead{} & \colhead{(km s$^{-1}$)} & \colhead{(km s$^{-1}$)} & \colhead{(km s$^{-1}$)} & \colhead{} & \colhead{(km s$^{-1}$)} & \colhead{(km s$^{-1}$)} & \colhead{}
}
\startdata
2MASS J00034394+8606422 & $7.26^{+0.2}_{-0.33}$ & $5.73 \pm 2.85$ & $10.8 \pm {0.28}$ & (8) & $18.77^{+0.95}_{-0.97}$ & $13.2 \pm {1.5}$ & (8) \\ 
2MASS J00312793+6139333 & $-34.44^{+0.2}_{-0.35}$ & $-38.02 \pm 0.69$ & \nodata & \nodata & $<10$ & \nodata & \nodata \\ 
2MASS J00381273+3850323 & $6.62^{+0.2}_{-0.32}$ & $3.42 \pm 1.58$ & \nodata & \nodata & $<10$ & \nodata & \nodata \\ 
2MASS J00452143+1634446 & $3.71^{+0.22}_{-0.37}$ & $555.81 \pm 4.96$ & $3.16 \pm {0.83}$ & (14) & $31.64^{+0.97}_{-1.04}$ & $32.82 \pm {0.17}$ & (4) \\ 
2MASS J00514593$-$1221458 & $15.66^{+0.24}_{-0.38}$ & $15.25 \pm 1.79$ & $15.29 \pm {0.11}$ & (20) & $15.85^{+1.21}_{-1.22}$ & \nodata & \nodata \\ 
2MASS J01154176+0059317 & $10.78^{+0.28}_{-0.59}$ & $11.7 \pm 0.41$ & \nodata & \nodata & $33.69^{+1.57}_{-1.12}$ & \nodata & \nodata \\ 
2MASS J01215816+0101007 & $10.28^{+0.22}_{-0.37}$ & $586.83 \pm 0.92$ & \nodata & \nodata & $<10$ & \nodata & \nodata \\ 
2MASS J01243124$-$0027556 & $-3.88^{+0.23}_{-0.41}$ & $577.03 \pm 1.68$ & $0.69 \pm {10.14}$ & (19) & $27.7^{+1.11}_{-1.04}$ & \nodata & \nodata \\
\enddata
\tablerefs{(1) \cite{Lepine:2003aa}, (2)
\cite{Furesz:2008aa}, (3), \cite{Reiners:2009aa}, (4) \cite{Blake:2010aa}, (5) \cite{West:2011aa}, (6) \cite{Deshpande:2012aa}, 
(7) \cite{Dahm:2012aa}, (8) \cite{Deshpande:2013aa}, (9) \cite{Newton:2014aa}, (10) \cite{Cottaar:2015aa}, (11) \cite{West:2015aa}, 
(12) \cite{Terrien:2015aa}, (13) \cite{Burgasser:2015ac}, (14) \cite{Faherty:2016aa}, (15) \cite{Reiners:2018aa}, (16) \cite{Gilhool:2018aa}, 
(17) \cite{Kesseli:2018aa}, (18) \cite{Kounkel:2019aa}, (19) \cite{Kiman:2019aa}, (20) \cite{Jonsson:2020aa}, (21) \cite{Gaia-Collaboration:2021aa}}
\tablenotetext{a}{Our weighted average measurements over all epochs.}
\tablenotetext{b}{RV from DR17 ASPCAP Pipeline.}
\tablenotetext{c}{We did not apply the effective temperature dependent correction to RV from \cite{Kounkel:2019aa} for 2MASS J04201611+2821325, 2MASS J04262939+2624137, 2MASS J04294568+2630468, 2MASS J04330945+2246487, and 2MASS J04363893+2258119. See Section~\ref{sec:rv} for details.}
\tablecomments{Measurements from individual spectra over individual or multiple epochs are combined using inverse uncertainty weighting (weight = $1/(\sigma_\text{upper}^2 + \sigma_\text{lower}^2)$); upper and lower uncertainties are also combined using inverse uncertainty-squared weighting. In cases where individual spectra have S/N $<$ 10, spectral data are combined first, then modeled.}
\end{deluxetable*}

\startlongtable
\begin{deluxetable*}{llrccccccc}
\tablecaption{Radial Velocities and Heliocentric Space Motions \label{table:uvw}}
\tablewidth{700pt}
\tabletypesize{\scriptsize}
\tablehead{
\colhead{} & \colhead{} & \colhead{} & \colhead{} & \colhead{} & \colhead{} & \colhead{} & \colhead{} & \colhead{} \\ 
\colhead{APOGEE ID} & \colhead{SpT} & \colhead{Adpoted RV} & \colhead{$U$} & \colhead{$V$} & \colhead{$W$} & \colhead{P[TD]/P[D]\tablenotemark{a}} & \colhead{Pop\tablenotemark{a}} & \colhead{BANYAN $\Sigma$} & \colhead{BANYAN $\Sigma$\tablenotemark{c}} \\ 
\colhead{} & \colhead{} & \colhead{(km s$^{-1}$)} & \colhead{(km s$^{-1}$)} & \colhead{(km s$^{-1}$)} & \colhead{(km s$^{-1}$)} & \colhead{} & \colhead{} & \colhead{Prob} & \colhead{Member}
} 
\startdata
2MASS J00034394+8606422\tablenotemark{b} & M6.0 & $7.26^{+0.2}_{-0.33}$ & $-16.2 \pm 0.2$ & $7.4 \pm 0.3$ & $1.6 \pm 0.1$ & 0.01 & D &  $>$99\% &  field \\ 
2MASS J00312793+6139333 & M7.0 & $-34.44^{+0.2}_{-0.35}$ & $-27.4 \pm 0.2$ & $-50.0 \pm 0.3$ & $29.8 \pm 0.1$ & 0.18 & D/TD &  $>$99\% &  field \\ 
2MASS J00381273+3850323 & M6.0 & $6.62^{+0.2}_{-0.32}$ & $-31.2 \pm 0.2$ & $-4.9 \pm 0.3$ & $5.3 \pm 0.1$ & 0.01 & D &  $>$99\% &  field \\ 
2MASS J00452143+1634446\tablenotemark{d} & L2.0 & $3.71^{+0.22}_{-0.37}$ & $-10.8 \pm 0.2$ & $-1.5 \pm 0.2$ & $1.5 \pm 0.3$ & 0.01 & D &  $>$99\% &  ARG \\ 
2MASS J00514593$-$1221458 & M6.0 & $15.66^{+0.24}_{-0.38}$ & $17.5 \pm 0.1$ & $13.1 \pm 0.1$ & $-9.7 \pm 0.4$ & 0.01 & D &  $>$99\% &  field \\ 
2MASS J01154176+0059317 & M6.0 & $10.78^{+0.28}_{-0.59}$ & $-35.8 \pm 0.4$ & $-7.4 \pm 0.3$ & $5.8 \pm 0.5$ & 0.01 & D &  $>$99\% &  field \\ 
2MASS J01215816+0101007 & M6.5 & $10.28^{+0.22}_{-0.37}$ & $-9.0 \pm 0.2$ & $-12.0 \pm 0.2$ & $-5.0 \pm 0.4$ & 0.01 & D &  $>$99\% &  field \\ 
2MASS J01243124$-$0027556 & M7.0 & $-3.88^{+0.23}_{-0.41}$ & $17.4 \pm 0.2$ & $-16.1 \pm 0.3$ & $-0.4 \pm 0.4$ & 0.01 & D &  $>$99\% &  field \\
\enddata
\tablenotetext{a}{Galactic thin disk (D), thick disk (TD), intermediate populations (D/TD) are assigned according to probability ratios 
$P(\mathrm{TD})/P(\mathrm{D}) < 0.1$, $P(\mathrm{TD})/P(\mathrm{D}) > 10$, and $0.1 < P(\mathrm{TD})/P(\mathrm{D}) < 10$, respectively, following \citet{Bensby:2003aa}.}
\tablenotetext{b}{Known or candidate binary.}
\tablenotetext{c}{BANYAN $\Sigma$ young moving group name abbreviation follows those defined in \citet{Gagne:2018ab}: AB Doradus (ABDMG), Argus (ARG), Carina-Near (CARN), Coma Berenices (CBER), Corona Australis (CRA), Hyades (HYA), Taurus (TAU), Upper Centaurus-Lupus (UCL), Upper Corona Australis (UCRA), and Upper Scorpius (USCO).}
\tablenotetext{d}{Reported in the literature or a new member of young moving groups.}
\tablenotetext{e}{Reported as a member of NGC 1333 in \cite{Cantat-Gaudin:2018aa, Yao:2018aa, Cantat-Gaudin:2020aa, Cantat-Gaudin:2020ab}, but we did not identify because NGC 1333 is not included in BANYAN $\Sigma$.}
\tablenotetext{f}{Reported as a member of the Castor Moving Group in \cite{Zuckerman:2013aa}, but we did not identify it because the Castor Moving Group is not included in BANYAN $\Sigma$.}
\end{deluxetable*}

\startlongtable
\begin{deluxetable*}{cccccc}
\tablecaption{Galactic Orbital Parameters \label{table:galactic_orbit}}
\tabletypesize{\small}
\tablehead{
\colhead{} & \colhead{} & \colhead{} & \colhead{} & \colhead{} & \colhead{} \\ 
\colhead{APOGEE ID} & \colhead{$R_\mathrm{min}$} & \colhead{$R_\mathrm{max}$} & \colhead{$Z_\mathrm{max}$} & \colhead{$e$} & \colhead{$i$} \\ 
\colhead{} & \colhead{(kpc)} & \colhead{(kpc)} & \colhead{(kpc)} & \colhead{} & \colhead{(deg)}
} 
\startdata
2MASS J00034394$+$8606422 & $8.0461\pm0.0009$ & $10.47\pm0.03$ & $0.0463\pm0.0009$ & $0.131\pm0.0017$ & $0.279\pm0.005$ \\ 
2MASS J00312793$+$6139333 & $5.002\pm0.008$ & $8.489\pm0.002$ & $0.3386\pm0.0008$ & $0.2584\pm0.0008$ & $2.865\pm0.014$ \\ 
2MASS J00381273$+$3850323 & $7.615\pm0.006$ & $9.7\pm0.03$ & $0.0659\pm0.0013$ & $0.1203\pm0.0012$ & $0.425\pm0.011$ \\ 
2MASS J00452143$+$1634446 & $7.897\pm0.013$ & $8.557\pm0.017$ & $0.023\pm0.003$ & $0.0401\pm0.0003$ & $0.16\pm0.02$ \\ 
2MASS J00514593$-$1221458 & $8.0163\pm0.0011$ & $11.293\pm0.012$ & $0.113\pm0.005$ & $0.1697\pm0.0005$ & $0.65\pm0.03$ \\ 
2MASS J01154176$+$0059317 & $7.047\pm0.019$ & $9.18\pm0.03$ & $0.071\pm0.01$ & $0.1315\pm0.0016$ & $0.5\pm0.07$ \\ 
2MASS J01215816$+$0101007 & $7.242\pm0.01$ & $8.3797\pm0.0009$ & $0.0494\pm0.0011$ & $0.0728\pm0.0007$ & $0.359\pm0.008$ \\ 
2MASS J01243124$-$0027556 & $6.883\pm0.016$ & $8.4725\pm0.0015$ & $0.0052\pm0.0015$ & $0.1035\pm0.0011$ & $0.038\pm0.011$ \\
\enddata
\tablenotetext{a}{The large uncertainties of Galactic orbital parameters of 2MASS J05350162$-$0521489 are due to its large proper motion uncertainty  ($\mu_{\alpha} \cos{\delta} = 0 \pm 44$~mas yr$^{-1}$ and $\mu_{\delta} = 0 \pm 40$~mas yr$^{-1}$; \citealp{Cutri:2014aa}).}
\end{deluxetable*}

\section{Analysis} \label{sec:analysis}

\subsection{$UVW$ Space Motions} \label{sec:uvw_velocity}
\begin{figure*}[!htbp]
\centering
\gridline{\fig{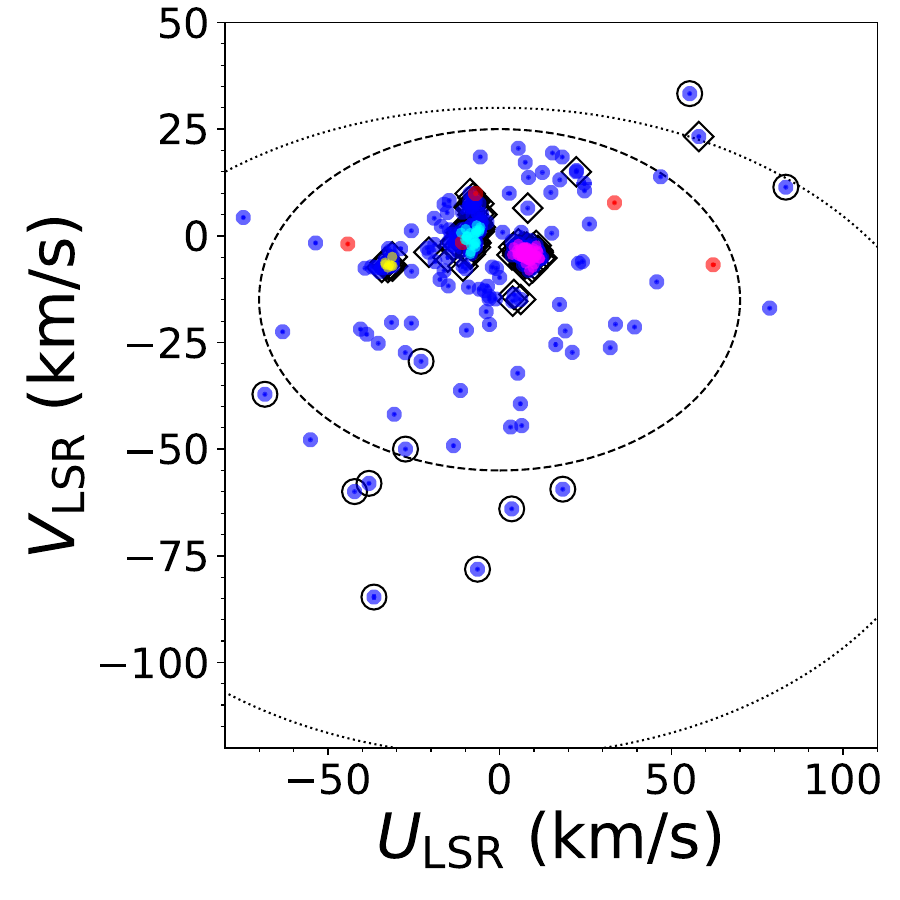}{0.45\textwidth}{}
\fig{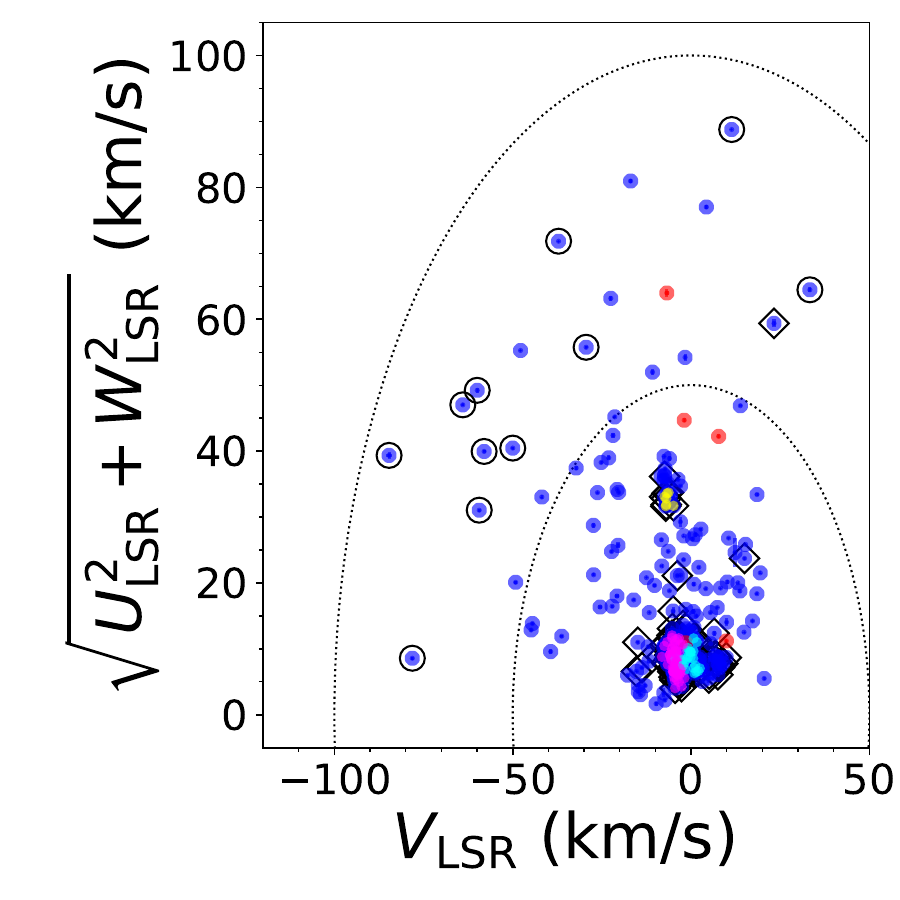}{0.45\textwidth}{}
}
\vspace{-1cm}
\gridline{\fig{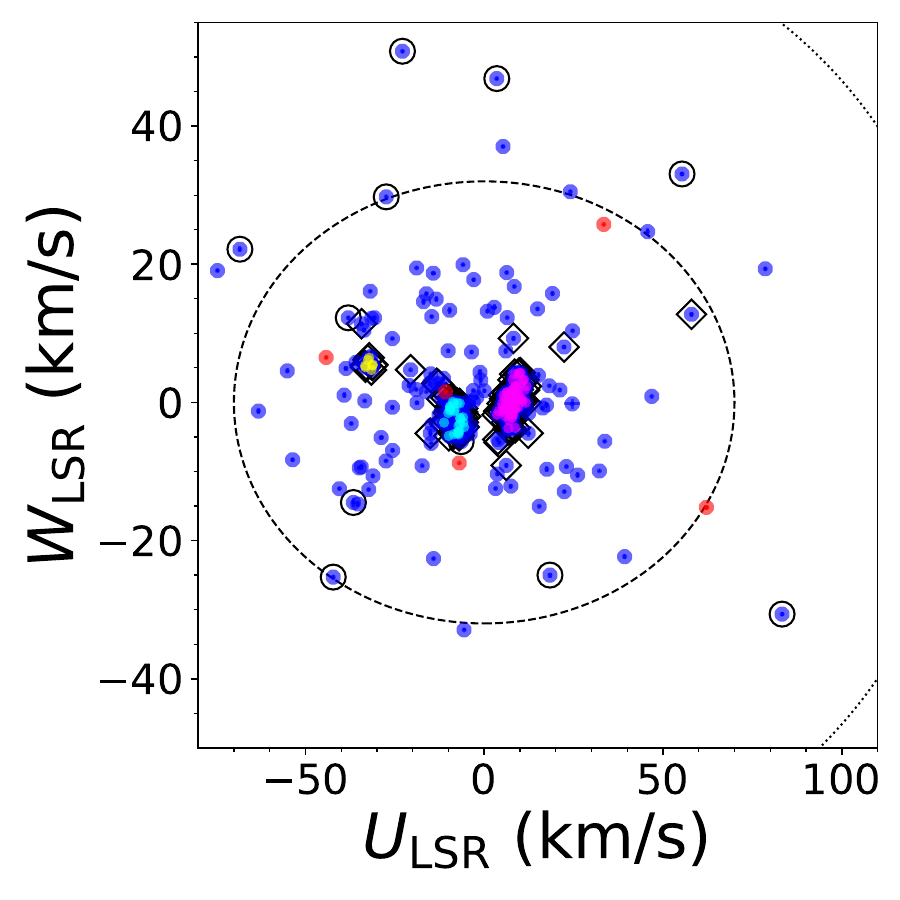}{0.45\textwidth}{}
\fig{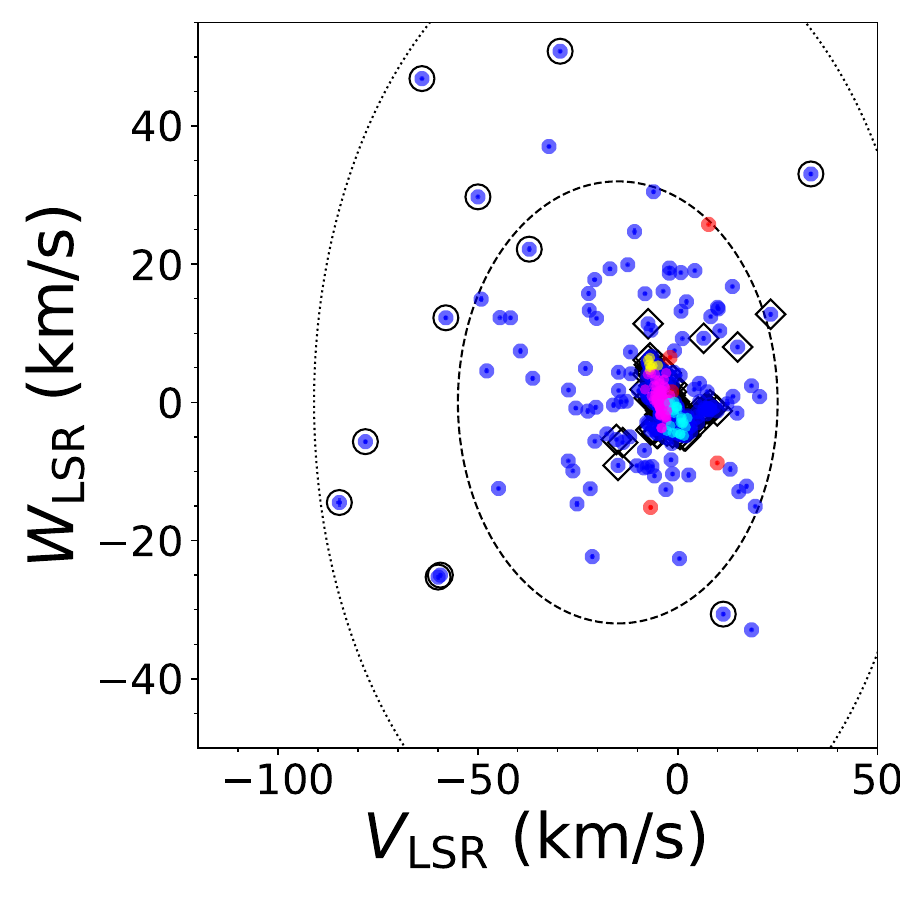}{0.45\textwidth}{}
}
\vspace{-1cm}
\caption{Space motions of our sample in the Local Standard of Rest (LSR). ($U_\mathrm{LSR}, V_\mathrm{LSR}$), ($U_\mathrm{LSR}, W_\mathrm{LSR}$), and ($V_\mathrm{LSR}, W_\mathrm{LSR}$) velocity pairs are shown along with the 2$\sigma$ uncertainty spheres for the thin disk (dashed lines) and thick disk (dotted lines) populations based on \citet{Bensby:2003aa}. M and L dwarfs are labeled as blue and red circles, respectively. 
The upper-right corner is a Toomre plot, with total velocities $v_\text{tot} = \sqrt{U_\mathrm{LSR}^2 + V_\mathrm{LSR}^2 + W_\mathrm{LSR}^2}$ indicated by dotted lines in steps of $50$ {\kms}. Young sources and intermediate thin/thick disk sources are highlighted with open diamonds and open circles, respectively. 
The major young cluster sources, Upper Scorpius, Taurus, and Hyades are labeled in magenta, cyan, and yellow, respectively.
\label{fig:uvw_velocity}}
\end{figure*}

We combined our RV measurements with \textit{Gaia} and ground-based astrometry to compute heliocentric $UVW$ space motions 
following \citet{Johnson:1987aa}. We adopt a right-handed coordinate system, with $U$ velocity pointing toward the Galactic center, $V$ velocity pointing in the direction of  Galactic rotation, and $W$ velocity pointing toward the Galactic North pole. We also computed velocities in the local standard of rest (LSR) assuming Solar $UVW$ LSR velocities of ($U_{\odot}$, $V_{\odot}$, $W_{\odot}$) = (11.1~{\kms}, 12.24~{\kms}, 7.5~{\kms}) from \cite{Schonrich:2010aa}.
These values are visualized in Figure~\ref{fig:uvw_velocity} and listed in Table~\ref{table:uvw}.
The mean $U_{LSR}$ = $-2.5 \pm 1.3$~{\kms} and $W_{LSR}$ = $= 1.1 \pm 0.6$~{\kms} velocities of our sample are consistent with zero, while the average $V_{LSR}$ = $-6.0 \pm 0.9$~{\kms} is slightly negative, as expected for asymmetric drift \citep{Stromberg:1924aa}. 

We followed \citet{Bensby:2003aa} to determine the probabilities of kinematic membership for each  star to thin disk, thick disk, or halo populations based on the LSR velocities.
We specifically separated thin disk, intermediate thin/thick disk, and thick disk membership by using the ratios $P[\text{TD}]/P[\text{D}] < 0.1$, $ 0.1 \leq P[\text{TD}]/P[\text{D}] \leq 10$, and $P[\text{TD}]/P[\text{D}] > 10$, respectively.  As expected, the majority of our sample\footnote{The Orion Nebular Cluster member 2MASS J05350162$-$0521489 was excluded from the kinematic analysis due to its large proper motion uncertainties.} 
is thin disk sources (246 sources), with 11 intermediate thin disk/thick disk members and no thick disk members.

We did not detect significant correlations between ($U_\mathrm{LSR},W_\mathrm{LSR}$) or ($V_\mathrm{LSR},W_\mathrm{LSR}$) velocity pairs (p-value = 0.47 and 0.08, respectively using the Wald Test with t-distribution; \citealp{Wald:1943aa,mckinney-proc-scipy-2010}), but did find a significant positive correlation for ($U_\mathrm{LSR},V_\mathrm{LSR}$) velocities (p-value $<$ 0.001, correlation coefficient R = 0.26), driven largely by our intermediate thin/thick disk members.
Exlcuding these sources significantly reduces the ($U_\mathrm{LSR},V_\mathrm{LSR}$) correlation (p-value = 0.016, R = 0.15). 
We also found significant positive correlations between total velocity-squared, 
$v_\mathrm{LSR}^2$ = $U_\mathrm{LSR}^2+V_\mathrm{LSR}^2+W_\mathrm{LSR}^2$ and absolute $|W_\mathrm{LSR}|$-velocity for the full sample (p-value $<$ 0.001, R = 0.65) and the thin disk subsample (p-value $<$ 0.001, R = 0.53).
$v_{LSR}^2$ correlates with the asymmetric drift \citep{Stromberg:1924aa}, while absolute $|W_\mathrm{LSR}|$-velocity has been used as a proxy of age \citep{Wielen:1977aa},
so this correlation is an indicator of age variation in the sample.
While the APOGEE sample selection is not volume-complete, these correlations are consistent with our prior analysis of the 20~pc UCD sample \citep{Hsu:2021aa}.

\subsection{Galactic Orbits} \label{sec:galactic_orbit}
Galactic orbits can identify sources with different spatial origins, including stars that had drifted radially inward or outward to the Solar radius. Starting with the LSR velocities and $XYZ$ Galactic spatial coordinates\footnote{$XYZ$ are the Galactic rectangular coordinates transformed from Galactic spherical coordinates longitude $l$ = 0$^\circ$ toward the Galactic center, latitude $b$ = 90$^\circ$ toward the Galactic North pole, and distance $d$ from the Sun. The heliocentric coordinates are defined as $X_h = d\cos{b}\cos{l}$, $Y_h = d\cos{b}\sin{l}$, and $Z_h = d\sin{b}$; while galactocentric coordinates are $X = X_h - R_{\odot}$, $Y = Y_h$, and $Z = Z_h + Z_{\odot}$, where ($R_{\odot}$, $Z_{\odot}$) = (8.43~kpc, 0.027~kpc) are the assumed Solar galactocentric cooordinates \citep{Bovy:2012aa, Chen:2001aa, Reid:2014aa}. 
This coordinate set is coaligned with $UVW$ velocities.}
of each source, Galactic orbits were computed using the \texttt{galpy} package \citep{Bovy:2015aa}, an efficient, ordinary differential equation solver that conserves energy and momentum for Galactic dynamics.
We used 
an axisymmetric potential from \citet{Miyamoto:1975aa}
and assumed an LSR azimuthal velocity, $v_{\phi}$ = 220~{\kms} \citep{Bovy:2012aa, Chen:2001aa, Reid:2014aa}. Each orbit was integrated from $-5$ to $+5$~Gyr in steps of 10~Myr, and 1000 orbit realizations were computed using Monte Carlo sampling of velocity uncertainties assuming normal distributions. 
We examined the specific orbital parameters of minimum and maximum Galactic cylindrical radius ($R_\text{min}$, $R_\text{max}$), maximum Galactic vertical height ($\lvert Z \rvert$), median orbital eccentricity ($e \equiv \langle R_\text{max} - R_\text{min} \rangle/\langle R_\text{max} + R_\text{min} \rangle$), and median orbital inclination ($\tan{i} \equiv \lvert Z/\sqrt{X^2 + Y^2} \rvert$).
These parameters are listed in Table~\ref{table:galactic_orbit}.

\begin{figure*}[!htbp]
\centering
\gridline{\fig{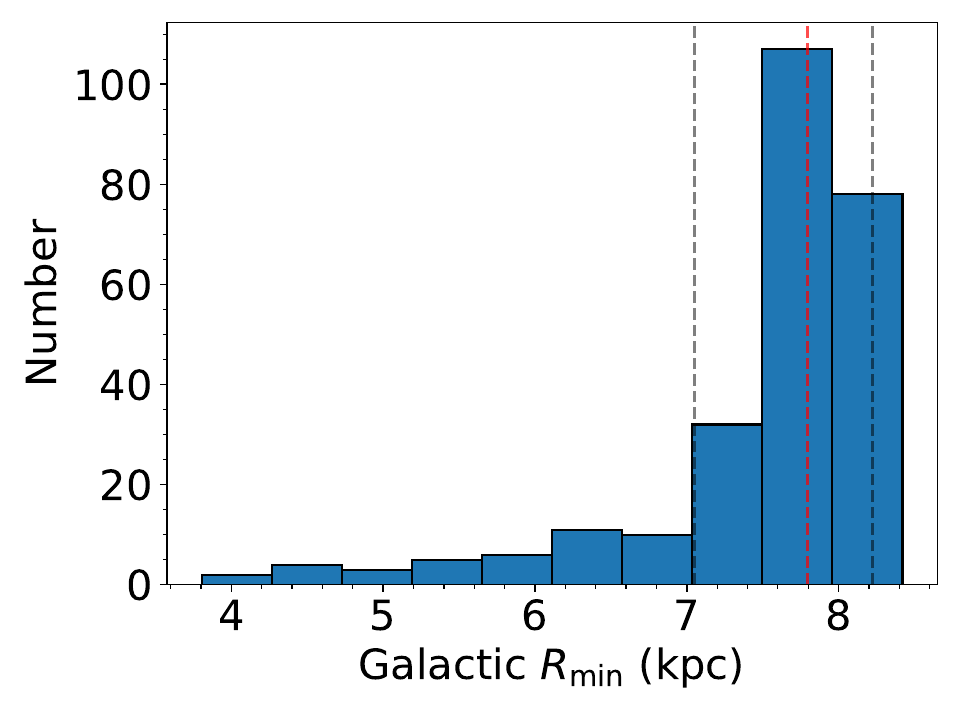}{0.33\textwidth}{}
\fig{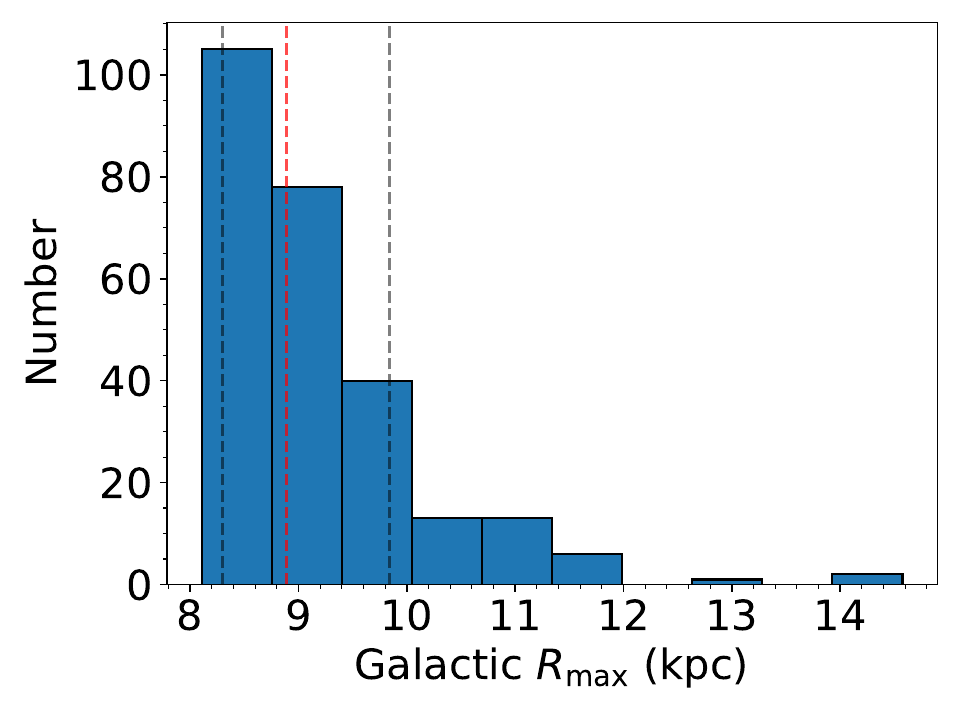}{0.33\textwidth}{}
\fig{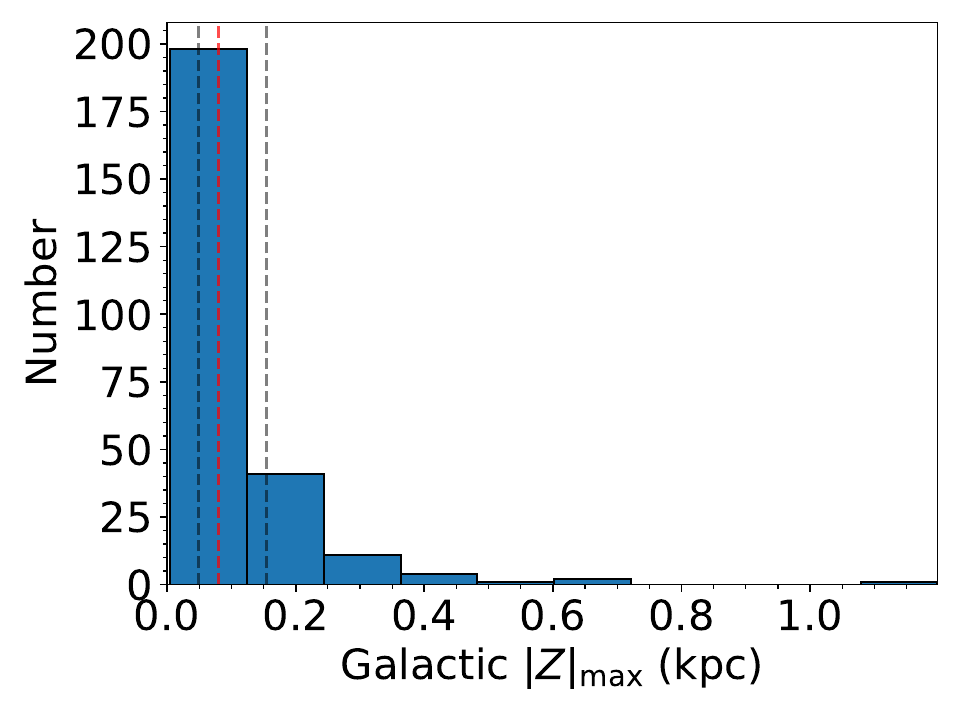}{0.33\textwidth}{}
}
\vspace{-1cm}
\gridline{\fig{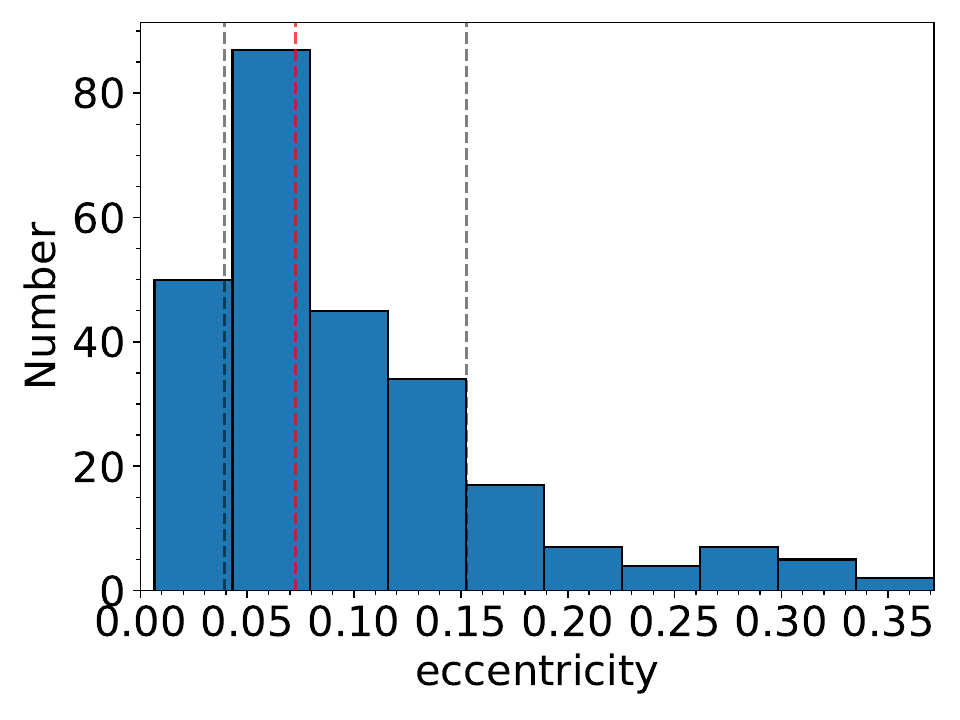}{0.33\textwidth}{}
\fig{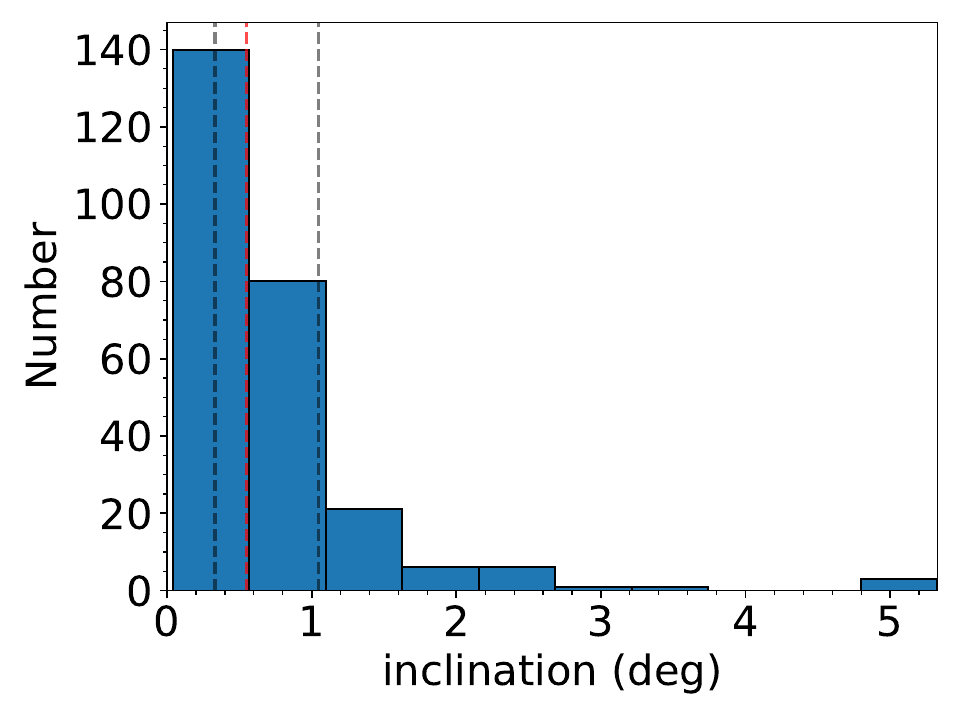}{0.33\textwidth}{}
\fig{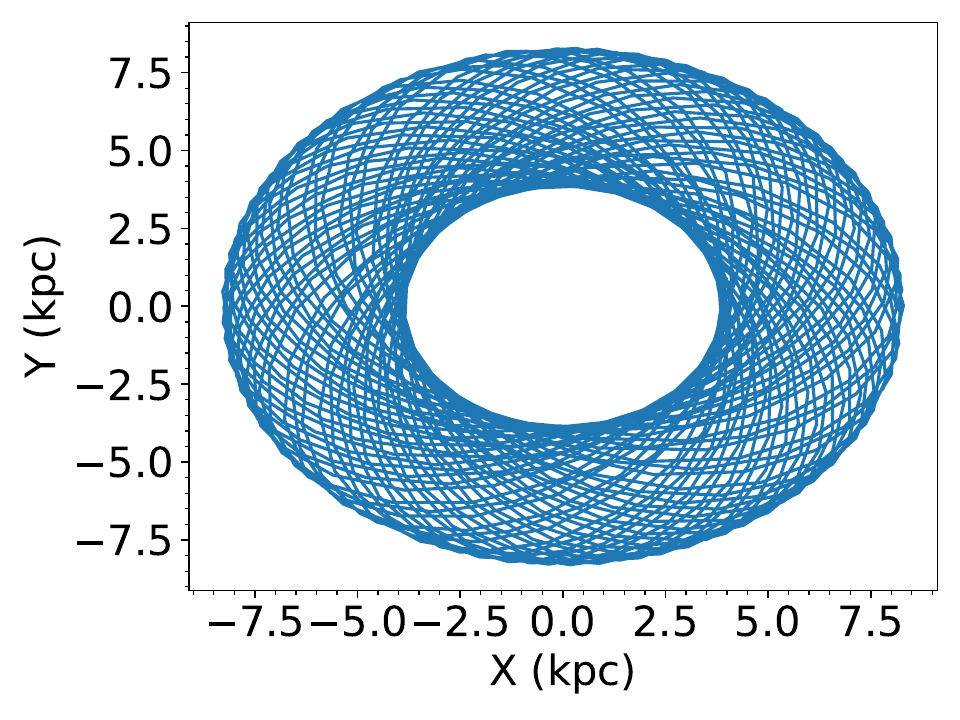}{0.33\textwidth}{}
}
\vspace{-1cm}
\caption{The distributions of inferred orbital parameters for our sample. 
\textit{upper left}: minimum Galactic radius $R_\text{min}$; 
\textit{upper middle}: maximum Galactic radius $R_\text{max}$; 
\textit{upper right}: minimum vertical displacement $\lvert Z \rvert$; 
\textit{lower left}: eccentricity $e$; 
\textit{lower middle}: inclination $i$; 
\textit{lower right}: The Galactic $XY$ orbit of the intermediate thin/thick disk source 2MASS J15211578+3420021, integrated between $-$5~Gyr to $+$5~Gyr. 
The median values and the 16$^\text{th}$/84$^\text{th}$ percentiles are shown in red and grey dashed lines, respectively. 
\label{fig:galactic_orbit}}
\end{figure*}

Figure~\ref{fig:galactic_orbit} shows the distributions of the derived orbital parameters. The majority of our sample exhibit circular ($\langle{e}\rangle = 0.07^{+0.08}_{-0.03}$) and planar orbits ($\langle{i}\rangle = 0.6^{\circ}~^{+0.8^{\circ}}_{-0.2^{\circ}}$), residing mostly at the Solar Galactic radius ($\langle{R_\text{min}}\rangle$ = 7.8$^{+0.4}_{-0.7}$~kpc, $\langle{R_\text{max}}\rangle$ = 8.9$^{+0.9}_{-0.6}$~kpc) and close to the Galactic Plane
($\langle\lvert Z \rvert\rangle$ =  0.08$^{+0.07}_{-0.03}$~kpc).
There are 22 sources that have $e > 0.2$, 10 of which are intermediate thin/thick disk or thick disk members.
There are also 13 sources with slightly non-planar orbits ($i > 2^{\circ}$), 8 of these being intermediate thin/thick disk members. 

\subsection{Cluster Membership} \label{sec:cluster_member}
We evaluated young association membership by comparing the 6D heliocentric spatial and velocity coordinates derived above 
to known nearby systems using the BANYAN $\Sigma$ web tool \citep{Gagne:2018ab}. Results are summarized in Table~\ref{table:uvw}.

Our APOGEE DR17 sample is highly biased toward young cluster members.
Out of 141 confirmed young sources in our sample, 114 are identified as kinematic members of young moving groups, with 138 of these previously reported in the literature, 3 new young sources, and 6 ruled out based on BANYAN $\Sigma$. 
The majority of cluster members are associated with 
Upper Scorpius (10$\pm$3~Myr, \citealp{Pecaut:2016aa}; 75 sources), 
Taurus (1--2~Myr, \citealp{Kenyon:1995aa}; 25 sources), and 
the Hyades Moving Group (750$\pm$100~Myr, \citealp{Brandt:2015aa}; 6 sources).  

We identify three moving group members not previously reported in the literature:
\begin{itemize}
    \item 2MASS~J05402570+2448090 (G 100-28; 67.2\% Argus moving group, 40-50~Myr, \citealp{Zuckerman:2019aa}; 30.9\% Carina Near moving group, $\sim$200~Myr, \citealp{Zuckerman:2006aa}), 
    \item 2MASS~J14093200+4138080 (LP 220-50; 99.6\% Argus moving group), and 
    \item 2MASS~J21272531+5553150 (LSPM J2127+5553; 99.3\% Carina Near moving group).
\end{itemize}

2MASS J05402570+2448090 is a well-studied active M dwarf binary that exhibits flares and H$_{\alpha}$ emission \citep{Pettersen:1983aa, Reid:1995aa, Gizis:2002ab, Lepine:2013aa, Gaidos:2014aa, Terrien:2015ab}. The secondary is separated at 0.4720$\arcsec$ \citep{Janson:2014aa, Winters:2021aa} with a short rotational period of 0.294~d \citep{Newton:2016aa}.
It was previously reported as a candidate member of Hyades \citep{Eggen:1993aa}, which is ruled out based on our RV of 23.1$^{+0.3}_{-0.2}$~{\kms}.
The uncertainty of its membership between Argus and Carina Near moving groups is based on the true systematic RV (best RV of 23.3 and 26.4~{\kms}, respectively), which requires future RV monitoring to sample its full orbit.

2MASS J14093200+4138080 has been reported as an active M6 dwarf in \cite{West:2015aa} (H$_{\alpha}$ equivalent width = 7.0$\pm$1.1~nm), but reported inactive in \cite{Newton:2017aa} (H$_{\alpha}$ equivalent width = $-7.2$$\pm$0.1~nm) and fast rotational period = 0.265~d \citep{Newton:2016aa}. No signatures of youth has been reported for 2MASS J14093200+4138080.

2MASS J21272531+5553150 has been reported as a candidate member of the Carina Near moving group (98.520\%) with BANYAN $\Sigma$ in \cite{Seli:2021aa} using \textit{Gaia} DR2 astrometry (no RV), so our RV places it as a highly likely member.

On the other hand, we rule out six sources previously associated with clusters which had previously lacked radial velocity data.
\begin{itemize}
    \item 2MASS J00381273+3850323 was reported as a member of the Hyades Moving Group, which was determined solely from \textit{Gaia} DR2 proper motions and parallax without the radial velocity \citep{Roser:2019aa, Lodieu:2019aa}. Our measured RV = $+6.6^{+0.2}_{-0.3}$~{\kms} is not consistent with the expected RV of Hyades ($+39^{+3}_{-4}$~{\kms}; \citealp{Gagne:2018ab}).
    \item 2MASS J04254894+1852479 was reported as a member of the $\beta$ Pic Moving Group, based on \textit{Gaia} DR2 proper motions and parallax only \citep{Gagne:2018ad}. Our measured RV = $+28.3 \pm 0.4$~{\kms} is  inconsistent with the expected RV of the $\beta$ Pic Moving Group ($+10 \pm 10$~{\kms}; \citealp{Gagne:2018ab}).
    \item 2MASS J07140394+3702459 (LSPM J0714+3702) was reported as a member of the Argus Moving Group and classified as M7$\beta$ (intermediate gravity class) by \cite{Gagne:2015aa}. 
    With the BANYAN $\Sigma$ web tool \citep{Gagne:2018ab}, the required RV for 2MASS J07140394+3702459 to be a member of Argus is 20.9~{\kms}, which is ruled out by our RV measurement of 35.3$\pm$0.2~{\kms}.
\end{itemize}

We also rule out three reported members of the Coma Berenices Cluster ($562^{+98}_{-84}$~Myr;  \citealp{Silaj:2014aa}), 
2MASS J12205439+2525568, 2MASS J12263913+2505546, and 2MASS J12265349+2543556, identified in \cite{Melnikov:2012aa} on the basis of astrometry and photometry alone. \cite{Melnikov:2012aa} used proper motions from the L\'{e}pine Shara Proper Motion (LSPM) catalog \citep{Lepine:2005aa}, and these sources were selected on the basis of insignificant proper motion, consistent with the small angular relative motion of the Coma Berenices Cluster at large, ($\mu_{\alpha}\cos\delta$, $\mu_{\delta}$) = ($-12.11 \pm 0.05$, $ -9.00 \pm 0.12$)~mas~yr$^{-1}$ \citep{Gaia-Collaboration:2018aa}.
The measured RV of 2MASS J12205439+2525568 = 14.81$\pm$0.29~{\kms} is highly different from the expected RV of the Coma Berenices Cluster = $-0.1 \pm 0.8$~{\kms} \citep{Gagne:2018ab}.
2MASS J12263913+2505546 and 2MASS J12265349+2543556 were ruled out largely based on improved astrometry from \textit{Gaia}.

These cases highlight the importance of RVs for kinematic cluster membership.
The 21 known young UCDs not matched by BANYAN $\Sigma$ are in clusters not included in this web tool\footnote{2MASS J08460531+1035309 is reported as a member of $\sim$3.4 Gyr old open cluster NGC 2682 \citep{Poovelil:2020aa}, but our RV = $43.7 \pm 0.3$~{\kms} is inconsistent compared to the expected RVs = 29.3--36.5~{\kms} \citep{Tarricq:2021aa}.}, as BANYAN $\Sigma$ only included young clusters and moving groups within 150~pc.
There are five sources in the 1--2~Myr cluster NGC~1333 identified by \textit{Gaia} DR2 astrometry \citep{Cantat-Gaudin:2018aa, Yao:2018aa, Cantat-Gaudin:2020aa, Cantat-Gaudin:2020ab}; our RV of 15.5--17.5~{\kms} is consistent with the expected cluster motion of $\sim$11.0--17.6~{\kms} \citep{Tarricq:2021aa}.
There are 11 sources in the 3 Myr cluster IC 348 (RV = 15.1--19.5~{\kms}; consistent with the expected RVs = 13.0--20.7~{\kms} in \citealp{Tarricq:2021aa}), and 4 sources in the $\sim$2 Myr Orion Nebula Cluster \citep{Yao:2018aa} (RV = 25.1--33.4~{\kms}; consistent with the expected RVs = 23.4--33.5~{\kms} in \citealp{Theissen:2022aa}).
2MASS J08294949+2646348 is reported as a member of the Castor Moving Group (200--$\sim$700~Myr, \citealp{Lopez-Santiago:2009aa, Zuckerman:2013aa})\footnote{The $XYZ$ positions and $UVW$ space velocities of the Castor Moving Group are widely spread \citep{Zuckerman:2013aa}, and we do not evaluate its true membership as it is beyond the scope of this work.}. 
We include all 141 sources in our ``young'' sample for our subsequent kinematic analysis.

Returning to our model fit parameters, we note that there is a distinct
difference between the {\logg} values between 
our young sources ($\langle${\logg}$\rangle$ = $3.55^{+0.42}_{-0.04}$~cm s$^{-2}$ dex) and 
our field sources ($\langle${\logg}$\rangle$ = $5.20^{+0.29}_{-0.56}$~cm s$^{-2}$ dex) based on the Sonora model fits.
This is consistent with expectations for the larger radii and lower masses expected for UCDs younger than $\sim$150~Myr.
We also found slightly higher {\vsini}s for 
the young sources ($\langle${\vsini}$\rangle$ = $22^{+34}_{-8}$~{\kms}) compared to 
the field objects ($\langle${\vsini}$\rangle$ $=14^{+13}_{-5}$~{\kms}), although these distributions overlap.
Again, this is line with expectations for the spin-down timescales of low-mass stars, which can be $\gtrsim$100~Myr for M $\sim$ 0.1~M$_\odot$ \citep{Barnes:2010aa, Reiners:2012aa, van-Saders:2013aa, Matt:2015aa}
Similar differences were seen for the BT-Settl model fits,
albeit with a much smaller sample size (48 young and 14 field objects).


\subsection{Kinematic Ages} \label{sec:kinematic_age}
Ensemble kinematics of a population provides age information, as stellar velocities become increasingly dispersed through dynamical interactions of Galactic structures \citep{Spitzer:1953aa,Wielen:1977aa,Aumer:2009aa,Ting:2019aa,Sharma:2021aa}. 
The increased dispersion over time has been historically captured in empirical age-velocity dispersion relations (AVRs), which can be inverted to derive mean kinematic ages for stellar samples \citep{Hsu:2021aa}.
Here, we considered two functional forms of the AVR in this study: the exponential relation of \citet{Wielen:1977aa}, and the power-law relation from \citet{Aumer:2009aa}. We followed the same analysis methodology as described in \citet{Hsu:2021aa}, and the results are summarized in Table~\ref{table:kinematic_age}.

\begin{deluxetable*}{lcccccccccl}
\tablecaption{Velocity Dispersions and Group Kinematic Ages\label{table:kinematic_age}}
\tablewidth{700pt}
\tabletypesize{\scriptsize}
\tablehead{
\colhead{Sample} & \colhead{N} & \colhead{$\langle U \rangle$} & \colhead{$\langle V \rangle$} & \colhead{$\langle W \rangle$} & \colhead{$\sigma_U$} & \colhead{$\sigma_V$} & \colhead{$\sigma_W$} & \colhead{$\sigma_\mathrm{tot}$} & \colhead{Age} & \colhead{Note} \\ 
\colhead{} & \colhead{} & \colhead{(km s$^{-1}$)} & \colhead{(km s$^{-1}$)} & \colhead{(km s$^{-1}$)} & \colhead{(km s$^{-1}$)} & \colhead{(km s$^{-1}$)} &  \colhead{(km s$^{-1}$)} &  \colhead{(km s$^{-1}$)} & \colhead{(Gyr)} & \colhead{}
} 
\startdata
All & 257 & $-2.5 \pm 1.3$ & $-6.0 \pm 0.9$ & $1.1 \pm 0.6$ & $20.9 \pm 0.1$ & $15.1 \pm 0.1$ & $10.0 \pm 0.1$ & $27.7 \pm 0.2$ & $1.25 \pm 0.10$ & Unweighted \\ 
 & &  &  &  & $32.8 \pm 0.2$ & $24.3 \pm 0.2$ & $14.8 \pm 0.1$ & $43.4 \pm 0.2$ & $3.41 \pm 0.04$ & $|W|$ Weighted \\ 
Thin Disk & 246 & $-2.3 \pm 1.2$ & $-4.4 \pm 0.7$ & $0.8 \pm 0.5$ & $19.3 \pm 0.1$ & $10.9 \pm 0.1$ & $8.0 \pm 0.1$ & $23.6 \pm 0.1$ & $0.74 \pm 0.08$ & Unweighted \\ 
 & &  &  &  & $29.3 \pm 0.2$ & $15.0 \pm 0.1$ & $11.1 \pm 0.1$ & $34.7 \pm 0.2$ & $2.01 \pm 0.03$ & $|W|$ Weighted \\ 
Not Young & 117 & $-5.9 \pm 2.6$ & $-10.1 \pm 2.0$ & $2.9 \pm 1.3$ & $28.4 \pm 0.2$ & $21.2 \pm 0.2$ & $14.3 \pm 0.1$ & $38.2 \pm 0.3$ & $3.30 \pm 0.19$ & Unweighted \\ 
 & &  &  &  & $35.6 \pm 0.3$ & $26.8 \pm 0.3$ & $16.5 \pm 0.2$ & $47.5 \pm 0.3$ & $4.13 \pm 0.05$ & $|W|$ Weighted \\  
Not Binary & 204 & $-1.4 \pm 1.5$ & $-6.5 \pm 1.1$ & $1.1 \pm 0.7$ & $21.2 \pm 0.1$ & $15.9 \pm 0.1$ & $10.3 \pm 0.1$ & $28.4 \pm 0.2$ & $1.36 \pm 0.11$ & Unweighted \\ 
 & &  &  &  & $33.0 \pm 0.2$ & $26.1 \pm 0.2$ & $15.6 \pm 0.1$ & $44.8 \pm 0.2$ & $3.66 \pm 0.04$ & $|W|$ Weighted \\  
Thin Disk & 106 & $-5.8 \pm 2.6$ & $-6.7 \pm 1.5$ & $2.4 \pm 1.1$ & $26.4 \pm 0.2$ & $15.6 \pm 0.1$ & $11.7 \pm 0.1$ & $32.8 \pm 0.3$ & $2.11 \pm 0.14$ & Unweighted \\ 
\& Not Young & &  &  &  & $32.1 \pm 0.3$ & $16.8 \pm 0.2$ & $12.8 \pm 0.1$ & $38.4 \pm 0.3$ & $2.58 \pm 0.04$ & $|W|$ Weighted \\ 
M Dwarfs & 102 & $-6.5 \pm 2.5$ & $-7.0 \pm 1.6$ & $2.4 \pm 1.1$ & $25.5 \pm 0.2$ & $15.8 \pm 0.1$ & $11.5 \pm 0.1$ & $32.1 \pm 0.3$ & $1.97 \pm 0.14$ & Unweighted \\ 
Thin Disk, Not Young & &  &  &  & $31.8 \pm 0.4$ & $17.1 \pm 0.2$ & $12.7 \pm 0.2$ & $38.3 \pm 0.3$ & $2.56 \pm 0.05$ & $|W|$ Weighted \\ 
L Dwarfs & 4 & $11.1 \pm 20.1$ & $2.2 \pm 3.4$ & $2.1 \pm 7.9$ & $37.5 \pm 7.3$ & $6.4 \pm 0.9$ & $14.4 \pm 3.4$ & $40.7 \pm 8.1$ & $4.3 \pm 2.4$ & Unweighted \\ 
Thin Disk, Not Young & &  &  &  & $41.7 \pm 8.0$ & $6.7 \pm 0.9$ & $12.0 \pm 3.6$ & $44.3 \pm 7.1$ & $3.6 \pm 1.2$ & $|W|$ Weighted \\ 
Thin Disk & 76 & $-5.9 \pm 3.3$ & $-7.2 \pm 1.9$ & $2.5 \pm 1.4$ & $28.5 \pm 0.3$ & $16.5 \pm 0.2$ & $12.1 \pm 0.1$ & $35.1 \pm 0.4$ & $2.58 \pm 0.17$ & Unweighted \\ 
Not Young or Binary & &  &  &  & $33.9 \pm 0.5$ & $17.7 \pm 0.2$ & $13.1 \pm 0.2$ & $40.4 \pm 0.4$ & $2.91 \pm 0.06$ & $|W|$ Weighted \\
Shallow\tablenotemark{a} & 175 & $-1.2 \pm 0.7$ & $-3.8 \pm 0.2$ & $0.2 \pm 0.2$ & $15.6 \pm 0.3$ & $5.8 \pm 0.1$ & $4.6 \pm 0.1$ & $17.3 \pm 0.3$ & $0.20 \pm 0.05$ & Unweighted \\ 
Wide Lower\tablenotemark{a} & 35 & $-32.4 \pm 0.8$ & $-26.8 \pm 1.9$ & $-9.0 \pm 0.6$ & $17.0 \pm 1.0$ & $40.7 \pm 1.9$ & $12.5 \pm 0.4$ & $39.4 \pm 1.2$ & $3.6 \pm 0.4$ & Unweighted \\ 
Wide Upper\tablenotemark{a} & 35 & $19.1 \pm 1.5$ & $9.0 \pm 0.6$ & $14.5 \pm 0.8$ & $32.7 \pm 1.4$ & $13.3 \pm 0.3$ & $17.5 \pm 0.5$ & $45.8 \pm 1.7$ & $5.6 \pm 0.6$ & Unweighted \\ 
\enddata
\tablecomments{Ages for unweighted velocities and $|W|$-weighted velocities are computed with the relation and parameters from \citet{Aumer:2009aa} and \citet{Wielen:1977aa}, respectively, following the implementation in \cite{Hsu:2021aa}.}
\tablenotetext{a}{Piece-wise linear fits to unweighted velocities, broken at $\sigma = \pm 1$
; see Section~\ref{sec:kinematic_age}.}
\end{deluxetable*}

We find an overall velocity dispersion of $\sigma_\text{tot} = 27.67 \pm 0.15$~{\kms}, which corresponds to a kinematic age of $\tau = 1.25 \pm 0.10$~Gyr using the \citet{Aumer:2009aa} relation. For the \citet{Wielen:1977aa} relation, the $W$-weighted velocity dispersion $\sigma_{W\text{-tot}} = 43.4 \pm 0.2$~{\kms} corresponds to a kinematic age of $\tau = 3.41 \pm 0.04$~Gyr. Compared to the late-M age of 4.0 $\pm$ 0.3~Gyr in \cite{Hsu:2021aa} based on the \citet{Aumer:2009aa} relation, our sample appears to have a younger average age, likely reflecting the sample bias toward young clusters.
Removing the 140
young cluster members increases the velocity dispersion to $\sigma_\text{tot} = 38.2 \pm 0.3$~{\kms}, corresponding to a
kinematic age of 3.30 $\pm$ 0.19~Gyr for the \citet{Aumer:2009aa} relation, in line with prior results \citep{Reiners:2009aa,Blake:2010aa,Burgasser:2015ac,Hsu:2021aa}. There is a also better agreement in this case with the $W$-weighted velocity dispersion and \citet{Wielen:1977aa} age of 4.13 $\pm$ 0.05~Gyr.
We also removed the 11 intermediate thin/thick disk and thick disk sources and 140 young sources, 
which resulted in an increased velocity dispersion and ``thin disk'' age of $2.11 \pm 0.14$~Gyr based on the \citet{Aumer:2009aa} relation.
For segregating these thin disk sources into M dwarfs (102 sources) and L dwarfs (4 sources) without young sources, we find similar velocity dispersions and kinematic ages ($1.97 \pm 0.14$~Gyr and $4.3 \pm 2.4$~Gyr, respectively), albeit with large uncertainties for the latter.
Finally, we examined removing the 42 RV variables from the thin disk sample; this had minimal influence on the inferred age ($1.36 \pm 0.11$~Gyr).

\begin{figure*}[!htbp]
\includegraphics[width=\textwidth, trim=10 10 0 0]{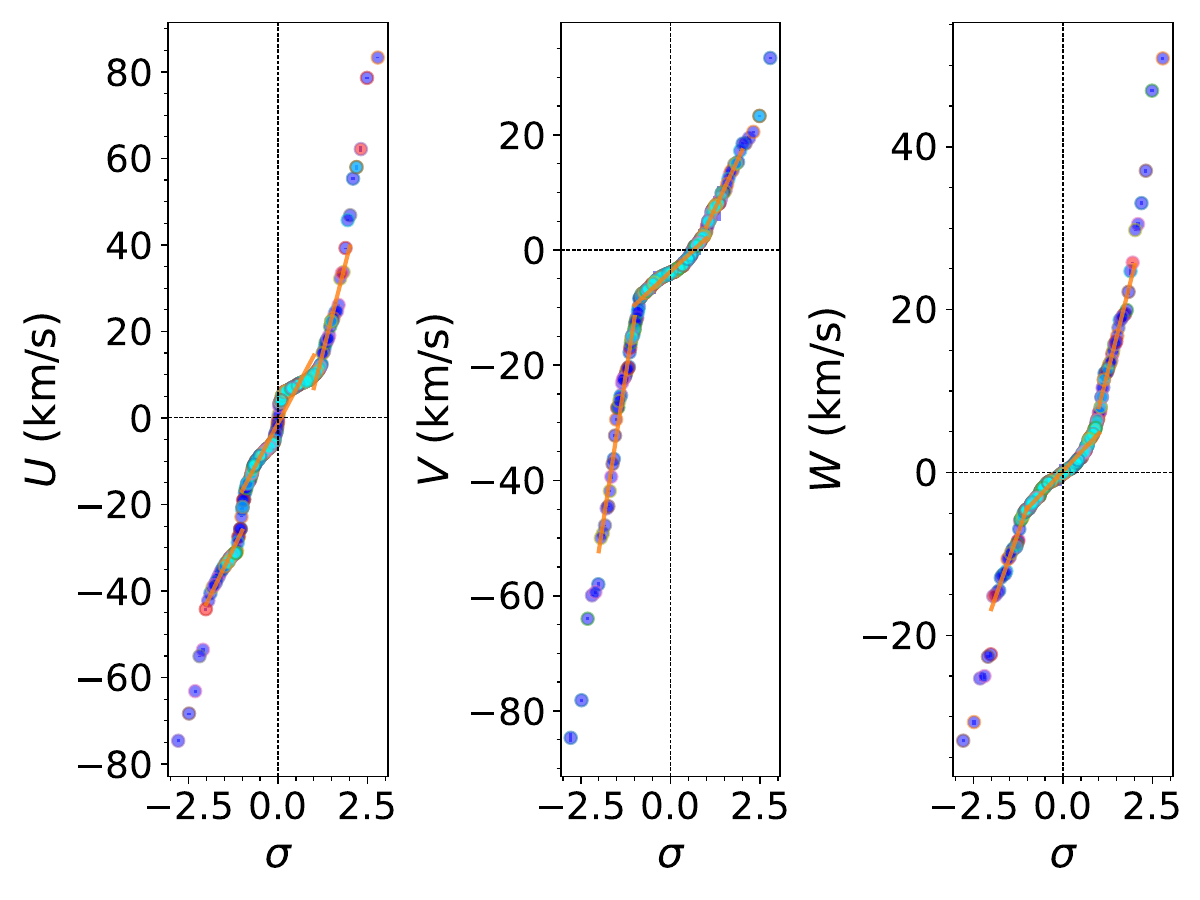}
\caption{Space velocity probit plots of the APOGEE sample. Individual velocities are indicated by blue and red circles for our M and L dwarfs, respectively, while a piece-wise linear fit broken at $\pm 1~\sigma$
is shown in orange solid/dashed lines for sources within/outside $1~\sigma$, respectively. 
Young sources are labeled in cyan (greenish-blue).
\label{fig:velocitydispersion}}
\end{figure*}

We can also discern distinct young and field populations using the velocity probability plot, or probit plot, that ranks the individual velocity components in steps of overall sample standard deviation. A normal distribution would be represented as a straight line in this diagram whose slope equals the sample dispersion \citep{Chambers:1983a, Reid:2002aa,Bochanski:2007aa}. 
Figure~\ref{fig:velocitydispersion} displays probit plots for each of the $UVW$ velocity components, all of which show two clear linear trends: a shallower ``core" sample and a steeper (and hence more dispersed) ``wide" sample.
A piece-wise linear fit to these trends broken at $\pm 1\sigma$
components yields total velocity dispersions of 
$\sigma_\mathrm{tot} = 17.3 \pm 0.3$~{\kms}, $\sigma_\mathrm{tot} = 39.4 \pm 1.2$~{\kms}, and $\sigma_\mathrm{tot} = 45.8 \pm 1.7$~{\kms} for the shallow, lower wide, and upper wide samples, respectively, corresponding to kinematic ages of 0.20 $\pm$ 0.05~Gyr, 3.6 $\pm$ 0.4~Gyr, and 5.6 $\pm$ 0.6~Gyr, based on the \citet{Aumer:2009aa} relation.
The shallow core sample is fully consistent with the thin disk sources (with young objects); the wide samples 
have older and similar ages, 
as expected for field age thin disk sources.

\subsection{Radial Velocity Variables and Candidate Multiple Systems} \label{sec:binaries}


One of the main stellar science goals of the APOGEE survey is to identify closely-separated binary systems, which are crucial for mass measurements and testing binary formation and evolution models. 
While the APOGEE ASPCAP pipeline is unable to provide robust RVs in the ultracool dwarf temperature regime, our RV precisions are sufficient to identify binaries at projected separations $\lesssim$1.1~AU from RV variability, assuming the total mass of 0.2~$M_{\odot}$, secondary mass of 5~M$_\mathrm{Jup}$ and RV precision of 0.3~{\kms}. 
Our sample contains 171 sources with at least two epochs of observations, 71 of which have four or more epochs.
Of the latter, 26 exhibit evidence of
significant RV variations ($p < 0.01$) based on a $\chi^2$ test,
and we consider these high probability binary systems. Among the 100 sources with 2 or 3 epochs of observations, 11 show significant RV variations, and we consider these promising binary candidates. All of the RV variables are listed in Table~\ref{table:binary_identify}.

\begin{deluxetable*}{ccccc}
\tablecaption{Radial Velocity Variations for Multi-epoch Observations\label{table:binary_identify}}
\tabletypesize{\normalsize}
\tablehead{
\colhead{} & \colhead{} & \colhead{} & \colhead{} & \colhead{} \\ 
\colhead{APOGEE ID} & \colhead{N$_\mathrm{obs}$\tablenotemark{a}} & \colhead{Mean RV} & \colhead{$\chi^2$} & \colhead{p-value} \\ 
\colhead{} & \colhead{} & \colhead{(km s$^{-1}$)} & \colhead{} & \colhead{}
}
\startdata
2MASS J00381273$+$3850323 & $6$ & $+6.69$ & $19.45$ & $0.002$ \\ 
2MASS J03282839$+$3116273 & $10$ & $+15.25$ & $39.8$ & $<0.001$ \\ 
2MASS J03284407$+$3120528 & $18$ & $+17.41$ & $108.18$ & $<0.001$ \\ 
2MASS J03290413$+$3056127 & $17$ & $+15.53$ & $60.28$ & $<0.001$ \\ 
2MASS J03291130$+$3117175 & $15$ & $+16.57$ & $97.11$ & $<0.001$ \\ 
2MASS J03293773$+$3122024 & $19$ & $+17.02$ & $59.11$ & $<0.001$ \\ 
2MASS J03413332$+$3157417 & $17$ & $+7.45$ & $50.42$ & $<0.001$ \\ 
2MASS J03413641$+$3216200 & $4$ & $+18.56$ & $19.43$ & $<0.001$ \\ 
2MASS J03440291$+$3152277 & $12$ & $+17.53$ & $39.87$ & $<0.001$ \\ 
2MASS J03440599$+$3215321 & $7$ & $+18.85$ & $32.63$ & $<0.001$ \\ 
2MASS J04161885$+$2752155 & $4$ & $+18.43$ & $32.63$ & $<0.001$ \\ 
2MASS J04254894$+$1852479 & $3$ & $+28.28$ & $11.13$ & $0.004$ \\ 
2MASS J05402570$+$2448090 & $4$ & $+22.93$ & $28.17$ & $<0.001$ \\ 
2MASS J09373349$+$5534057 & $20$ & $+0.8$ & $109.65$ & $<0.001$ \\ 
2MASS J09453388$+$5458511 & $10$ & $-3.78$ & $22.9$ & $0.006$ \\ 
2MASS J09522188$-$1924319 & $3$ & $-17.09$ & $53.22$ & $<0.001$ \\ 
2MASS J10225090$+$0032169 & $3$ & $+34.09$ & $1172.27$ & $<0.001$ \\ 
2MASS J10541102$-$8505023 & $2$ & $-10.55$ & $45.52$ & $<0.001$ \\ 
2MASS J11203609$+$0704135 & $3$ & $-14.47$ & $11.74$ & $0.003$ \\ 
2MASS J11232934$+$0154040 & $3$ & $-0.13$ & $14.16$ & $0.001$ \\ 
2MASS J12080810$+$3520281 & $2$ & $+29.19$ & $8.46$ & $0.004$ \\ 
2MASS J12261350$+$5605445 & $5$ & $+4.13$ & $20.64$ & $<0.001$ \\ 
2MASS J12265349$+$2543556 & $8$ & $-0.25$ & $29.59$ & $<0.001$ \\ 
2MASS J12481860$-$0235360 & $14$ & $+31.27$ & $39.9$ & $<0.001$ \\ 
2MASS J13122681$+$7245338 & $10$ & $-3.47$ & $27.96$ & $0.001$ \\ 
2MASS J13202007$+$7213140 & $12$ & $-27.34$ & $35.62$ & $<0.001$ \\ 
2MASS J13232423$+$5132272 & $7$ & $-5.39$ & $17.91$ & $0.006$ \\ 
2MASS J13495109$+$3305136 & $13$ & $-9.51$ & $44.38$ & $<0.001$ \\ 
2MASS J13564148$+$4342587 & $3$ & $-19.5$ & $13.18$ & $0.001$ \\ 
2MASS J14005977$+$3226109 & $16$ & $-17.89$ & $70.77$ & $<0.001$ \\ 
2MASS J15010818$+$2250020 & $5$ & $+6.55$ & $44.72$ & $<0.001$ \\ 
2MASS J16002844$-$2209228 & $3$ & $-3.56$ & $19.69$ & $<0.001$ \\ 
2MASS J16090451$-$2224523 & $3$ & $-5.15$ & $43.57$ & $<0.001$ \\ 
2MASS J16114261$-$2525511 & $2$ & $-4.78$ & $10.0$ & $0.002$ \\ 
2MASS J16271825$+$3538347 & $4$ & $-8.56$ & $48.52$ & $<0.001$ \\ 
2MASS J22551142$+$1442456 & $5$ & $-14.27$ & $18.55$ & $0.001$ 
\enddata
\tablenotetext{a}{We regard sources with numbers of observations $<$ 4 promising candidates; see Section~\ref{sec:binaries} for more details.}
\end{deluxetable*}

The majority of the RV variables have too few epochs to fully sample a complete orbit, and hence only partial constraints can be made on orbital parameters. 
We attempted to make these constraints for each RV variable with at least four epochs of observation using \texttt{The Joker} \citep{Price-Whelan:2017aa}, a Monte Carlo rejection sampler that quantifies single-line RV orbits in terms of period ($P$), velocity variation semi-amplitude ($K$), eccentricity ($e$), systemic velocity ($v_0$), mean anomaly ($M_0$), and argument of periastron ($\omega$), and identifies a family of orbits consistent with the measurements. 
We ran \texttt{The Joker} using its default settings.
For each system, we initially selected limiting ranges for the minimum and maximum orbital period ($P_\text{min}$ and $P_\text{max}$), the maximum RV semi-amplitude ($K_0$), and the number of input samplers (10$^5$ $\leq$ $N_{samp}$ $\leq$ 5$\times$10$^6$). 
Initial estimates of $v_0$ and $K$ were determined from the mean and standard deviation of RV measurements), and both of these quantities were assumed to follow normal distributions with scale factors $\sigma_K$, $\sigma_v$ = 1--4~{\kms} constrained from the $\Delta$RV variations in the observed RV time series. 
The period distribution was assumed to follow $\mathcal{P}(P) \propto {P}^{-1}$ following \cite{Uehara:2016aa, Price-Whelan:2017aa}; and \citet{Kipping:2018aa}. 
The eccentricity distribution was assumed to follow a Beta distribution
\begin{equation}
\mathcal{P}(e) = \frac{\Gamma(a + b)}{\Gamma(a)+\Gamma(a)} e^{a-1}\big[ 1 - e \big]^{b-1},
\end{equation} 
where $\Gamma$ is the Gamma function and $a = 0.867$ and $b = 3.03$ fixed following \citet{Kipping:2013aa}.
The distributions of mean anomaly and periastron angle were assumed to be uniformly distributed between 0 and 2$\pi$.
The standard deviation of the RV semi-amplitude $\sigma_K$ prior, was assumed to be a normal distribution defined as 
\begin{equation}
    \sigma_K^2 = \sigma_{K, 0}^2 \, \left(\frac{P}{1~{\rm year}}\right)^{-2/3} \,
            \left(1 - e^2\right)^{-1}.
\end{equation}
This form was chosen because the RV semi-amplitude of the primary is $K_1 = \sqrt{\frac{G}{(1-e^2)}} m_2 \sin{i} (m_1+m_2)^{-1/2} a^{-1/2} = G^{1/3}(2\pi)^{-1/3} (1-e^2)^{-1/2} m_2 \sin{i} (m_1 + m_2)^{-2/3} P^{-1/3}$, with $G$ the gravitational constant, and $m_1$ and $m_2$ the masses of the primary and secondary, $i$ the orbital inclination $i$, and $a$ the semi-major axis $a$. As such, the adopted prior for $\sigma_K$ has the advantage that the RV semi-amplitude $K$ has a fixed form for a given primary mass independent of period and eccentricity \citep{Price-Whelan:2020aa}.

\begin{longrotatetable}
\startlongtable
\begin{deluxetable*}{cclcccrrcccc}
\tablecaption{Binary Candidates and Orbital Parameters \label{table:binary_thejoker}}
\tabletypesize{\footnotesize}
\tablehead{
\colhead{APOGEE ID} & \colhead{$N_\mathrm{obs}$} & \colhead{$P_\mathrm{Prior}$ } & \colhead{$K_\mathrm{0, Prior}$} & \colhead{$\sigma_{K_0}$} & \colhead{$\sigma_{v}$} & \colhead{$N_\mathrm{sam, in}$} & \colhead{$P$} & \colhead{$K_1$} & \colhead{$e$} & \textit{Gaia} & $\Delta$BIC \\ 
\colhead{} & \colhead{} & \colhead{(day)} & \colhead{(km s$^{-1}$)} & \colhead{(km s$^{-1}$)} & \colhead{(km s$^{-1}$)} & \colhead{} & \colhead{(day)} & \colhead{(km s$^{-1}$)} & \colhead{} & \colhead{RUWE} & \colhead{}
} 
\startdata
\hline
\multicolumn{12}{c}{$\Delta$BIC $>$ 10: Very strong/Highly probable multiple systems} \\
\hline
2MASS J03282839$+$3116273 & 10 & 1--300 & 3.0 & 2.0 & 2.0 & 1000000 & $24.1^{+86.9}_{-21.8}$ & $1.0^{+0.3}_{-0.2}$ & $0.175^{+0.245}_{-0.132}$ & 0.995 & 26 \\ 
2MASS J03284407$+$3120528 & 18 & 1--1000 & 6.0 & 3.0 & 3.0 & 10000000 & $4.2^{+21.8}_{-3.0}$ & $1.3^{+0.2}_{-0.2}$ & $0.195^{+0.259}_{-0.192}$ & 1.093 & 97 \\ 
2MASS J03290413$+$3056127 & 17 & 1--1000 & 3.0 & 2.0 & 2.0 & 10000000 & $73.3^{+2.4}_{-35.5}$ & $0.6^{+0.1}_{-0.1}$ & $0.265^{+0.251}_{-0.181}$ & 0.886 & 43 \\ 
2MASS J03291130$+$3117175 & 15 & 1--1000 & 5.0 & 3.0 & 3.0 & 100000000 & $15.2^{+0.0}_{-0.0}$ & $1.4^{+0.4}_{-0.2}$ & $0.503^{+0.093}_{-0.178}$ & 1.170 & 77 \\ 
2MASS J03293773$+$3122024 & 19 & 1--500 & 6.0 & 3.0 & 3.0 & 10000000 & $7.4^{+17.5}_{-5.8}$ & $0.7^{+0.2}_{-0.1}$ & $0.187^{+0.266}_{-0.137}$ & 1.051 & 30 \\ 
2MASS J03413332$+$3157417 & 17 & 1--500 & 5.0 & 3.0 & 3.0 & 10000000 & $10.5^{+69.6}_{-8.7}$ & $0.6^{+0.1}_{-0.2}$ & $0.144^{+0.218}_{-0.111}$ & 4.749 & 27 \\ 
2MASS J03413641$+$3216200 & 4 & 1--500 & 4.0 & 3.0 & 3.0 & 10000000 & $4.3^{+3.3}_{-2.6}$ & $2.0^{+1.3}_{-0.7}$ & $0.201^{+0.281}_{-0.155}$ & 1.081 & 57 \\ 
2MASS J03440291$+$3152277 & 12 & 1--500 & 4.0 & 3.0 & 3.0 & 10000000 & $4.5^{+2.2}_{-2.7}$ & $0.9^{+0.4}_{-0.2}$ & $0.384^{+0.255}_{-0.259}$ & 1.130 & 24 \\ 
2MASS J03440599$+$3215321 & 7 & 1--500 & 4.0 & 3.0 & 3.0 & 10000000 & $75.0^{+182.9}_{-70.4}$ & $1.0^{+0.6}_{-0.3}$ & $0.168^{+0.252}_{-0.127}$ & 0.948 & 25 \\ 
2MASS J04161885$+$2752155 & 4 & 1--500 & 3.0 & 3.0 & 3.0 & 10000000 & $33.0^{+169.9}_{-30.1}$ & $1.3^{+1.0}_{-0.4}$ & $0.182^{+0.259}_{-0.141}$ & 1.157 & 55 \\ 
2MASS J05402570$+$2448090 & 4 & 1--800 & 4.0 & 3.0 & 3.0 & 1000000 & $27.8^{+49.5}_{-24.8}$ & $2.3^{+1.2}_{-0.9}$ & $0.186^{+0.268}_{-0.143}$ & \nodata & 52 \\  
2MASS J09453388$+$5458511 & 10 & 1--800 & 4.0 & 3.0 & 2.0 & 5000000 & $15.1^{+119.8}_{-12.7}$ & $1.0^{+0.5}_{-0.5}$ & $0.315^{+0.298}_{-0.244}$ & 0.947 & 20 \\ 
2MASS J12261350$+$5605445 & 5 & 1--500 & 4.0 & 3.0 & 3.0 & 10000000 & $14.9^{+48.3}_{-10.5}$ & $1.4^{+0.9}_{-0.5}$ & $0.227^{+0.299}_{-0.174}$ & 1.253 & 78 \\ 
2MASS J12265349$+$2543556 & 8 & 1--100 & 3.0 & 3.0 & 2.0 & 10000000 & $12.6^{+2.7}_{-9.1}$ & $0.9^{+0.3}_{-0.2}$ & $0.172^{+0.236}_{-0.132}$ & 0.906 & 50 \\ 
2MASS J12481860$-$0235360 & 14 & 1--100 & 3.0 & 3.0 & 2.0 & 10000000 & $9.4^{+4.5}_{-1.8}$ & $0.8^{+0.4}_{-0.3}$ & $0.346^{+0.282}_{-0.263}$ & 0.993 & 31 \\ 
2MASS J13122681$+$7245338 & 10 & 1--2000 & 4.0 & 3.0 & 3.0 & 10000000 & $7.9^{+1046.4}_{-5.9}$ & $1.3^{+0.8}_{-0.4}$ & $0.188^{+0.267}_{-0.144}$ & 1.273 & 18 \\ 
2MASS J13202007$+$7213140 & 12 & 20--150 & 3.0 & 2.0 & 3.0 & 5000000 & $52.6^{+4.1}_{-16.0}$ & $0.9^{+0.6}_{-0.5}$ & $0.338^{+0.268}_{-0.213}$ & 1.332 & 53 \\ 
2MASS J13232423$+$5132272 & 7 & 1--50 & 3.0 & 2.0 & 2.0 & 1000000 & $3.6^{+8.0}_{-2.0}$ & $0.7^{+0.4}_{-0.3}$ & $0.207^{+0.332}_{-0.154}$ & 1.090 \& 0.918 & 23 \\ 
2MASS J13495109$+$3305136 & 13 & 1--100 & 5.0 & 3.0 & 3.0 & 10000000 & $2.1^{+2.1}_{-0.7}$ & $2.0^{+1.0}_{-0.7}$ & $0.25^{+0.32}_{-0.194}$ & 0.927 & 40 \\ 
2MASS J14005977$+$3226109 & 16 & 1--600 & 3.0 & 3.0 & 3.0 & 5000000 & $1.4^{+321.0}_{-0.3}$ & $1.5^{+1.0}_{-0.4}$ & $0.419^{+0.258}_{-0.255}$ & 0.986 & 48 \\ 
2MASS J15010818$+$2250020 & 5 & 5--50 & 3.0 & 3.0 & 2.0 & 1000000 & $23.3^{+1.3}_{-11.0}$ & $1.2^{+0.7}_{-0.3}$ & $0.189^{+0.284}_{-0.142}$ & 1.661 & 92 \\ 
2MASS J16271825$+$3538347 & 4 & 60--200 & 5.0 & 4.0 & 3.0 & 5000000 & $69.7^{+17.6}_{-7.4}$ & $3.4^{+2.1}_{-1.3}$ & $0.524^{+0.175}_{-0.238}$ & 0.976 & 41 \\  
2MASS J22551142$+$1442456 & 5 & 1--50 & 3.0 & 3.0 & 2.0 & 100000 & $2.8^{+4.8}_{-1.1}$ & $1.6^{+0.9}_{-0.5}$ & $0.178^{+0.256}_{-0.135}$ & 1.091 & 18 \\
\hline
\multicolumn{12}{c}{$\Delta$BIC $\leq$ 10: Strong candidate multiple systems} \\
\hline
2MASS J00381273$+$3850323 & 6 & 1--1000 & 3.0 & 3.0 & 3.0 & 10000000 & $191.6^{+441.8}_{-174.2}$ & $0.9^{+0.7}_{-0.3}$ & $0.183^{+0.267}_{-0.141}$ & 1.053 & 10 \\ 
2MASS J04230607$+$2801194 & 6 & 1--500 & 3.0 & 3.0 & 3.0 & 10000000 & $8.1^{+25.6}_{-5.7}$ & $0.7^{+0.5}_{-0.3}$ & $0.192^{+0.274}_{-0.148}$ & 1.036 & 7 \\ 
2MASS J09373349$+$5534057 & 20 & 1--100 & 4.0 & 3.0 & 2.0 & 5000000 & $5.7^{+13.5}_{-3.9}$ & $0.4^{+0.3}_{-0.2}$ & $0.271^{+0.297}_{-0.199}$ & 1.103 \& \nodata & 7 \\ 
\enddata
\end{deluxetable*}
\end{longrotatetable}

\begin{figure}
\centering
\includegraphics[width=\linewidth, trim=0 0 10 0]{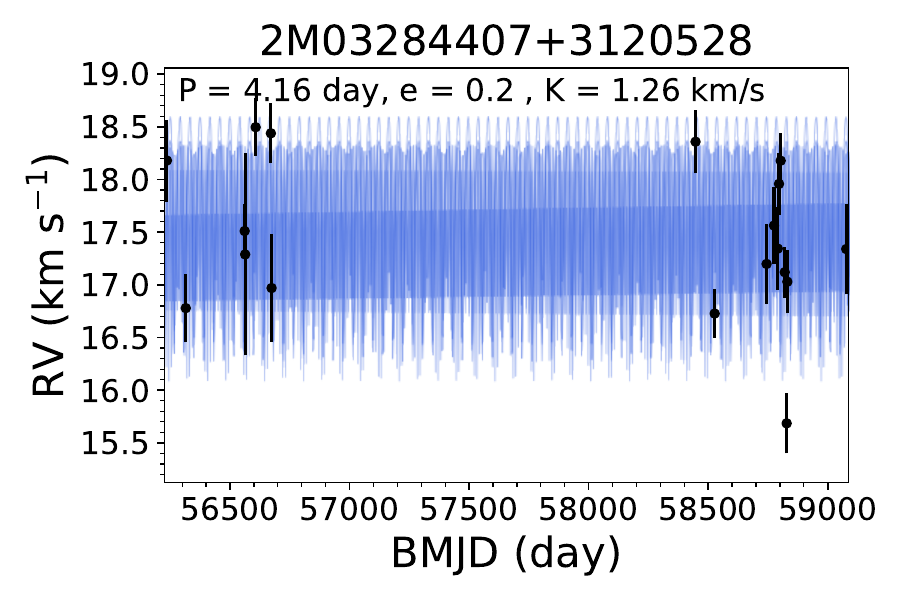}
\includegraphics[width=\linewidth, trim=0 0 10 0]{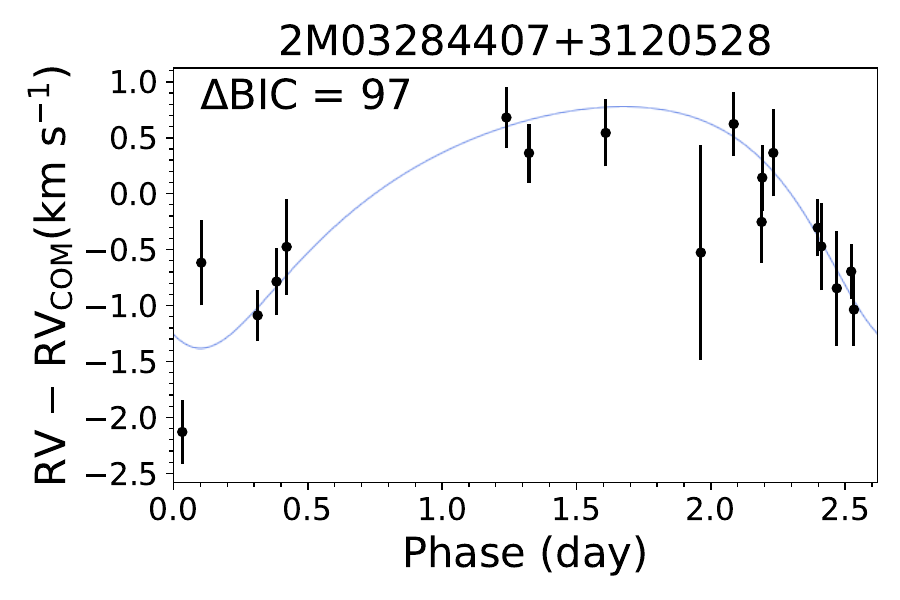}
\caption{Binary orbital fit for 2MASS J03284407$+$3120528 using the Monte Carlo rejection sampler \texttt{The Joker}.
\textit{top}: RV time series of 2MASS J03284407$+$3120528 (black dots) and possible orbital solutions (blue lines), labeled with the corresponding parameter estimates. 
\textit{bottom}: Phase-folded RV time series (systematic RV corrected RVs vs. phase in days; black dots) with the median orbital solution estimated (blue line) from the Monte Carlo rejection sampler \texttt{The Joker}. The $\Delta$BIC (Bayesian information criterion) between the median orbital solution and a flat line centered at the systematic RV is labeled.
The inferred orbital parameters are $P_\mathrm{fit}$ = $4.2^{+21.8}_{-3.0}$~day, $K_\mathrm{1,fit}$ = $1.3 \pm 0.2$~km s$^{-1}$~km s$^{-1}$, and $e_\mathrm{fit}$ = $0.20^{+0.26}_{-0.19}$. 
}
\label{fig:binary_orbit_fit}
\end{figure}

Results for these fits are provided in Table~\ref{table:binary_thejoker}, and individual fits to all RV variables are provided in Appendix~\ref{appendix:binary_orbit}.
To assess the robustness of these fits, we computed the relative Bayesian Information Criterion $\Delta$BIC between the best orbital solution and a constant RV, where BIC = $\chi^2 + k \ln{n}$, $\chi^2$ is the chi-square quality of fit (see Eqn.~3), $k$ is the number of model parameters (6 for the full orbit and 1 for a constant RV), and $n$ the number of RV measurements \citep{schwarz1978}.
$\Delta$BIC ranges of 0--2, 2--6, 6--10, and $>$10 correspond to insignificant, positive, strong, and very strong evidence against the null hypothesis (constant RV), respectively \citep{Kass:1995aa}.
Among the 25 sources with significant RV variations
we found three to be strong,
and 22 sources to be very strong; the last set we consider to be highly probable binary candidates.
Figure~\ref{fig:binary_orbit_fit} illustrates an example orbit fit for the binary candidate with the largest $\Delta$BIC and $K_1$, 2MASS J03284407$+$3120528, which has 18 epochs of observations, from which we were able to constrain a period $P_\mathrm{fit}$ = $4.2^{+21.8}_{-3.0}$~day, an RV semi-amplitude  $K_\mathrm{fit}$ = $1.3 \pm 0.2$~km s$^{-1}$, and an eccentricity $e_\mathrm{fit}$ = $0.20^{+0.26}_{-0.19}$.
There are nine highly probable binaries with estimated orbit periods less than 10~days (Table~\ref{table:binary_thejoker}):
2MASS~J14005977+3226109 ($P = 1.4^{+321.9}_{-0.3}$~day\footnote{This orbit is poorly constrained at the upper bound as long-period solutions are also possible.}), 2MASS~J13495109+3305136 ($P = 2.1^{+2.1}_{-0.7}$~day), 2MASS~J22551142+1442456 ($P = 2.8^{+4.8}_{-1.1}$~day), 2MASS~J13232423+5132272 ($P = 3.6^{+8.0}_{-2.0}$~day), 2MASS~J03284407+3120528 ($P = 4.2^{+21.8}_{-3.0}$~day), 2MASS~J03413641+3216200 ($P = 4.3^{+3.3}_{-2.6}$~day), 2MASS~J03440291+3152277 ($P = 4.5^{+2.2}_{-2.7}$~day), 2MASS~J03293773+3122024 ($P = 7.4^{+17.5}_{-5.8}$~day), and 2MASS~J12481860$-$0235360 ($P = 9.4^{+4.5}_{-1.8}$~day).
We note that the shortest-period confirmed binary in our sample is 2MASS J03505737+1818069 (LP 413$-$53), which has only one epoch of APOGEE data, but follow-up observations have determined this source to be a 0.71~day period UCD binary \citep{Hsu:2023aa}.
The remaining highly probable binaries have estimated orbit periods between 10--100 days.
It is important to note that the sparse sampling of these binary candidates results in large uncertainties on period (median uncertainty = 29~day)
and eccentricity (median uncertainty = 0.21), and more complete sampling of the RV orbit
is required to both confirm and robustly constrain orbital parameters.

\subsection{Rotation Periods, Projected Radii, and Inclinations} \label{sec:rsini}

The rotational broadening values inferred from our fits are directly related to rotation period, size, and inclination. While these cannot be disentangled from the spectral measurements alone, some constraints can be made on size and viewing geometry when
an independent measure of rotation period, such as photometric variability, is available.
The projected radius $R\sin{i}$ in particular can be inferred directly from {\vsini} and period as
\begin{equation}
    \frac{R\sin{i}}{R_{\odot}} = 0.0198\frac{P}{\rm day}\frac{v\sin{i}}{\rm km/s}.
\end{equation}
Thanks to the \textit{K2} Mission \citep{Howell:2014aa}, the \textit{Transiting Exoplanet Survey Satellite} (\textit{TESS}; \citealt{Ricker:2015aa}), and ground-based photometric monitoring programs, rotational periods have been measured for several UCDs in our APOGEE sample, in particular for members of the Upper Scorpius cluster.
Noting that roughly one-half of our sample is kinematically associated with nearby young clusters with known ages, these measurements can be used to examine radius evolution as a function of time to test 
evolutionary models \citep{Jackson:2010aa}.

\startlongtable
\begin{deluxetable*}{llccccccc}
\tablecaption{Inferred Projected Radii, Inclination and Literature Period Measurements \label{table:inclination}}
\tablewidth{700pt}
\tabletypesize{\small}
\tablehead{
\colhead{APOGEE ID} & \colhead{Cluster\tablenotemark{a}} & \colhead{Age\tablenotemark{b}} & \colhead{$v\sin{i}$} & \colhead{Period\tablenotemark{c}} & \colhead{$R\sin{i}$} & \colhead{inclination} & \colhead{Period Ref} \\ 
\colhead{} & \colhead{} & \colhead{(Myr)} & \colhead{(km s$^{-1}$)} & \colhead{(day)}& \colhead{(R$_{\odot}$)} & \colhead{(deg)} & \colhead{}
} 
\startdata
2MASS J00452143$+$1634446 & ARG & 40--50 & $31.6\pm1.0$ & $0.1\pm0.004$ & $0.06\pm0.0$ & $23\pm3$ & (14) \\ 
2MASS J01243124$-$0027556 & field & \nodata & $27.7\pm1.2$ & 0.555 & $0.3\pm0.03$ & \nodata & (2) \\ 
2MASS J03040207$+$0045512 & HYA & 750 $\pm$ 100 & $20.0\pm1.0$ & 1.293 & $0.51\pm0.06$ & \nodata & (6) \\ 
2MASS J04110642$+$1247481 & HYA & 750 $\pm$ 100 & $14.2\pm1.1$ & 0.897 & $0.25\pm0.02$ & \nodata & (5) \\ 
2MASS J04214435$+$2024105 & HYA & 750 $\pm$ 100 & $32.8\pm1.4$ & 0.34 & $0.22\pm0.02$ & \nodata & (9), (11) \\ 
2MASS J04214955$+$1929086 & HYA & 750 $\pm$ 100 & $44.4\pm1.6$ & 0.205 & $0.18\pm0.01$ & \nodata & (5) \\ 
2MASS J04254894$+$1852479 & field & \nodata & $31.0\pm1.0$ & 0.419 & $0.26\pm0.02$ & \nodata & (1) \\ 
2MASS J04305718$+$2556394 & TAU & 1--2 & $16.0\pm1.0$ & 1.157 & $0.36\pm0.03$ & $60\pm19$ & (13) \\ 
2MASS J04322329$+$2403013 & TAU & 1--2 & $11.1\pm1.0$ & 3.364 & $0.73\pm0.08$ & \nodata & (13) \\ 
2MASS J04330945$+$2246487 & TAU & 1--2 & $16.0\pm1.0$ & 3.492 & $1.11\pm0.1$ & \nodata & (13) \\ 
2MASS J04340619$+$2418508 & TAU & 1--2 & $29.5\pm1.8$ & 0.711 & $0.41\pm0.03$ & $60\pm18$ & (13) \\ 
2MASS J04350850$+$2311398 & TAU & 1--2 & $7.3\pm1.2$ & 1.498 & $0.21\pm0.04$ & \nodata & (13) \\ 
2MASS J04351354$+$2008014 & HYA & 750 $\pm$ 100 & $23.7\pm1.1$ & 0.37 & $0.17\pm0.01$ & \nodata & (9), (11) \\ 
2MASS J04354183$+$2234115 & TAU & 1--2 & $48.2\pm1.2$ & 0.688 & $0.65\pm0.04$ & \nodata & (13) \\ 
2MASS J04361038$+$2259560 & TAU & 1--2 & $10.4\pm1.0$ & 2.933 & $0.6\pm0.07$ & \nodata & (13) \\ 
2MASS J04363893$+$2258119 & TAU & 1--2 & $16.8\pm1.0$ & 0.964 & $0.32\pm0.03$ & $52\pm13$ & (13) \\ 
2MASS J04385871$+$2323595 & TAU & 1--2 & $51.2\pm1.1$ & 0.664 & $0.67\pm0.04$ & \nodata & (13) \\ 
2MASS J04440164$+$1621324 & TAU & 1--2 & $14.3\pm1.2$ & 2.172 & $0.61\pm0.07$ & \nodata & (13) \\ 
2MASS J04464498$+$2436404 & HYA & 750 $\pm$ 100 & $17.9\pm1.0$ & 0.66 & $0.23\pm0.01$ & \nodata & (9), (11) \\ 
2MASS J05352501$-$0509095 & ORION & $\lesssim$3 & $17.9\pm1.1$ & 1.48 & $0.52\pm0.04$ & \nodata & (1) \\ 
2MASS J05353193$-$0531477 & ORION & $\lesssim$3 & $15.8\pm1.1$ & 4.36 & $1.36\pm0.12$ & \nodata & (1) \\ 
2MASS J05402570$+$2448090 & ARG & 40--50 & $30.4\pm1.0$ & 0.294 & $0.18\pm0.01$ & $70\pm9$ & (6) \\ 
2MASS J07464256$+$2000321 & field & \nodata & $34.6\pm1.0$ & 0.086 & $0.06\pm0.0$ & $36\pm5$ & (3) \\ 
2MASS J08072607$+$3213101 & field & \nodata & $13.7\pm1.0$ & 0.345 & $0.09\pm0.01$ & $61\pm11$ & (6) \\ 
2MASS J08294949$+$2646348 & CMG & 200 & $11.9\pm1.0$ & 0.459 & $0.11\pm0.01$ & $57\pm11$ & (6) \\ 
2MASS J10372897$+$3011117 & field & \nodata & $13.1\pm1.1$ & 1.012 & $0.26\pm0.03$ & \nodata & (6) \\ 
2MASS J13022083$+$3227103 & field & \nodata & $25.7\pm1.0$ & 0.4 & $0.2\pm0.02$ & \nodata & (6) \\ 
2MASS J13564148$+$4342587 & field & \nodata & $16.2\pm1.1$ & 0.477 & $0.15\pm0.01$ & \nodata & (6) \\ 
2MASS J14320849$+$0811313 & field & \nodata & $9.2\pm1.1$ & 0.757 & $0.14\pm0.02$ & \nodata & (6) \\ 
2MASS J15010818$+$2250020 & field & \nodata & $65.3\pm1.1$ & 0.082 & $0.11\pm0.01$ & $69\pm8$ & (4) \\ 
2MASS J15555600$-$2045187 & USCO & 10 $\pm$ 3 & $19.9\pm1.1$ & 1.7 & $0.67\pm0.05$ & \nodata & (7) \\ 
2MASS J15560104$-$2338081 & USCO & 10 $\pm$ 3 & $13.0\pm1.3$ & 1.505 & $0.39\pm0.03$ & \nodata & (7) \\ 
2MASS J15560497$-$2106461 & USCO & 10 $\pm$ 3 & $92.8\pm1.5$ & 0.26 & $0.48\pm0.03$ & \nodata & (1) \\ 
2MASS J15591135$-$2338002 & USCO & 10 $\pm$ 3 & $16.1\pm1.5$ & 1.216 & $0.39\pm0.04$ & \nodata & (1) \\ 
2MASS J15592591$-$2305081 & USCO & 10 $\pm$ 3 & $23.5\pm1.0$ & 0.62 & $0.29\pm0.02$ & $70\pm10$ & (7) \\ 
2MASS J16003023$-$2334457 & USCO & 10 $\pm$ 3 & $73.6\pm1.3$ & 0.448 & $0.65\pm0.04$ & \nodata & (7) \\ 
2MASS J16014955$-$2351082 & USCO & 10 $\pm$ 3 & $38.7\pm1.8$ & 0.527 & $0.4\pm0.02$ & \nodata & (7) \\ 
2MASS J16020429$-$2050425 & USCO & 10 $\pm$ 3 & $57.1\pm1.2$ & 0.422 & $0.48\pm0.03$ & \nodata & (7) \\ 
2MASS J16044303$-$2318258 & USCO & 10 $\pm$ 3 & $79.3\pm1.9$ & 0.208 & $0.33\pm0.02$ & \nodata & (1) \\ 
2MASS J16063110$-$1904576 & USCO & 10 $\pm$ 3 & $16.7\pm1.1$ & 2.301 & $0.76\pm0.09$ & \nodata & (7) \\ 
2MASS J16090168$-$2740521 & USCO & 10 $\pm$ 3 & $59.3\pm1.4$ & 0.306 & $0.36\pm0.02$ & \nodata & (7) \\ 
2MASS J16090197$-$2151225 & USCO & 10 $\pm$ 3 & $47.8\pm1.5$ & 0.269 & $0.25\pm0.01$ & $62\pm11$ & (7) \\ 
2MASS J16090451$-$2224523 & USCO & 10 $\pm$ 3 & $14.1\pm1.0$ & 2.181 & $0.61\pm0.06$ & \nodata & (7) \\ 
2MASS J16093019$-$2059536 & USCO & 10 $\pm$ 3 & $17.9\pm1.0$ & 1.593 & $0.57\pm0.03$ & \nodata & (7) \\ 
2MASS J16095107$-$2722418 & USCO & 10 $\pm$ 3 & $56.1\pm1.4$ & 0.543 & $0.6\pm0.03$ & \nodata & (7) \\ 
2MASS J16095217$-$2136277 & USCO & 10 $\pm$ 3 & $44.1\pm1.0$ & 0.702 & $0.61\pm0.03$ & \nodata & (7) \\ 
2MASS J16095852$-$2345186 & USCO & 10 $\pm$ 3 & $34.7\pm1.7$ & 1.411 & $0.96\pm0.07$ & \nodata & (7) \\ 
2MASS J16095990$-$2155424 & USCO & 10 $\pm$ 3 & $22.0\pm1.8$ & 0.874 & $0.38\pm0.03$ & \nodata & (7) \\ 
2MASS J16100541$-$1919362 & USCO & 10 $\pm$ 3 & $16.2\pm1.0$ & 2.552 & $0.81\pm0.07$ & \nodata & (1) \\ 
2MASS J16105499$-$2126139 & USCO & 10 $\pm$ 3 & $57.8\pm1.2$ & 0.52 & $0.59\pm0.03$ & \nodata & (7) \\ 
2MASS J16111711$-$2217173 & USCO & 10 $\pm$ 3 & $63.9\pm1.5$ & 0.362 & $0.46\pm0.03$ & \nodata & (7) \\ 
2MASS J16113837$-$2307072 & USCO & 10 $\pm$ 3 & $36.6\pm1.1$ & 0.72 & $0.52\pm0.03$ & \nodata & (7) \\ 
2MASS J16114261$-$2525511 & USCO & 10 $\pm$ 3 & $55.3\pm1.1$ & 0.629 & $0.69\pm0.04$ & \nodata & (7) \\ 
2MASS J16115439$-$2236491 & USCO & 10 $\pm$ 3 & $50.3\pm1.5$ & 0.484 & $0.48\pm0.03$ & \nodata & (7) \\ 
2MASS J16122703$-$2013250 & USCO & 10 $\pm$ 3 & $30.9\pm1.0$ & 0.888 & $0.54\pm0.03$ & \nodata & (7) \\ 
2MASS J16124692$-$2338408 & USCO & 10 $\pm$ 3 & $60.4\pm1.4$ & 0.284 & $0.34\pm0.02$ & \nodata & (7) \\ 
2MASS J16124726$-$1903531 & USCO & 10 $\pm$ 3 & $29.0\pm1.1$ & 1.188 & $0.68\pm0.05$ & \nodata & (7) \\ 
2MASS J16131211$-$2305031 & USCO & 10 $\pm$ 3 & $23.5\pm1.0$ & 1.154 & $0.54\pm0.04$ & \nodata & (7) \\ 
2MASS J16132665$-$2230348 & USCO & 10 $\pm$ 3 & $19.9\pm1.9$ & 1.532 & $0.6\pm0.06$ & \nodata & (7) \\ 
2MASS J16132809$-$1924524 & USCO & 10 $\pm$ 3 & $23.1\pm1.0$ & 1.512 & $0.69\pm0.06$ & \nodata & (7) \\ 
2MASS J16134027$-$2233192 & USCO & 10 $\pm$ 3 & $14.1\pm1.0$ & 1.716 & $0.48\pm0.02$ & \nodata & (1) \\ 
2MASS J16134079$-$2219459 & USCO & 10 $\pm$ 3 & $19.4\pm1.0$ & 1.336 & $0.51\pm0.03$ & \nodata & (7) \\ 
2MASS J16141974$-$2428404 & USCO & 10 $\pm$ 3 & $57.4\pm1.3$ & 0.346 & $0.39\pm0.02$ & \nodata & (7) \\ 
2MASS J16143287$-$2242133 & USCO & 10 $\pm$ 3 & $18.0\pm1.0$ & 1.823 & $0.65\pm0.03$ & \nodata & (7) \\ 
2MASS J16152516$-$2144013 & USCO & 10 $\pm$ 3 & $16.0\pm1.0$ & 1.744 & $0.55\pm0.04$ & \nodata & (7) \\ 
2MASS J16155507$-$2444365 & USCO & 10 $\pm$ 3 & $16.0\pm1.0$ & 2.007 & $0.63\pm0.05$ & \nodata & (7) \\ 
2MASS J16235470$-$2438319 & USCO & 10 $\pm$ 3 & $19.8\pm1.0$ & 1.765 & $0.69\pm0.05$ & \nodata & (7) \\ 
2MASS J16262152$-$2426009 & USCO & 10 $\pm$ 3 & $23.1\pm1.1$ & 2.497 & $1.14\pm0.08$ & \nodata & (7) \\ 
2MASS J16265619$-$2213519 & USCO & 10 $\pm$ 3 & $58.7\pm1.4$ & 0.284 & $0.33\pm0.02$ & \nodata & (7) \\ 
2MASS J16272658$-$2425543 & ROPH & $<$2 & $12.3\pm1.4$ & 2.884 & $0.7\pm0.09$ & $69\pm10$ & (7) \\ 
2MASS J16281707$+$1334204 & field & \nodata & $15.8\pm1.1$ & 0.603 & $0.19\pm0.02$ & \nodata & (6) \\ 
2MASS J16281808$-$2428358 & ROPH & $<$2 & $16.2\pm1.1$ & 0.796 & $0.25\pm0.02$ & $43\pm22$ & (7) \\ 
2MASS J16311879$+$4051516 & field & \nodata & $14.8\pm1.0$ & 0.512 & $0.15\pm0.02$ & \nodata & (6) \\ 
2MASS J16402068$+$6736046 & field & \nodata & $16.0\pm1.0$ & 0.378 & $0.12\pm0.01$ & $74\pm7$ & (6) \\ 
2MASS J17071830$+$6439331 & field & \nodata & $25.4\pm1.0$ & 0.151 & $0.08\pm0.0$ & $48\pm7$ & (15) \\ 
2MASS J21272531$+$5553150 & CARN & $\sim$200 & $17.3\pm1.0$ & 0.54 & $0.18\pm0.02$ & \nodata & (6) \\ 
2MASS J21381698$+$5257188 & field & \nodata & $40.6\pm1.0$ & 0.183 & $0.15\pm0.01$ & \nodata & (6) \\ 
2MASS J22021125$-$1109461 & ABDMG & 149$^{+51}_{-19}$ & $21.3\pm1.0$ & 0.428 & $0.18\pm0.01$ & \nodata & (1)
\enddata
\tablenotetext{a}{Young moving group name abbreviation mostly follows those defined in \cite{Gagne:2018ab}: 
AB Doradus (ABDMG), 
Argus (ARG), 
Castor (Castor),
Corona Australis (CRA),
Carina Near (CARN),
Hyades (HYA), 
$\rho$ Ophiuchi (ROPH),
Taurus (TAU), 
and Upper Scorpius (USCO). Orion Nebula Cluster is labeled as (ORION).}
\tablenotetext{b}{The age of field dwarfs is assumed between 1--10~Gyr. See Section~\ref{sec:rsini} for details.}
\tablenotetext{c}{We assume 5\% uncertainty for each period without reported uncertainty. See Section~\ref{sec:rsini} for details.}
\tablerefs{(1) \cite{Watson:2006aa}, (2) \cite{Ivezic:2007aa}, (3) \cite{Berger:2009aa}, (4) \cite{Crossfield:2014aa}, (5) \cite{Douglas:2019aa}, (6) \cite{Newton:2016aa}, (7) \cite{Rebull:2018aa}, (8) \cite{Reiners:2018aa}, (9) \cite{Douglas:2019aa}, (10) \cite{Vos:2019aa}, (11) \cite{Freund:2020aa}, (12) \cite{Nardiello:2020aa}, (13) \cite{Rebull:2020aa}, (14) \cite{Vos:2020aa}, (15) \cite{Rockenfeller:2006aa} }
\end{deluxetable*}

\begin{figure}
\centering
\includegraphics[width=\linewidth, trim=20 0 10 10]{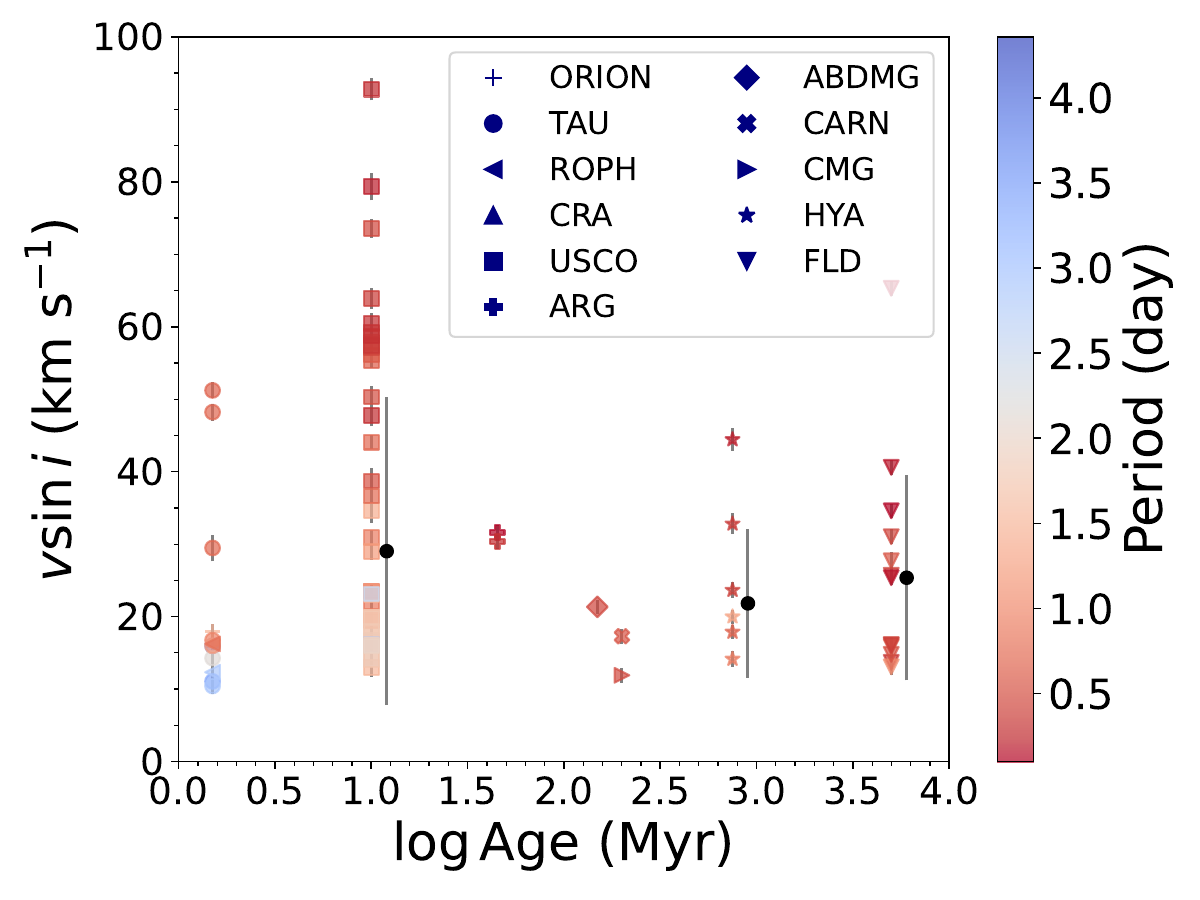}
\includegraphics[width=\linewidth, trim=20 10 10 10]{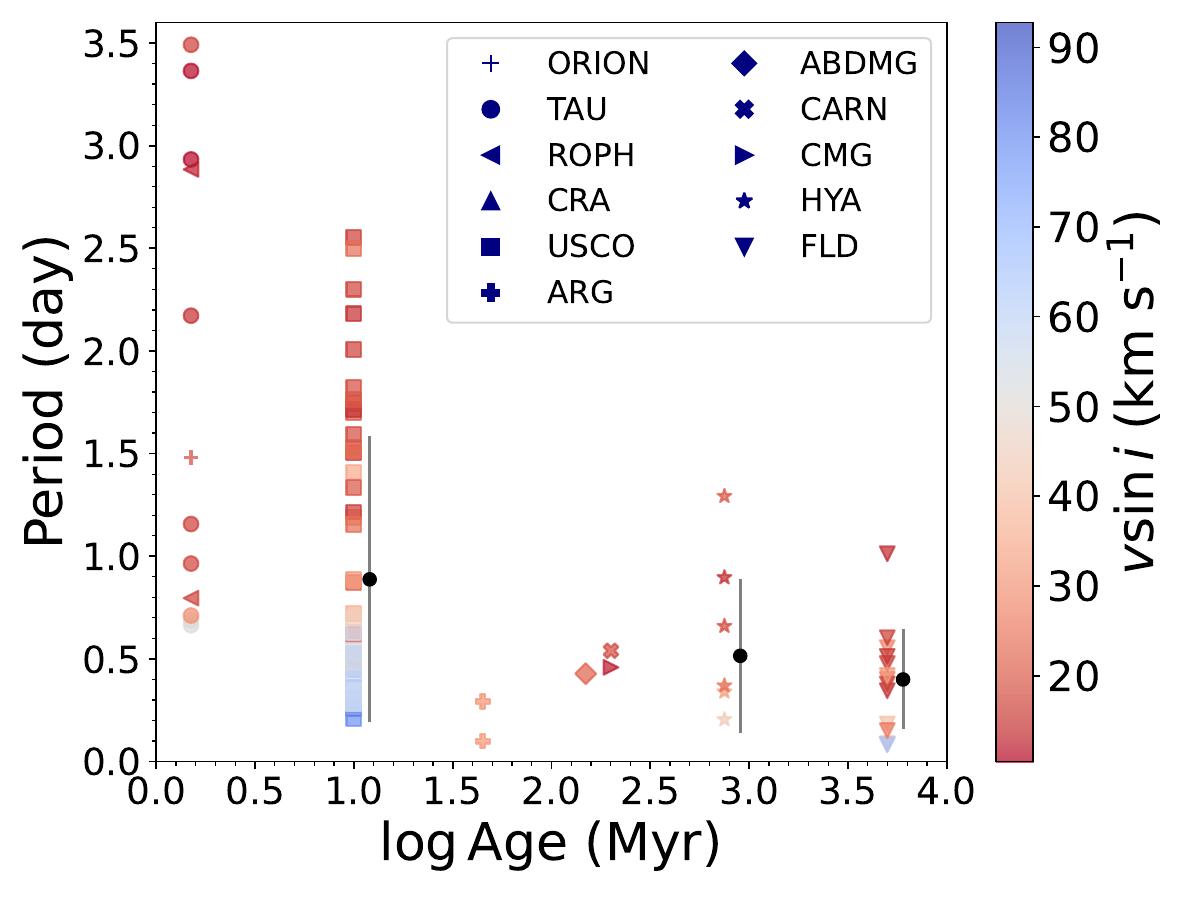}
\caption{Projected rotational velocities and rotational periods as a function of ages. \textit{Top}: The projected rotational velocities ($v\sin{i}$s) for each source are color-coded with their rotational periods with shapes corresponding to their clusters, including Orion (ORION; thin plus), Taurus (TAU; circle), $\rho$ Ophiuchi (ROPH; left triangle), Corona Australis (CRA; upper triangle), Upper Scorpius (USCO; square), Argus (ARG; thick plus), AB Doradus (ABDMG; diamond), Carina Near (CARN; cross), Castor (CMG; right triangle), Hyades (HYA; star), and field objects (FLD; lower triangle).  For sources in the Taurus, Upper Scorpius and Hyades, and field objects. 
The median, 16$^\mathrm{th}$, and 84$^\mathrm{th}$ percentiles for the subsamples are plotted in black dots with a slight shift to the right with respect to their ages.
\textit{Bottom}: The rotational periods as a function of ages, color-coded with their projected rotational velocities. The label conventions are the same as those on the left panel.
}
\label{fig:vsini_period_age}
\end{figure}

We have compiled period measurements from the literature for 78 APOGEE sources, listed in Table~\ref{table:inclination}.
These include 64 young cluster members and 14 field objects.
For rotation periods reported without uncertainties, we assumed relative uncertainties of 5\%. 
We note that for these rotation period measurements, the young sources (median period of 0.88 days) have longer rotation periods than the field sources (median period of 0.49 days; Figure~\ref{fig:vsini_period_age}). 

Figure~\ref{fig:vsini_period_age} illustrates our sample with both $v\sin{i}$ and rotational periods at different ages.
The spin-up trend in rotational periods has been reported in previous studies \citep{Popinchalk:2021aa, Vos:2022aa}.
Our slower $v\sin{i}$ trend toward older age (10~Myr vs. $>$100~Myr) in our sample is consistent with previous studies \citep{Zapatero-Osorio:2006aa} using their sample of masses between $\sim$30--70~M$_\mathrm{Jup}$. 
Note that our sample has limited measurements for objects between Upper Scorpius (10~Myr) and Hyades (750~Myr) ages, and the ages are mostly unavailable for late-M dwarfs with $v\sin{i}$ measurements \citep{Crossfield:2014aa, Vos:2017aa, Jeffers:2018aa, Hsu:2021aa}.

\begin{figure}
\centering
\includegraphics[width=\linewidth, trim=20 10 10 10]{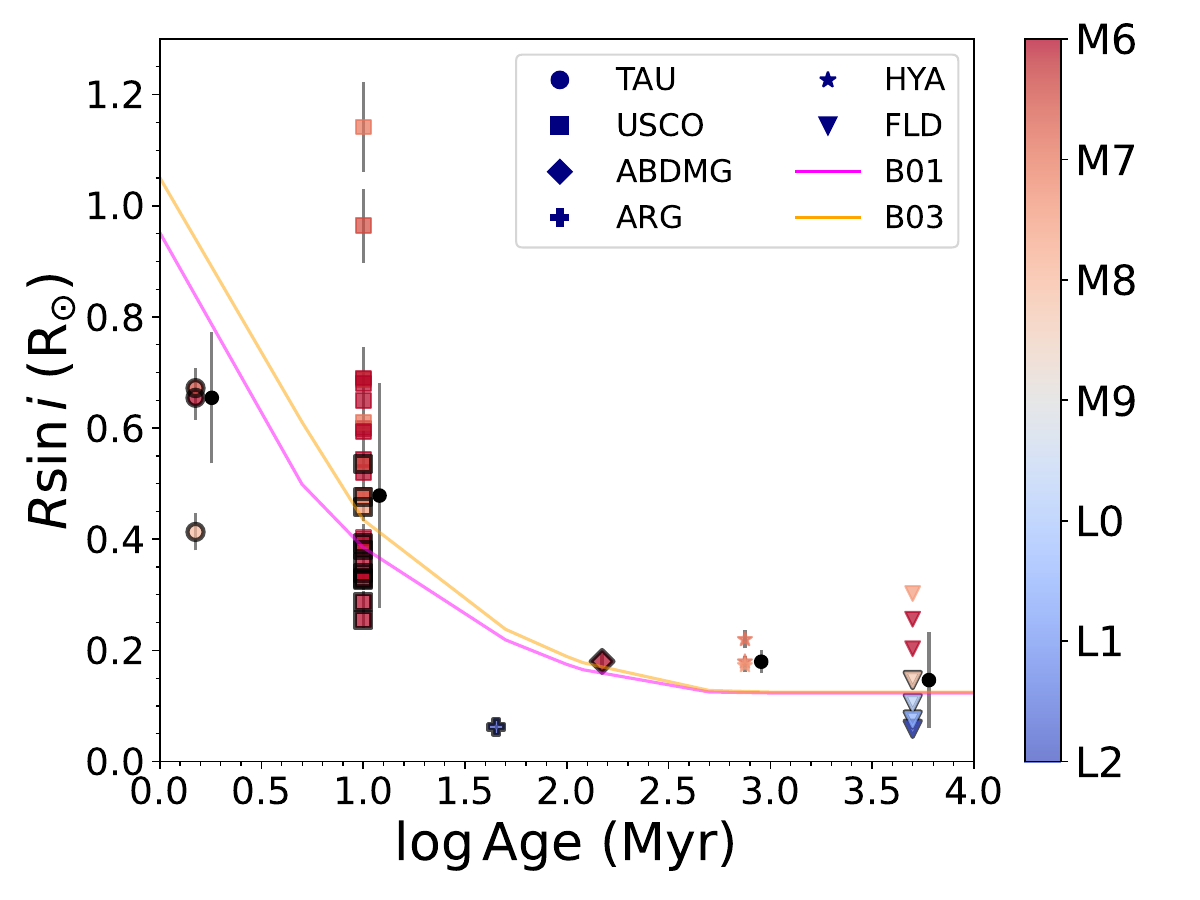}
\caption{Projected radii as a function of ages. The Projected radii ($R\sin{i}$) for each source are color-coded with their spectral types with shapes corresponding to their clusters, including Taurus (TAU; circle), Upper Scorpius (USCO; square), Argus (ARG; plus), AB Doradus (ABDMG; diamond), Hyades (HYA; star), and field objects (FLD; lower triangle).  For sources in the Taurus, Upper Scorpius and Hyades, and field objects, the median, 16$^\mathrm{th}$, and 84$^\mathrm{th}$ percentiles for the subsamples are plotted in black dots with a slight shift to the right with respect to their ages.
Sources with physically possible inclinations are labeled in black open symbols (9 sources; see Section~\ref{sec:rsini} for details).
The theoretical radius tracks assuming mass $M$ = 0.1~M$_{\odot}$ from \cite{Burrows:2001aa} and \cite{Baraffe:2003aa} models are illustrated in magenta and orange lines, respectively.
}
\label{fig:rsini}
\end{figure}

While opposite our previously observed {\vsini} trend (larger rotations speeds for younger sources; Figure~\ref{fig:vsini_period_age}), the larger radii of young objects are an important factor.
We computed the projected radii of each source using Eqn.~6, propagating uncertainties by the Monte Carlo method.
Since {\vsini} values close to our detection limit could potentially yield overestimated projected radii, we conservatively constrained our analysis to sources with {\vsini} $>$ 20~{\kms}, which encompasses 41 sources in total, including 33 young sources (25 sources in the Upper Scorpius cluster) and 7 field objects.
The resulting $R\sin{i}$ values as a function of
age are shown in Figure~\ref{fig:rsini}, 

Overall, the projected radii are larger for the younger sources (median $R\sin{i}$ = 0.47~R$_{\odot}$) compared to the field objects (median $R\sin{i}$ = 0.15~R$_{\odot}$), and show a consistent decline with age from Taurus at 1--2 Myr, to Upper Scorpius at 10$\pm$3~Myr 
to the Hyades at 750$\pm$100~Myr \citep{Brandt:2015aa} to field ages (assumed as 5~Gyr).
The observed projected radii are qualitatively consistent with the evolutionary models, with predicted values from \citet{Baraffe:2003aa} for corresponding ages provided in Table~\ref{tab:inferred_radii}.

However, we find that over half (26 out of 41 sources) of the observed projected radii over the model radii are larger than 1, with the median radii from theoretical models underestimated by 25\%.
While our sample is yet small, our inferred $R\sin{i}$s imply that the radius inflation extends to the very low-mass stars and (young) brown dwarfs ($M \sim$ 0.05--0.1~M$_{\odot}$ using the \citealp{Baraffe:2003aa} models).

\replaced{We note that}{The} radius inflation of higher-mass M dwarfs ($M$ = 0.1--0.6~M$_{\odot}$) is widely reported in the literature ($\Delta{R}/R \sim$ \replaced{14 $\pm$ 2$\%$}{5--25\%}; e.g. \citealt{Stassun:2012aa, Jackson:2014aa, Mann:2015aa, Venuti:2017aa, Jackson:2018aa, Kesseli:2018aa, Parsons:2018aa, Khata:2020aa, Swayne:2024aa, Kiman:2024aa}).
For these higher-mass M dwarfs, the radius inflation is found to be independent of age, metallicity or multiplicity \citep{Mann:2015aa}. 
There is correlation between radius inflation and magnetic activities reported in \cite{Stassun:2012aa, Kiman:2024aa} for single M dwarfs for $M$ = 0.5--0.6~M$_{\odot}$, but no clear trend has been identified in lower mass regime between $M$ = 0.1--0.5~M$_{\odot}$.
Possible culprits in the theoretical models could be due to opacities and convective modeling (convective overshoot, mixing length) \citep{Mann:2015aa}.

For very-low-mass stars and brown dwarfs ($M$ $\lesssim$ 0.1~M$_{\odot}$), the radius inflation issue was also reported but with a much smaller sample size. 
Although very-low-mass individual eclipsing binaries might present consistent \citep{Triaud:2020aa, Davis:2024aa} or inflated \citep{Casewell:2020ab, Buzard:2022aa} radii compared to theoretical models, the observed radii as a population could appear to inflate by $\sim$10--30\% \citep{Kesseli:2018aa, Triaud:2020aa, Canas:2022aa, Carmichael:2023aa, Davis:2024aa}.
Therefore, our empirical projected radii also support that the radius inflation issue extends to masses $M$ $\sim$ 0.05--0.1~M$_{\odot}$, using our sample of 41 very-low-mass objects compared to the literature measurements ($\lesssim$50).
While there might be additional systematic uncertainties for {\vsini} and rotational periods, the theoretical radii are likely to be underestimated. The difference between \cite{Burrows:2001aa} and \cite{Baraffe:2003aa} models can be more than 40\%.
We emphasize that our observed sample of projected radii is small for the majority of the clusters, and a larger sample of projected radii is needed to quantify the underestimated radii across different ages and spectral types.

\begin{deluxetable*}{lccccccl}
\tablecaption{Projected and Estimated Radii for Variable UCDs \label{tab:inferred_radii}} 
\tabletypesize{\small} 
\tablehead{ 
\colhead{Cluster} &
\colhead{Age} & 
\colhead{SpT} & 
\colhead{$R_\mathrm{Burrows\, evol}$\tablenotemark{a}} &
\colhead{$R_\mathrm{Baraffe\, evol}$\tablenotemark{a}} &
\colhead{$N$} &
\colhead{$\langle{R\sin{i}}\rangle$} &
\colhead{Ref} \\
\colhead{} &
\colhead{} &
\colhead{} &
\colhead{(R$_\mathrm{\odot}$)} &
\colhead{(R$_\mathrm{\odot}$)} &
\colhead{} &
\colhead{(R$_\mathrm{\odot}$)} &
\colhead{}
}
\startdata 
\hline
Taurus & 1--2~Myr & M6--M8 & 0.265--0.468 & 0.384--0.665 & 3 & $0.654^{+0.012}_{-0.164}$ & (1) \\
Upper Scorpius & 10$\pm$3~Myr & M6--M7.5 & 0.243--0.344 & 0.261--0.404 & 25 & $0.48^{+0.20}_{-0.14}$ & (2) \\
Argus & 40--50~Myr & L2 & 0.135--0.152 & 0.143--0.153 & 1 & $0.062 \pm 0.003$ & (3) \\
AB Doradus & 149$^{+51}_{-19}$~Myr & M6 & 0.129--0.136 & 0.149--0.158 & 1  & $0.18 \pm 0.01$ & (4) \\
Hyades & 750$\pm$100~Myr & M7--M7.5 & 0.102--0.105 & 0.114--0.118 & 3 & $0.180^{+0.027}_{-0.005}$ &  (5) \\
field & 1--10~Gyr & M6--L0 & 0.094--0.111 & 0.098--0.125\tablenotemark{b} & 7 & $0.15^{+0.11}_{-0.07}$ &  (6) \\
\enddata
\tablenotetext{a}{Based on the \cite{Burrows:2001aa} or \cite{Baraffe:2003aa} evolutionary models for the {\teff} range of the sample sources (from the sample spectral types) and individual cluster ages.}
\tablenotetext{b}{The highest effective temperature available for \cite{Baraffe:2003aa} is 2786~K at 0.1~M$_{\odot}$, corresponding to 0.125~R$_{\odot}$.}
\tablerefs{(1) \cite{Kenyon:1995aa}, (2) \cite{Pecaut:2016aa}, (3) \cite{Zuckerman:2019aa}, (4) \cite{Bell:2015aa}, (5) \cite{Brandt:2015aa}, (6) See references in the main text}
\end{deluxetable*}

\section{Summary} \label{sec:sum}

We summarize our main results as follows:

\begin{enumerate}
\item We constructed a spectroscopically-classified UCD sample of 258 ultracool dwarfs with 931 epochs from APOGEE high-resolution near-infrared spectra from SDSS DR17, with spectral types ranging from M6 to L2, with distance ranging from 3.6~pc to 414~pc. 
These include sources with new classifications presented here based on low-resolution optical spectroscopy.
We also constructed a broader
sample of 444 unclassified UCD candidates within 400~pc of the Sun selected through photometric and astrometric selection criteria using 2MASS and \textit{Gaia} EDR3,
which collectively encompass 2474 epochs of APOGEE observations.
\item We employed a Markov Chain Monte Carlo forward-modeling routine to fit the APOGEE data to four sets of model atmospheres. 
We found that BT-Settl models \citep{Baraffe:2015aa} fit best for higher temperature sources ({\teff} $\gtrsim$ 2800~K, spectral type $\lesssim$ M9) and Sonora models \citep{Marley:2018aa, Marley:2021aa} fit best for lower temperature sources, even sources hotter than the nominal 2400~K parameter limit of this model set.
Given the narrow spectral range deployed in these fits and missing opacities in the models (e.g., FeH in BT-Settl, condensate clouds in Sonora), we discourage use of inferred {\teff} and {\logg} values from these fits, which do not significantly impact inferred RV and {\vsini} values. 
\item We measured RVs and {\vsini}s with median precisions of 0.4~{\kms} and 1.1~{\kms}, respectively. Most of our RVs and {\vsini}s are consistent with the measurements available in the literature, with the outliers from unresolved or confirmed binaries, low-resolution spectral measurements, additional correction in RVs \citep{Kounkel:2019aa}, and differences between Sonora and other models. 
{\vsini} measurements have a resolution floor of 10~{\kms} and extend up to 92.8~{\kms}, with a median {\vsini} = 17~{\kms} that is consistent with prior samples of late-M and early L dwarfs in the literature. 
\item Combining our RVs with \textit{Gaia} astrometry, we inferred heliocentric and Local Standard of Rest $UVW$ velocities for each of our sources. 
From these kinematics, we identified 11 sources as members of the intermediate thin/thick disk Galactic population; the remainder are thin disk members.
We also computed Galactic orbits; most of the sample have circular and planar orbits (e $\leq$ 0.1, $i~\leq~2\%$) as expected for thin disk members.
\item Using the BANYAN $\Sigma$ tool with $UVW$ velocities and $XYZ$ positions, we found that roughly one-half of our sample (141 sources) are kinematic members of nearby young clusters or moving groups, with most previously reported in the literature.
We identified three new young cluster kinematic members: 
2MASS J05402570+2448090 (G 100-28; 67.2\% Argus moving group, 30.9\% Carina Near moving group), 
2MASS J14093200+4138080 (LP 220-50; 99.6\% Argus moving group), 
and 2MASS J21272531+5553150 (LSPM J2127$+$5553; 99.3\% Carina Near moving group).
We also ruled out six previously identified young cluster candidates that lacked RV measurement:
2MASS J00381273+3850323 (Hyades Moving Group; \citealp{Roser:2019aa, Lodieu:2019aa}),
2MASS J04254894+1852479 ($\beta$ Pic Moving Group; \citealp{Gagne:2018ad});
2MASS J07140394+3702459 (Argus moving group; \citealt{Gagne:2015aa}), 
and 
2MASS J12205439+2525568, 2MASS J12263913+2505546, and 2MASS J12265349+2543556 (Coma Berenices Cluster; \citealt{Melnikov:2012aa}).
\item We computed kinematic ages for our sample using empirical age-velocity dispersion relations from \citet{Wielen:1977aa} and \citet{Aumer:2009aa}. After removal of young sources in our sample, we found a total velocity dispersion of 38.2 $\pm$ 0.3~{\kms}, corresponding to a kinematic age of 3.30 $\pm$ 0.19~Gyr, consistent with measurements of local 20~pc late-M dwarfs \citep{Hsu:2021aa}.
\item For 171 sources with multiple APOGEE epochs, we identified 37 sources that show statistically significant radial velocity variability. Of these, 23 with more than four observing epochs were found to be highly probable binaries, based on statistical comparison between a constant RV model and orbital parameters inferred from the Monte Carlo rejection sampler \texttt{The Joker}. 
These include nine candidate binaries with estimated orbit periods $P$ $<$ 10~days, adding to the previously identified short-period binary in this sample,  2MASS J03505737+1818069 (LP 413$-$53, $P$ = 0.71~days; \citealt{Hsu:2023aa}).
\item Combining photometric variability period measurements from the \textit{Kepler}/K2 mission and ground-based studies, we computed the projected radii ($R~\sin{i}$) for 78 sources, including 64 young and 14 field objects. The $R~\sin{i}$s show a general decline with age from 1-100~Myr, consistent with the expected contraction of young stars. 
We also computed inclinations for this subsample using model radii from \cite{Baraffe:2003aa}, only 15 of which proved to be physical.
We found that the median radii from theoretical models are underestimated by 25\%, which implies that the radius inflation issue of M dwarfs widely reported in the literature extends to M6--L2 dwarfs.
\end{enumerate}

This study significantly expands the low-mass stellar sample for which robust radial and rotational velocities have been inferred from high-resolution spectra, improving assessment of Galactic population and young cluster membership, refining kinematic ages as a function of temperature and mass, and identifying new candidate low-mass binaries for direct mass measurement. There are nevertheless several improvements that can be made to this work to ensure accurate measure of these quantities. 

Foremost is improvement in the theoretical stellar atmosphere and evolutionary models, particularly for the temperatures below 3000~K. Older model sets such as BT-Settl struggle with missing opacities for late-M and L dwarfs in the infrared, mostly notably FeH. Recent models with improved molecular opacities such as Sonora perform better, but are limited in {\teff} range and may be missing the necessary cloud opacities that shape L dwarf spectra. 

Prior work has demonstrated that incorrect line opacities can lead to systematic RV offsets, particularly among the coldest brown dwarfs \citep{Hsu:2021aa, Tannock:2022aa}; and this may explain the systematic RV shift seen among the lowest-mass members of young clusters \citep{Cottaar:2014aa,Cook:2014aa,Kounkel:2019aa}.

In addition to improvements in the models, validating sources that show RV variations require follow-up observations, both to confirm this variability and to infer orbital parameters necessary to 
periods, eccentricities, and ultimately masses. 
Finally, these small-sample studies are essential for laying the scientific groundwork for next-generation high-resolution spectroscopic surveys that have the sensitivity to target UCDs. Most notable of these is the MaunaKea Spectroscopic Explorer (MSE), whose proposed 11.25~m telescope will be matched to a high-resolution spectrometer providing equivalent resolution spectra as APOGEE in the red optical (0.5-0.9~$\mu$m) for thousands of sources per pointing down to AB magnitudes of 20 \citep{Saunders:2016aa, McConnachie:2016aa, The-MSE-Science-Team:2019aa}.
At these sensitivities, MSE will reach thousands of UCDs down to the T spectral class ({\teff} $\sim$ 1000~K) and reach larger samples of both young cluster brown dwarfs and older (and metal-poor) brown dwarfs in the Galactic thick disk and halo. 
By continuing to improve high-resolution spectral fits to local UCDs, we can ensure the maximum science yield of MSE and other future spectroscopic surveys.

\pagebreak

\facilities{Sloan (APOGEE), Shane (Kast Double spectrograph)}

\software{
\texttt{apogee} \citep{Bovy:2016aa},
\texttt{Astropy} \citep{Astropy-Collaboration:2013aa, Astropy-Collaboration:2018aa},
\texttt{corner} \citep{Foreman-Mackey:2016aa},
\texttt{emcee} \citep{Foreman-Mackey:2013aa},
\texttt{galpy} \citep{Bovy:2015aa},
\texttt{IPython} \citep{Perez:2007aa},
\texttt{matplotlib} \citep{Hunter:2007aa},
\texttt{numpy} \citep{van-der-Walt:2011aa},
\texttt{pandas} \citep{mckinney-proc-scipy-2010, reback2020pandas},
\texttt{PyMC3} \citep{Salvatier:2016aa},
\texttt{SciPy} \citep{Virtanen:2020aa},
\texttt{SMART} \citep{Hsu:2021ab, Hsu:2021aa},
\texttt{SPLAT} \citep{Burgasser:2017ac},
\texttt{The Joker} \citep{Price-Whelan:2017aa}.
}

\acknowledgments 
The authors thank the referee, Kelle Cruz, for her useful review which has significantly improved the original manuscript.
CCH and AJB acknowledge funding and the SDSS/Faculty and Student Team (FAST) Initiative program.
Funding for the Sloan Digital Sky Survey IV has been provided by the Alfred P. Sloan Foundation, the U.S. Department of Energy Office of Science, and the Participating Institutions. 
SDSS-IV acknowledges support and resources from the Center for High Performance Computing at the University of Utah. The SDSS website is www.sdss.org.
SDSS-IV is managed by the Astrophysical Research Consortium for the Participating Institutions of the SDSS Collaboration including the Brazilian Participation Group, the Carnegie Institution for Science, Carnegie Mellon University, Center for Astrophysics | Harvard \& Smithsonian, the Chilean Participation Group, the French Participation Group, Instituto de Astrof\'isica de Canarias, The Johns Hopkins University, Kavli Institute for the Physics and Mathematics of the Universe (IPMU) / University of Tokyo, the Korean Participation Group, LawrenceBerkeley National Laboratory, Leibniz Institut f\"ur Astrophysik Potsdam (AIP),  Max-Planck-Institut f\"ur Astronomie (MPIA Heidelberg), Max-Planck-Institut f\"ur Astrophysik (MPA Garching), Max-Planck-Institut f\"ur Extraterrestrische Physik (MPE), National Astronomical Observatories of China, New Mexico State University, New York University, University of Notre Dame, Observat\'ario Nacional / MCTI, The Ohio State University, Pennsylvania State University, Shanghai Astronomical Observatory, United Kingdom Participation Group, Universidad Nacional Aut\'onoma de M\'exico,University of Arizona, University of Colorado Boulder, University of Oxford, University of Portsmouth, University of Utah, University of Virginia, University of Washington, University of Wisconsin, Vanderbilt University, and Yale University.
This research was supported in part through the computational resources and staff contributions provided for the Quest high performance computing facility at Northwestern University which is jointly supported by the Office of the Provost, the Office for Research, and Northwestern University Information Technology.
This work used computing resources provided by Northwestern University and the Center for Interdisciplinary Exploration and Research in Astrophysics (CIERA). This research was supported in part through the computational resources and staff contributions provided for the Quest high performance computing facility at Northwestern University which is jointly supported by the Office of the Provost, the Office for Research, and Northwestern University Information Technology.





\appendix
\restartappendixnumbering








\section{DR17 Full UCD Sample and Measurements}\label{appendix:full_sample}
\restartappendixnumbering

In Section~\ref{sec:sample}, we presented\deleted{ the construction of} our \replaced{``full''}{gold} APOGEE DR17 sample \replaced{based on photometric and astrometric data}{with sources that have spectral type determinations in this work or in the literature}. 

We also constructed a second, more comprehensive ``full'' sample by matching APOGEE sources to \textit{Gaia} EDR3 data, and selected a subsample based on the \textit{Gaia} $G-G_\mathrm{RP}$/spectral type relation of \citet{Kiman:2019aa} and the \textit{Gaia} color-magnitude distribution of spectroscopically classified UCDs.
We conservatively selected sources with 
\begin{itemize}
    \item \textit{Gaia} parallax $\pi > 2.5$~mas (distances $<$ 400~pc),
    \item \textit{Gaia} M$_{G} > 10.5$,
    \item \textit{Gaia} $ 1.32 < G-G_\text{RP} < 1.9$, and 
    \item Galactic latitude $\left| b \right| > 15^{\circ}$.
\end{itemize}
The absolute magnitude and color criteria correspond to spectral types $\gtrsim$M4--M5, while the Galactic latitude criterion aims to reduce contamination from reddened background sources. 
Additionally, we used 2MASS photometry and an empirically-determined set of 2MASS $J$, $K$ and \textit{Gaia} $G$, $G_\mathrm{RP}$ color and magnitude criteria based on our APOGEE gold sample to remove reddened contaminant sources:
\begin{itemize}
    \item $6.5 <$ M$_J$,
    \item $ 0.82 < J - K < 1.8$, 
    \item 3.3 $< G - J <$ 5.2, and
    \item $ ( G - J ) - \frac{5}{6} \times (G - G_\mathrm{RP}) > 2.167$.
\end{itemize}
The last criterion aims to remove embedded young sources, although known young sources with published spectral classifications are included in our gold sample.
Including the APOGEE SNR $>$ 10 constraint, our full sample contains 2474 spectra of 702 sources with an additional 1543 spectra of 444 sources 
(Table~\ref{table:sample_full}).
We note that the full sample includes 239 sources from the gold sample; 19 gold sample sources fall outside the full sample selection criteria because they
lack \textit{Gaia} astrometry (4 sources), are near the Galactic plane (14 sources), or
being a young binary (1 source; 2MASS J16183317$-$2517504). 

Similar to our gold sample, the majority of the full sample has distances larger than 30~pc (90\%) and APOGEE spectral observations taken over multiple epochs (51\%). A total of 115 sources in the full sample satisfy the criteria with at least four observations and SNR $>$ 10.

This Appendix provides the source information for this sample in Table~\ref{table:sample_full} and best-fit parameters in Table~\ref{table:indmeasurement_full}, which include additional 444 candidate UCDs and 1543 epochs. The short summary of RV and $v\sin{i}$ measurements for the full sample is available in Table~\ref{table:rv_vsini_summary_full_sample}.
We note that the best model for each source is determined by the lower $\chi^2$ generally (see Section~\ref{sec:modeling} for details).

\begin{longrotatetable}
\begin{deluxetable*}{cllllrrrrrrrr}
\tablecaption{APOGEE DR17 Sample\label{table:sample_full}}
\tablewidth{700pt}
\tabletypesize{\small}
\tablehead{
\colhead{} & \colhead{} & \colhead{} & \colhead{} & \colhead{} & \colhead{} & \colhead{} & \colhead{} & \colhead{} &
\colhead{}\\
\colhead{APOGEE ID} & 
\colhead{RA} & \colhead{Dec} & 
\colhead{$Gaia$ eDR3 Source ID} & 
\colhead{SpT} & \colhead{$N_\text{obs}$} & \colhead{2MASS $H$} & 
\colhead{$\mu_{\alpha}$} & \colhead{$\mu_{\delta}$} & \colhead{$\pi$}\\ 
\colhead{} & \colhead{(deg)} & \colhead{(deg)} & \colhead{} & \colhead{} & \colhead{} & \colhead{(mag)} & \colhead{(mas yr$^{-1}$)} & \colhead{(mas yr$^{-1}$)}& \colhead{(mas)}
} 
\startdata
2MASS J00091064$-$2116214 & 2.294345 & $-$21.272621 & 2364486119011986432 & M5.1 & 3 & $11.53 \pm 0.021$ & $-108.87 \pm 0.09$ & $-51.74 \pm 0.09$ & $21.54 \pm 0.09$ \\ 
2MASS J00111055$-$2021034 & 2.793974 & $-$20.350962 & 2364839646360029824 & M5.9 & 3 & $11.616 \pm 0.026$ & $71.88 \pm 0.07$ & $-44.11 \pm 0.05$ & $14.16 \pm 0.07$ \\ 
2MASS J00590731$-$3326248 & 14.780473 & $-$33.440243 & 5027425005705789952 & M5.3 & 19 & $14.253 \pm 0.058$ & $-2.71 \pm 0.14$ & $16.25 \pm 0.11$ & $9.35 \pm 0.2$ \\ 
2MASS J01120002+1502170 & 18.000108 & $+$15.038074 & 2591201534008672256 & M5.6 & 3 & $11.351 \pm 0.02$ & $166.12 \pm 0.15$ & $-316.99 \pm 0.09$ & $35.08 \pm 0.1$ \\ 
2MASS J01152739+0705532 & 18.864165 & $+$7.098118 & 2576842599344771584 & M4.9 & 2 & $11.362 \pm 0.024$ & $220.69 \pm 0.05$ & $-12.51 \pm 0.04$ & $28.58 \pm 0.05$ \\ 
2MASS J01234279+1802177 & 20.928321 & $+$18.038263 & 2593501540534738560 & L1.3 & 5 & $11.765 \pm 0.023$ & $-31.91 \pm 0.13$ & $-228.91 \pm 0.07$ & $15.55 \pm 0.11$ \\ 
2MASS J01252949$-$0715591 & 21.372875 & $-$7.266439 & 2478324776947441920 & M7.3 & 2 & $11.452 \pm 0.031$ & $26.99 \pm 0.09$ & $-9.0 \pm 0.04$ & $16.42 \pm 0.07$ \\ 
2MASS J01254187+1339222 & 21.42448 & $+$13.656189 & 2590169844209007488 & M5.1 & 2 & $11.193 \pm 0.034$ & $316.22 \pm 0.06$ & $18.05 \pm 0.04$ & $32.87 \pm 0.05$ \\
\enddata
\tablecomments{The astrometry, proper motions, and parallaxes are compiled from \textit{Gaia} EDR3 \citep{Gaia-Collaboration:2021ac}.}
\tablecomments{The full table is presented in a machine-readable table format.}
\end{deluxetable*}
\end{longrotatetable}

\begin{longrotatetable}
\begin{deluxetable*}{clllrrrrrrrrrrrr}
\tablecaption{Spectral Model Fit Parameters for Full Sample\label{table:indmeasurement_full}}
\tablewidth{650pt}
\tabletypesize{\scriptsize}
\tablehead{
\colhead{} & \colhead{} & \colhead{} & \colhead{} & \colhead{} & \colhead{} & \colhead{} & \colhead{} & \colhead{} &
\colhead{} & \colhead{} & \colhead{} & \colhead{} & \colhead{} & \colhead{} & \colhead{} \\
\colhead{APOGEE ID} & 
\colhead{Plate} & \colhead{Loc.} & 
\colhead{Fiber} & 
\colhead{JD} & \colhead{bary.\tablenotemark{a}} & \colhead{S/N} & \colhead{RV\tablenotemark{b}} & 
\colhead{$\langle$RV$\rangle$\tablenotemark{b, c}} & \colhead{$v\sin i$\tablenotemark{b}}& \colhead{$\langle$$v\sin i$$\rangle$\tablenotemark{b, c, d}} & \colhead{$T_\mathrm{eff}$\tablenotemark{f}} & \colhead{$\langle$$T_\mathrm{eff}$$\rangle$\tablenotemark{c, f}} &  \colhead{$\log{g}$\tablenotemark{f}} & \colhead{$\langle$$\log{g}$$\rangle$\tablenotemark{c, f}} & \colhead{Mdl\tablenotemark{e}} \\ 
\colhead{} & \colhead{} & \colhead{ID} & \colhead{ID} & \colhead{(day)} & \colhead{(km s$^{-1}$)} & \colhead{} & \colhead{(km s$^{-1}$)} & \colhead{(km s$^{-1}$)} & \colhead{(km s$^{-1}$)} &
\colhead{(km s$^{-1}$)} & \colhead{(K)} & \colhead{(K)} & \colhead{(cm s$^{-2}$)} & \colhead{(cm s$^{-2}$)} & \colhead{}
} 
\startdata
2M0009$-$2116 & 11933 & 6123 & 230 & 2458796.580 & $-22.73$ & 66 & $7.0^{+0.2}_{-0.3}$ & $7.5^{+0.3}_{-0.5}$ & $27.3^{+1.0}_{-1.0}$ & $26.5^{+1.0}_{-1.0}$ & $2399.7^{+0.4}_{-0.2}$ & $2399.7^{+0.2}_{-0.1}$ & $4.36^{+0.02}_{-0.02}$ & $4.38^{+0.01}_{-0.01}$ & S \\ 
\nodata & 11933 & 6123 & 218 & 2458803.574 & $-24.65$ & 64 & $7.7^{+0.6}_{-0.7}$ & \nodata & $25.3^{+1.0}_{-1.0}$ & \nodata & $2399.7^{+0.4}_{-0.2}$ & \nodata & $4.41^{+0.01}_{-0.01}$ & \nodata & S \\ 
\nodata & 11933 & 6123 & 221 & 2458824.536 & $-28.14$ & 56 & $8.6^{+0.7}_{-0.3}$ & \nodata & $27.2^{+1.0}_{-1.0}$ & \nodata & $2399.7^{+0.3}_{-0.2}$ & \nodata & $4.32^{+0.03}_{-0.02}$ & \nodata & S \\ 
2M0011$-$2021 & 11932 & 6123 & 39 & 2458774.618 & $-14.45$ & 37 & $12.3^{+1.0}_{-0.8}$ & $13.6^{+0.3}_{-0.5}$ & $36.8^{+1.0}_{-1.3}$ & $37.6^{+1.0}_{-1.2}$ & $2952.2^{+8.4}_{-12.8}$ & $3048.6^{+4.0}_{-4.5}$ & $5.25^{+0.03}_{-0.04}$ & $5.3^{+0.02}_{-0.02}$ & B \\ 
\nodata & 11932 & 6123 & 39 & 2458794.586 & $-21.95$ & 32 & $18.0^{+0.3}_{-0.3}$ & \nodata & $46.0^{+1.9}_{-1.4}$ & \nodata & $3072.4^{+5.1}_{-5.3}$ & \nodata & $5.33^{+0.04}_{-0.03}$ & \nodata & B \\ 
\nodata & 11932 & 6123 & 40 & 2458802.571 & $-24.26$ & 41 & $9.1^{+0.3}_{-0.3}$ & \nodata & $38.5^{+1.8}_{-2.0}$ & \nodata & $3061.3^{+10.5}_{-11.3}$ & \nodata & $5.34^{+0.06}_{-0.03}$ & \nodata & B \\ 
2M0059$-$3326 & 10211 & 5550 & 165 & 2458384.723 & $-1.04$ & 10 & $24.5^{+0.3}_{-0.4}$ & $24.8^{+0.3}_{-0.4}$ & $20.1^{+1.3}_{-1.2}$ & $17.1^{+1.0}_{-1.1}$ & $2903.9^{+9.3}_{-8.9}$ & $2992.2^{+3.7}_{-3.9}$ & $5.43^{+0.04}_{-0.03}$ & $5.47^{+0.01}_{-0.01}$ & B \\ 
\nodata & 10211 & 5550 & 171 & 2458412.640 & $-12.06$ & 10 & $24.6^{+0.3}_{-0.4}$ & \nodata & $14.2^{+1.8}_{-1.5}$ & \nodata & $3061.5^{+9.4}_{-9.8}$ & \nodata & $5.48^{+0.04}_{-0.02}$ & \nodata & B \\
\enddata
\tablenotetext{a}{Barycentric correction.}
\tablenotetext{b}{Systematic uncertainties of 0.19~{\kms} and 0.95~{\kms} are added to RV and {\vsini} measurements, respectively. See Sections~\ref{sec:rv} and ~\ref{sec:vsini} for details.}
\tablenotetext{c}{Weighted average over all epochs.}
\tablenotetext{d}{Averaged {\vsini} $<$ 10~km~s$^{-1}$ is below our {\vsini} detection limit; see Section~{\ref{sec:vsini}} for details.}
\tablenotetext{e}{Models used: S = Sonora 2018 \citep{Marley:2018aa}; B = BT-Settl \citep{Baraffe:2015aa}.}
\tablenotetext{f}{Readers should not take our {\teff} and {\logg} measurements as accurate. We report these values only for the reproducibility of our work. See Section~\ref{sec:teff_logg} for details.}
\tablecomments{Measurements from individual spectra over individual or multiple epochs are combined using inverse uncertainty weighting (weight = $1/(\sigma_\text{upper}^2 + \sigma_\text{lower}^2)$); upper and lower uncertainties are also combined using inverse uncertainty-squared weighting. In cases where individual spectra have S/N $<$ 10, spectral data are combined first, then modeled.}
\end{deluxetable*}
\end{longrotatetable}

\startlongtable
\begin{deluxetable}{ccc}
\tablecaption{Summary of RV and $v\sin{i}$ Measurements of APOGEE DR17 Full Sample\label{table:rv_vsini_summary_full_sample}}
\tablewidth{700pt}
\tabletypesize{\small}
\tablehead{
\colhead{} & \colhead{} & \colhead{} \\
\colhead{APOGEE ID} & 
\colhead{$\langle$RV$\rangle$} & \colhead{$\langle v\sin{i} \rangle$} \\
\colhead{} & \colhead{(km s$^{-1}$)} & \colhead{(km s$^{-1}$)}
} 
\startdata
2MASS J00091064$-$2116214 & $7.5^{+0.3}_{-0.5}$ & $26.5^{+1.0}_{-1.0}$ \\ 
2MASS J00111055$-$2021034 & $13.6^{+0.3}_{-0.5}$ & $37.6^{+1.0}_{-1.2}$ \\ 
2MASS J00590731$-$3326248 & $24.8^{+0.3}_{-0.4}$ & $17.1^{+1.0}_{-1.1}$ \\ 
2MASS J01152739+0705532 & $9.8^{+0.3}_{-0.5}$ & $14.0^{+1.0}_{-1.2}$ \\ 
2MASS J01234279+1802177 & $-9.5^{+0.3}_{-0.4}$ & $<$10 \\ 
2MASS J01252949$-$0715591 & $12.0^{+0.3}_{-0.5}$ & $23.5^{+1.0}_{-1.1}$ \\ 
2MASS J01254187+1339222 & $17.0^{+0.3}_{-0.6}$ & $18.1^{+1.0}_{-1.2}$ \\ 
2MASS J02025502$-$7259155 & $13.6^{+0.3}_{-0.3}$ & $70.6^{+1.1}_{-1.1}$ \\
\enddata
\tablecomments{The full table is presented in a machine-readable table format.}
\end{deluxetable}

\section{Binary Candidate Orbital Fit}\label{appendix:binary_orbit}
In Section~\ref{sec:binaries}, we presented the orbit fits for our binary candidates based on the Monte Carlo rejection sampler \texttt{The Joker}. Here, we present the fits for all of the sources listed in Table~\ref{table:binary_thejoker}.

\nopagebreak




\begin{figure*} 
\centering 
\plottwo{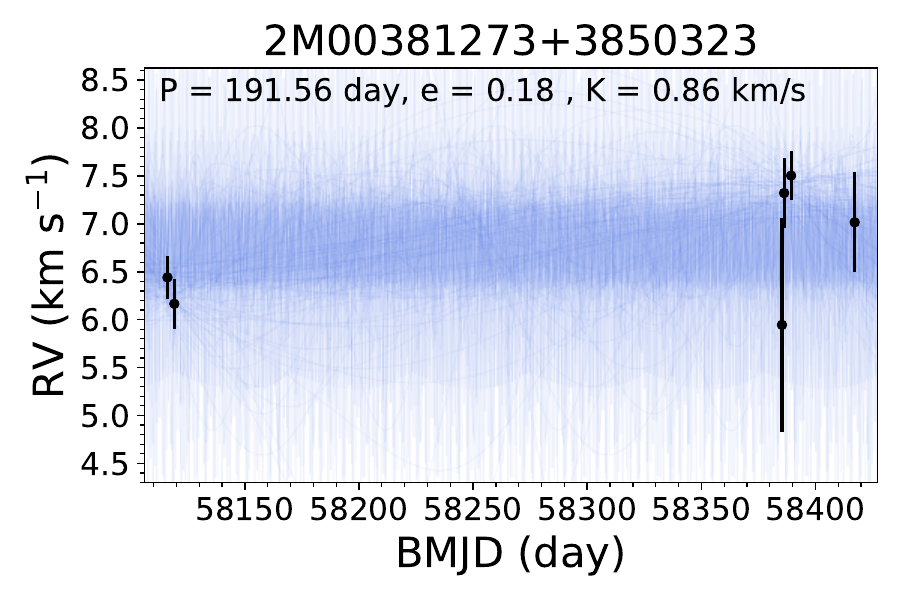}{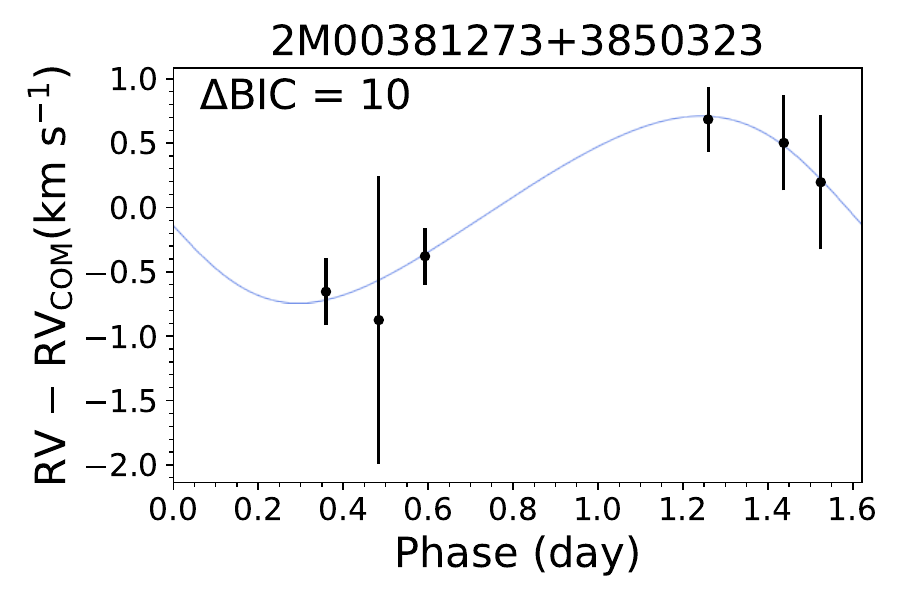} 
\caption{Same as Figure~\ref{fig:binary_orbit_fit} for 00381273$+$3850323. The best-fit orbital parameters are $P_\mathrm{fit}$ = $192^{+442}_{-174}$~day, $K_\mathrm{fit}$ = $0.9^{+0.7}_{-0.3}$~km s$^{-1}$, and $e_\mathrm{fit}$ = $0.18^{+0.27}_{-0.14}$. } 
\end{figure*} 

\begin{figure*} 
\centering 
\plottwo{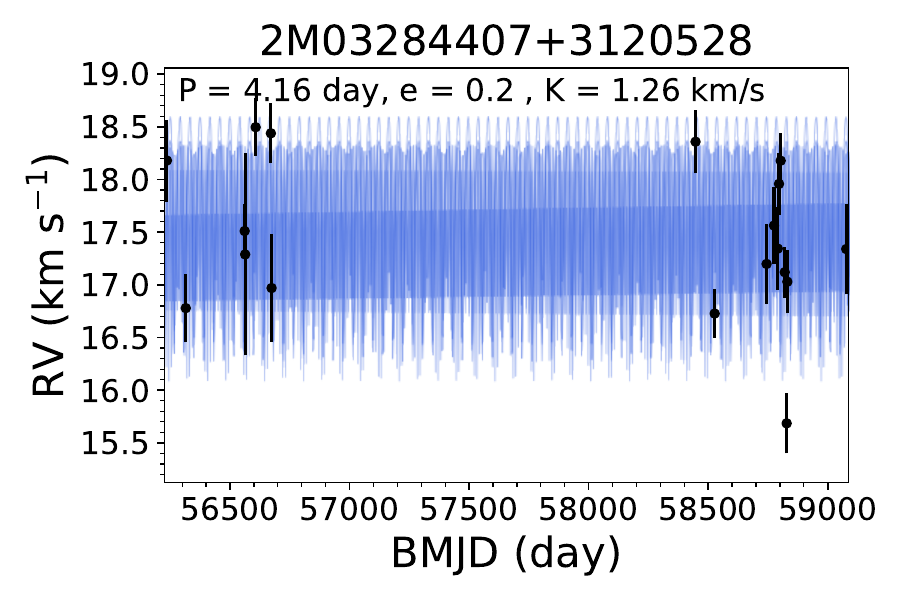}{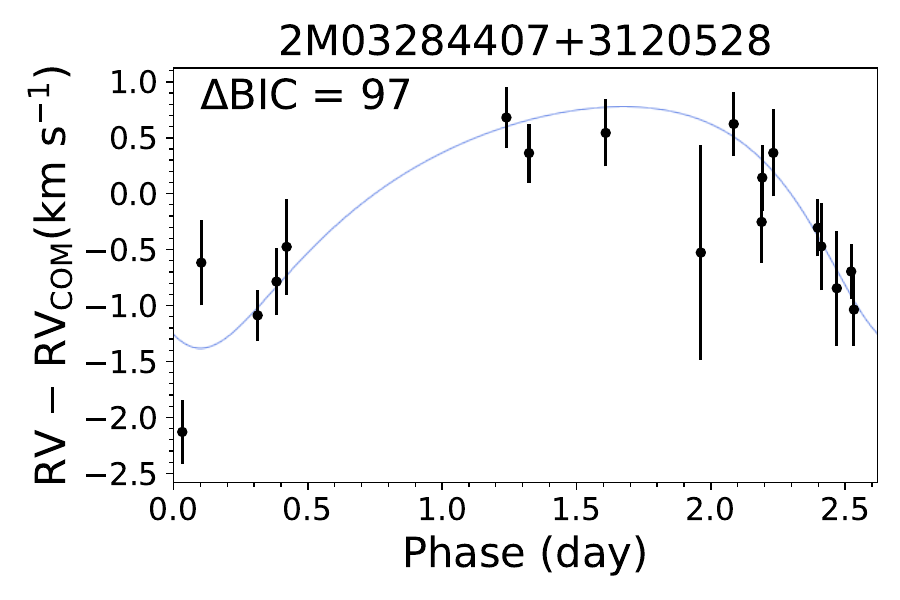} 
\caption{Same as Figure~\ref{fig:binary_orbit_fit} for 03284407$+$3120528. The best-fit orbital parameters are $P_\mathrm{fit}$ = $4^{+22}_{-3}$~day, $K_\mathrm{fit}$ = $1.3^{+0.2}_{-0.2}$~km s$^{-1}$, and $e_\mathrm{fit}$ = $0.2^{+0.26}_{-0.19}$. } 
\end{figure*} 

\begin{figure*} 
\centering 
\plottwo{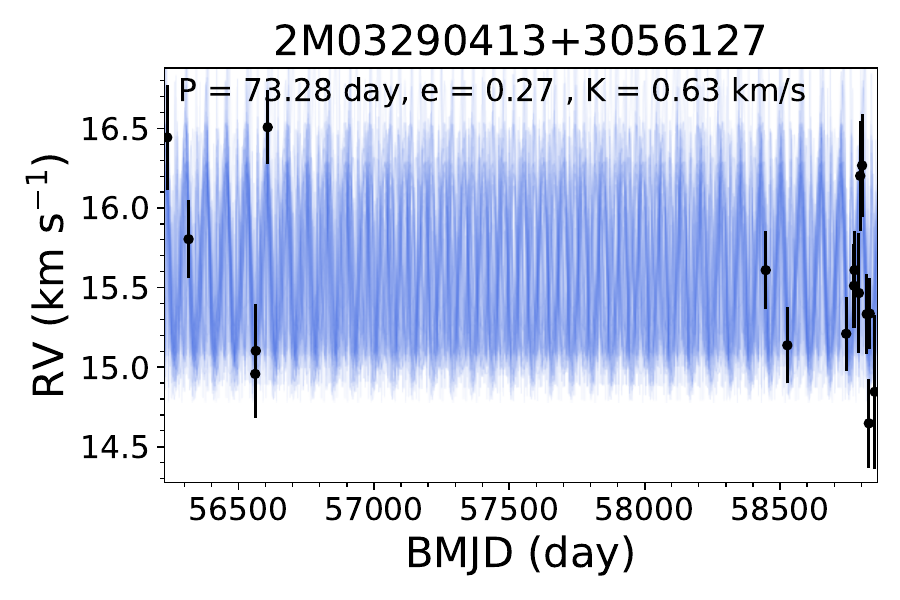}{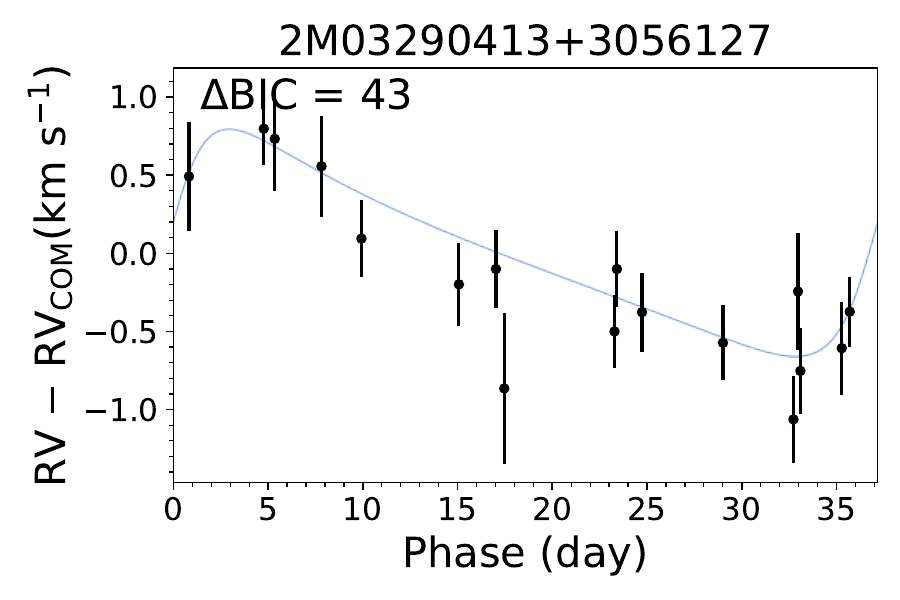} 
\caption{Same as Figure~\ref{fig:binary_orbit_fit} for 03290413$+$3056127. The best-fit orbital parameters are $P_\mathrm{fit}$ = $73^{+2}_{-36}$~day, $K_\mathrm{fit}$ = $0.6^{+0.1}_{-0.1}$~km s$^{-1}$, and $e_\mathrm{fit}$ = $0.27^{+0.25}_{-0.18}$. } 
\end{figure*} 

\begin{figure*} 
\centering 
\plottwo{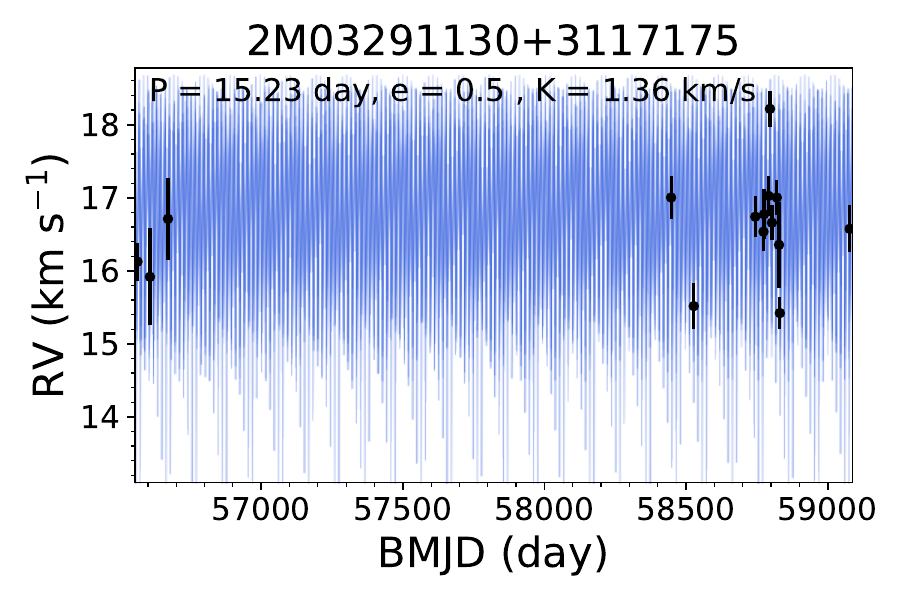}{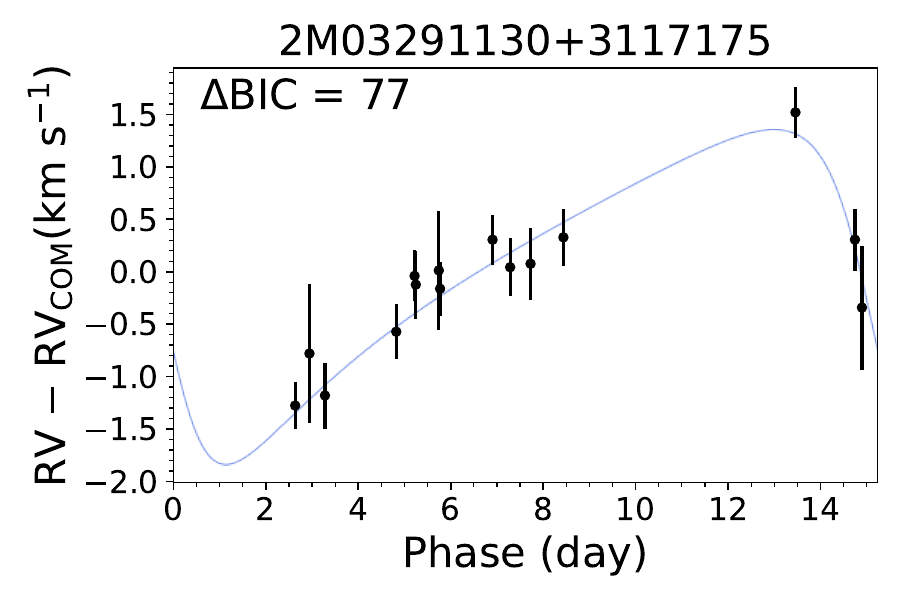} 
\caption{Same as Figure~\ref{fig:binary_orbit_fit} for 03291130$+$3117175. The best-fit orbital parameters are $P_\mathrm{fit}$ = $15^{+0}_{-0}$~day, $K_\mathrm{fit}$ = $1.4^{+0.4}_{-0.2}$~km s$^{-1}$, and $e_\mathrm{fit}$ = $0.5^{+0.09}_{-0.18}$. } 
\end{figure*} 

\begin{figure*} 
\centering 
\plottwo{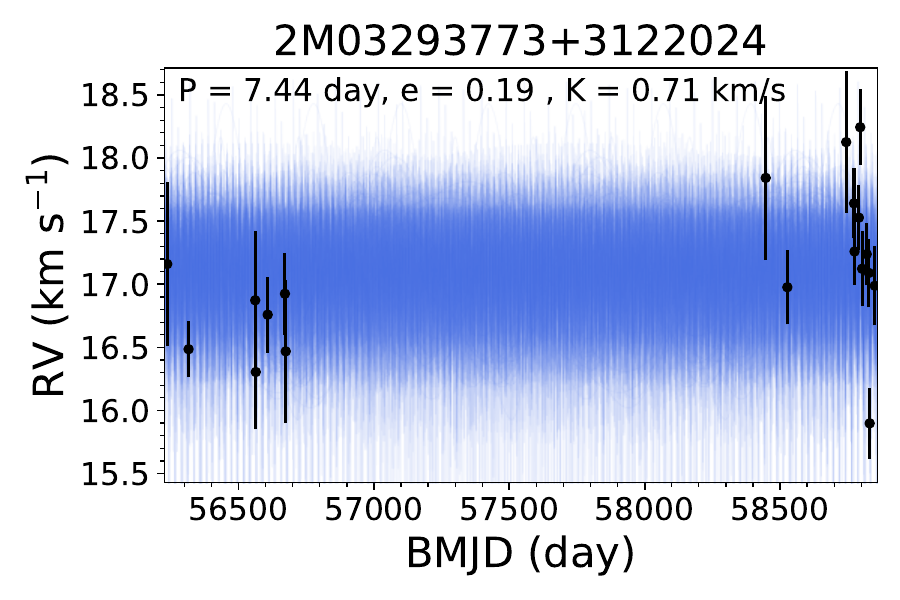}{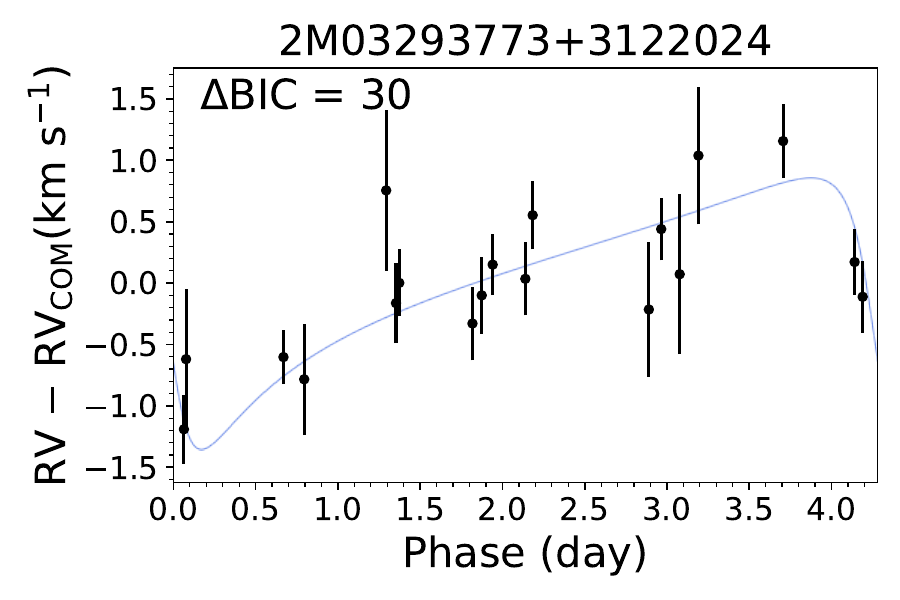} 
\caption{Same as Figure~\ref{fig:binary_orbit_fit} for 03293773$+$3122024. The best-fit orbital parameters are $P_\mathrm{fit}$ = $7^{+17}_{-6}$~day, $K_\mathrm{fit}$ = $0.7^{+0.2}_{-0.1}$~km s$^{-1}$, and $e_\mathrm{fit}$ = $0.19^{+0.27}_{-0.14}$. } 
\end{figure*} 

\begin{figure*} 
\centering 
\plottwo{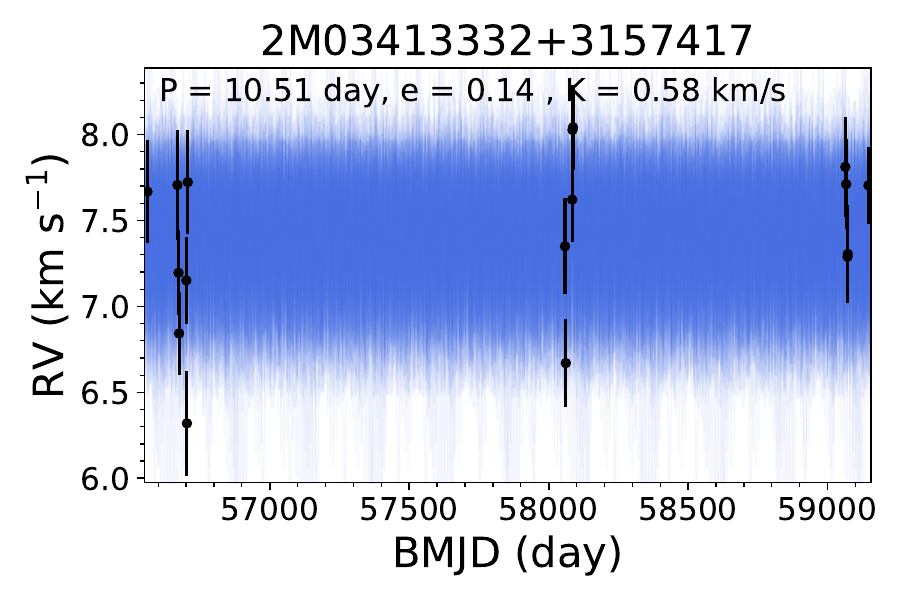}{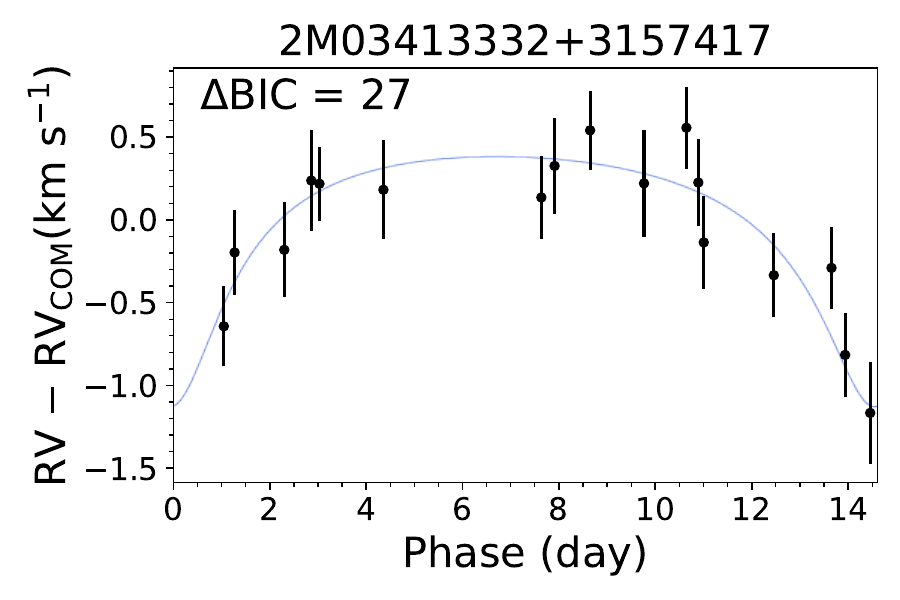} 
\caption{Same as Figure~\ref{fig:binary_orbit_fit} for 03413332$+$3157417. The best-fit orbital parameters are $P_\mathrm{fit}$ = $11^{+70}_{-9}$~day, $K_\mathrm{fit}$ = $0.6^{+0.1}_{-0.2}$~km s$^{-1}$, and $e_\mathrm{fit}$ = $0.14^{+0.22}_{-0.11}$. } 
\end{figure*} 

\begin{figure*} 
\centering 
\plottwo{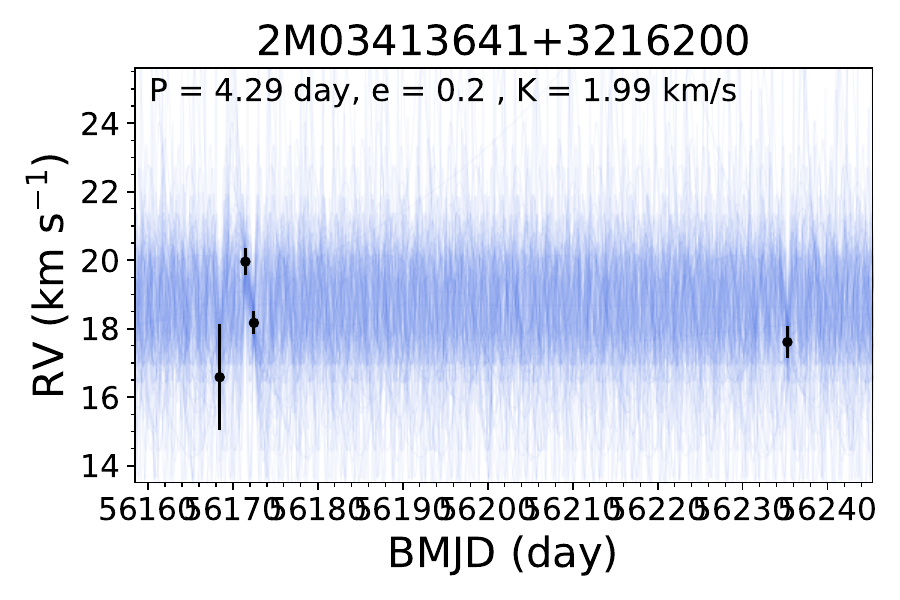}{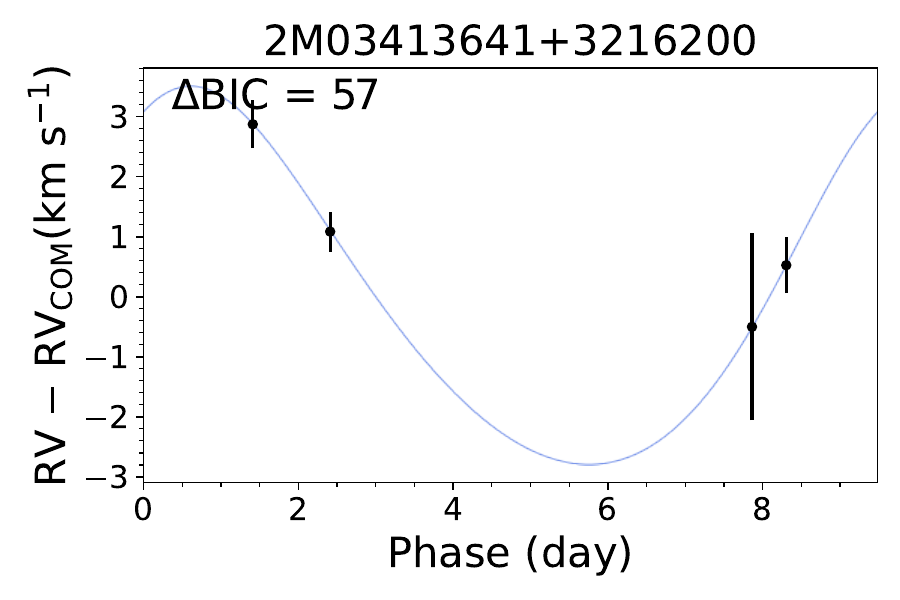} 
\caption{Same as Figure~\ref{fig:binary_orbit_fit} for 03413641$+$3216200. The best-fit orbital parameters are $P_\mathrm{fit}$ = $4^{+3}_{-3}$~day, $K_\mathrm{fit}$ = $2.0^{+1.3}_{-0.7}$~km s$^{-1}$, and $e_\mathrm{fit}$ = $0.2^{+0.28}_{-0.16}$. } 
\end{figure*} 

\begin{figure*} 
\centering 
\plottwo{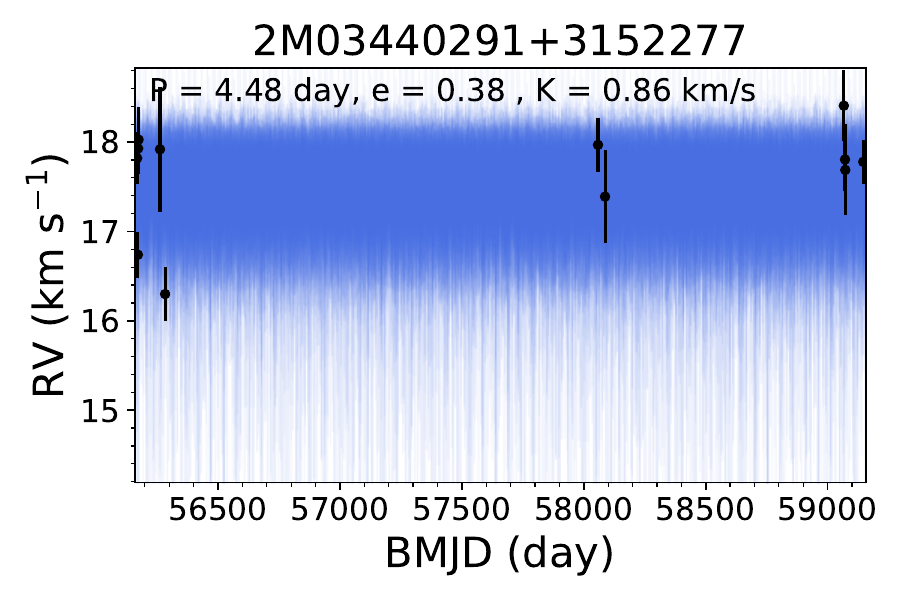}{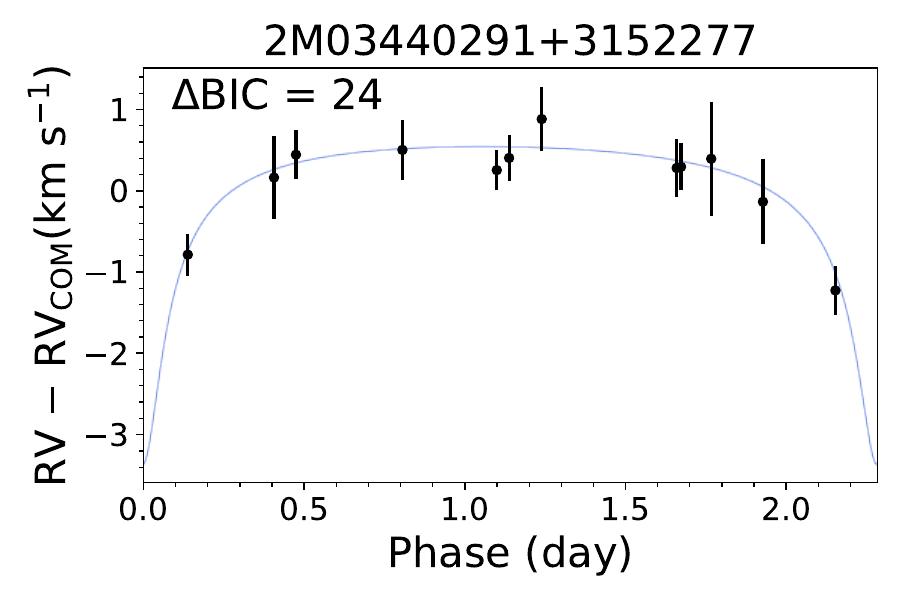} 
\caption{Same as Figure~\ref{fig:binary_orbit_fit} for 03440291$+$3152277. The best-fit orbital parameters are $P_\mathrm{fit}$ = $4^{+2}_{-3}$~day, $K_\mathrm{fit}$ = $0.9^{+0.4}_{-0.2}$~km s$^{-1}$, and $e_\mathrm{fit}$ = $0.38^{+0.25}_{-0.26}$. } 
\end{figure*} 

\begin{figure*} 
\centering 
\plottwo{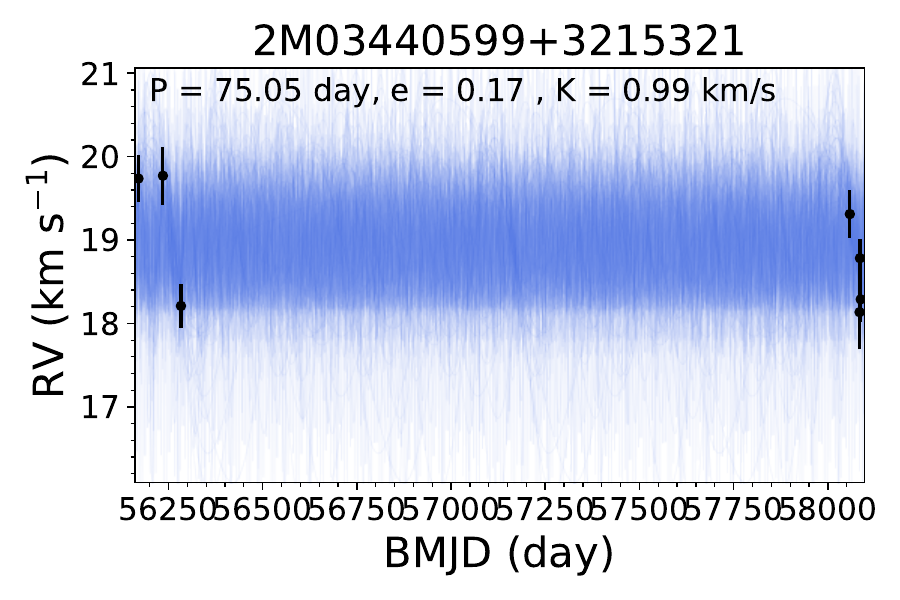}{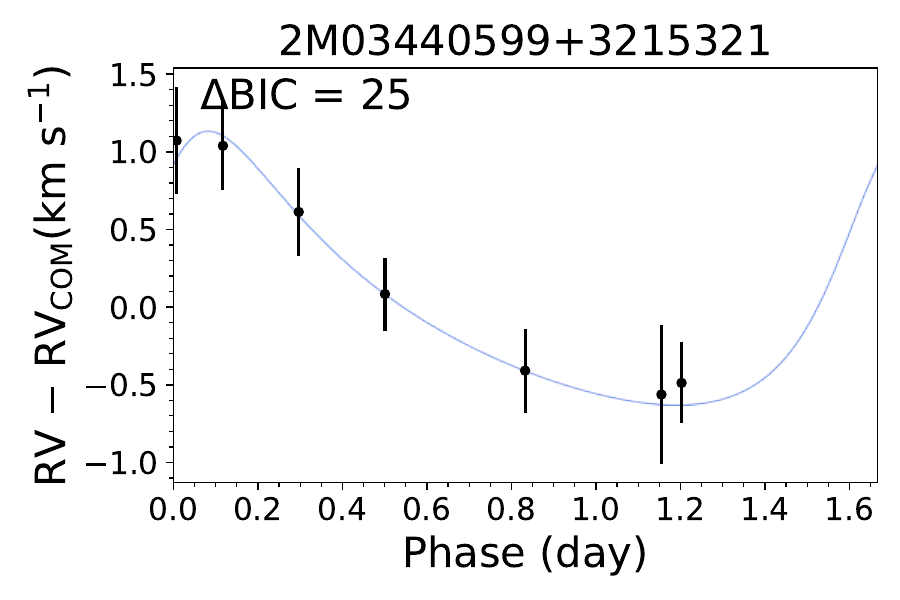} 
\caption{Same as Figure~\ref{fig:binary_orbit_fit} for 03440599$+$3215321. The best-fit orbital parameters are $P_\mathrm{fit}$ = $75^{+183}_{-70}$~day, $K_\mathrm{fit}$ = $1.0^{+0.6}_{-0.3}$~km s$^{-1}$, and $e_\mathrm{fit}$ = $0.17^{+0.25}_{-0.13}$. } 
\end{figure*} 

\begin{figure*} 
\centering 
\plottwo{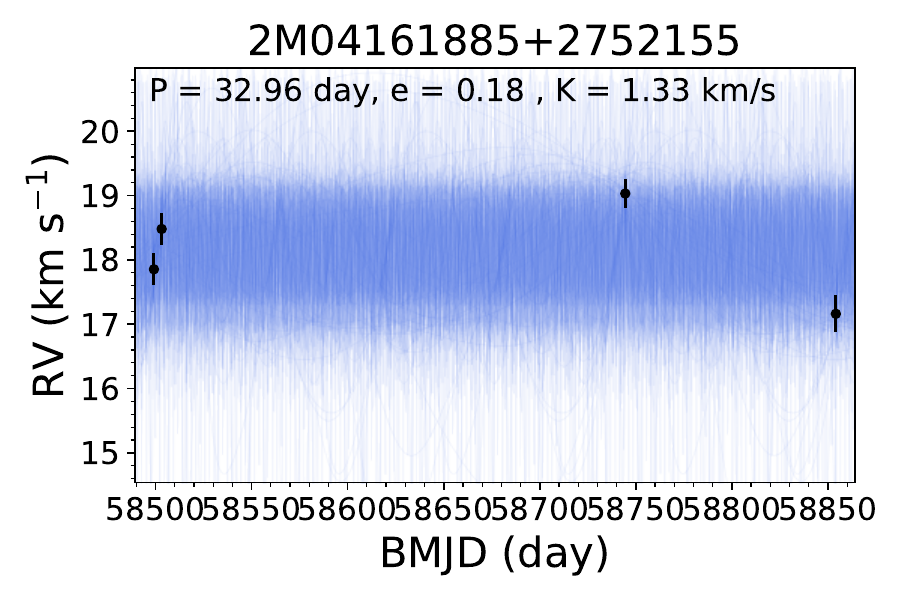}{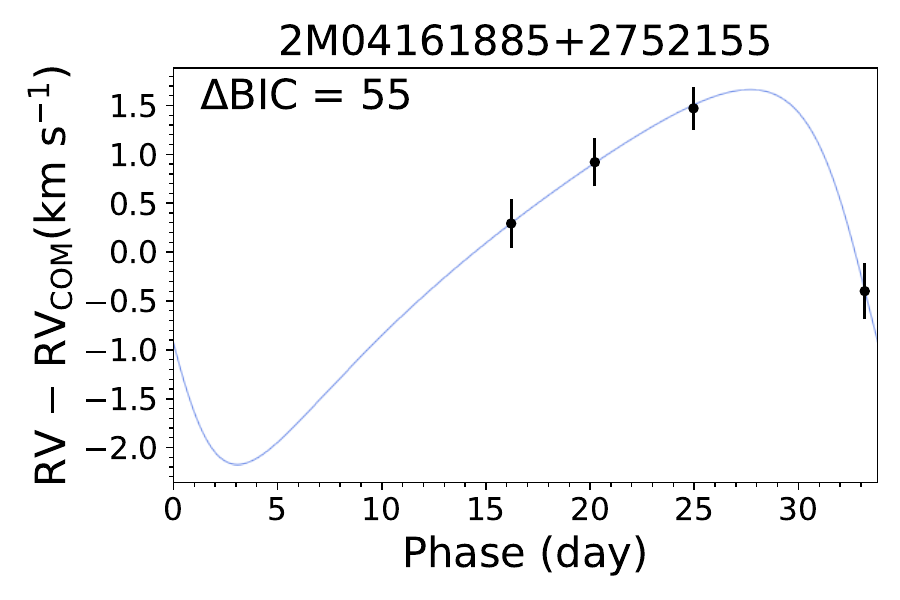} 
\caption{Same as Figure~\ref{fig:binary_orbit_fit} for 04161885$+$2752155. The best-fit orbital parameters are $P_\mathrm{fit}$ = $33^{+170}_{-30}$~day, $K_\mathrm{fit}$ = $1.3^{+1.0}_{-0.4}$~km s$^{-1}$, and $e_\mathrm{fit}$ = $0.18^{+0.26}_{-0.14}$. } 
\end{figure*} 

\begin{figure*} 
\centering 
\plottwo{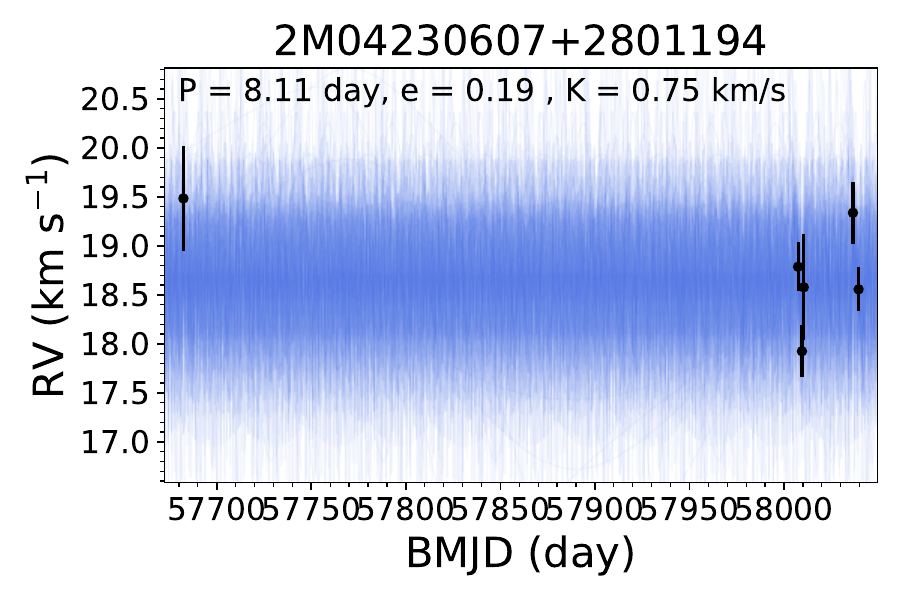}{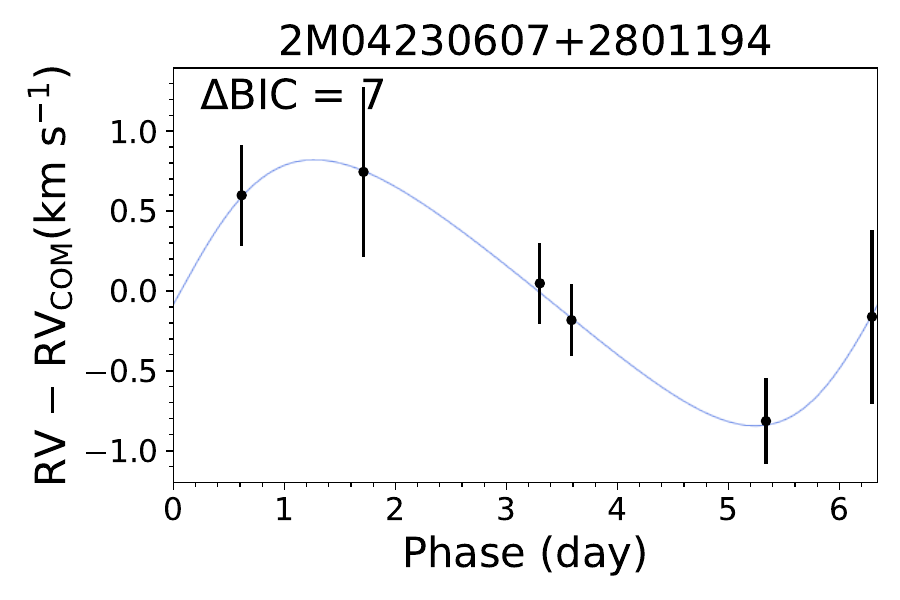} 
\caption{Same as Figure~\ref{fig:binary_orbit_fit} for 04230607$+$2801194. The best-fit orbital parameters are $P_\mathrm{fit}$ = $8^{+26}_{-6}$~day, $K_\mathrm{fit}$ = $0.7^{+0.5}_{-0.3}$~km s$^{-1}$, and $e_\mathrm{fit}$ = $0.19^{+0.27}_{-0.15}$. } 
\end{figure*} 

\begin{figure*} 
\centering 
\plottwo{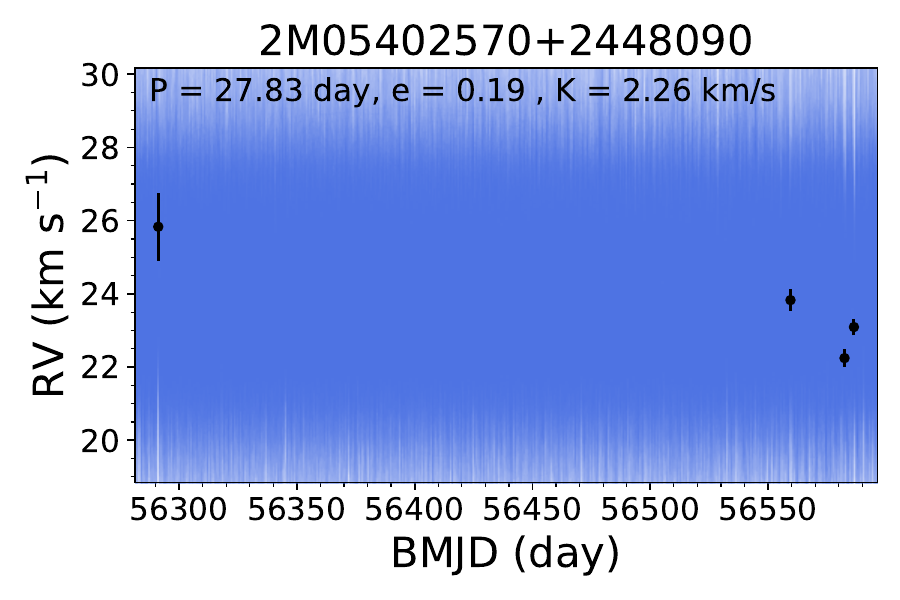}{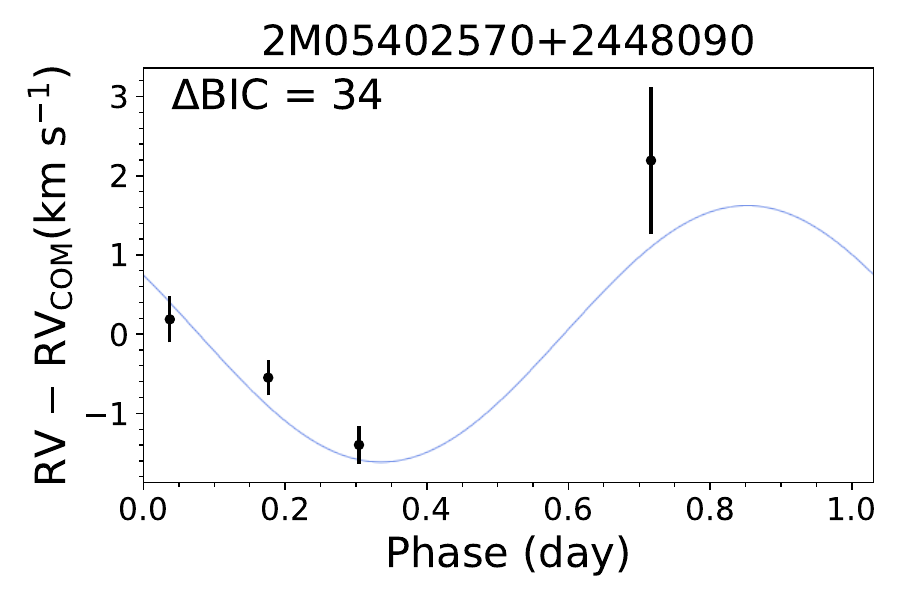} 
\caption{Same as Figure~\ref{fig:binary_orbit_fit} for 05402570$+$2448090. The best-fit orbital parameters are $P_\mathrm{fit}$ = $28^{+49}_{-25}$~day, $K_\mathrm{fit}$ = $2.3^{+1.2}_{-0.9}$~km s$^{-1}$, and $e_\mathrm{fit}$ = $0.19^{+0.27}_{-0.14}$. } 
\end{figure*} 

\begin{figure*} 
\centering 
\plottwo{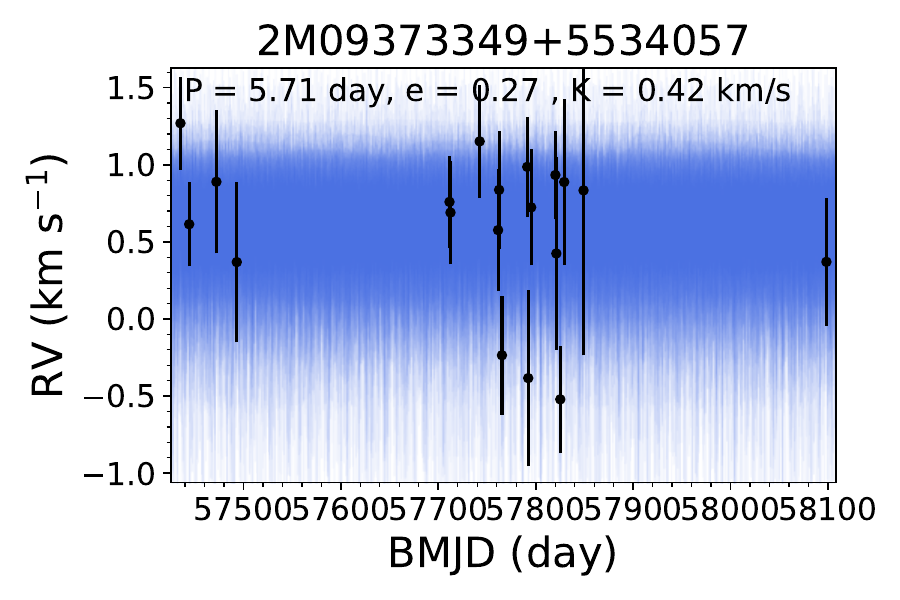}{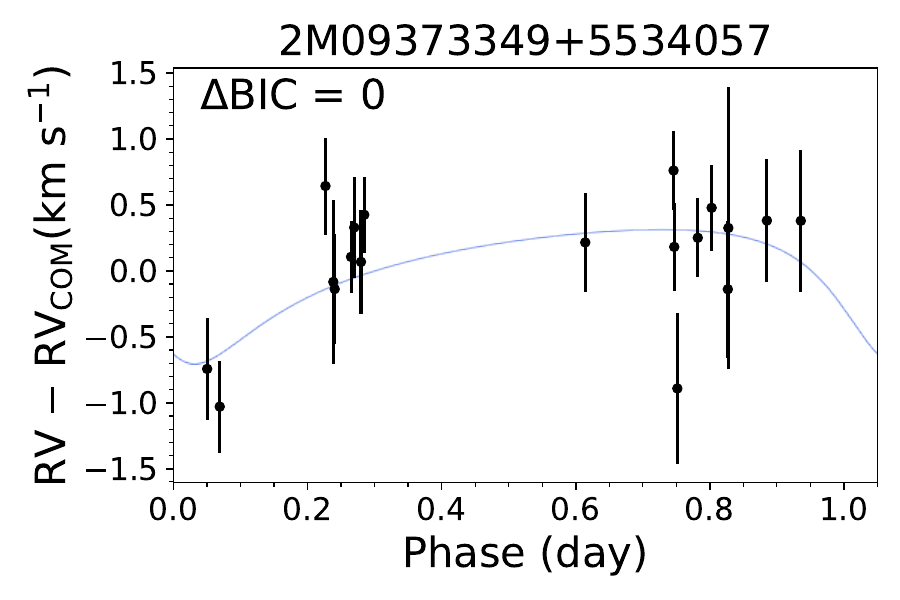} 
\caption{Same as Figure~\ref{fig:binary_orbit_fit} for 09373349$+$5534057. The best-fit orbital parameters are $P_\mathrm{fit}$ = $6^{+13}_{-4}$~day, $K_\mathrm{fit}$ = $0.4^{+0.3}_{-0.2}$~km s$^{-1}$, and $e_\mathrm{fit}$ = $0.27^{+0.3}_{-0.2}$. } 
\end{figure*} 

\begin{figure*} 
\centering 
\plottwo{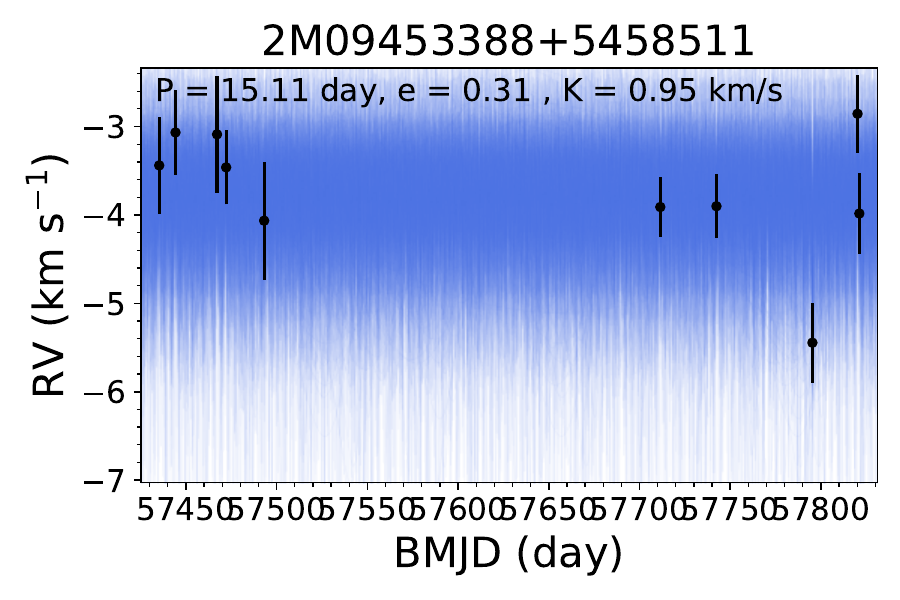}{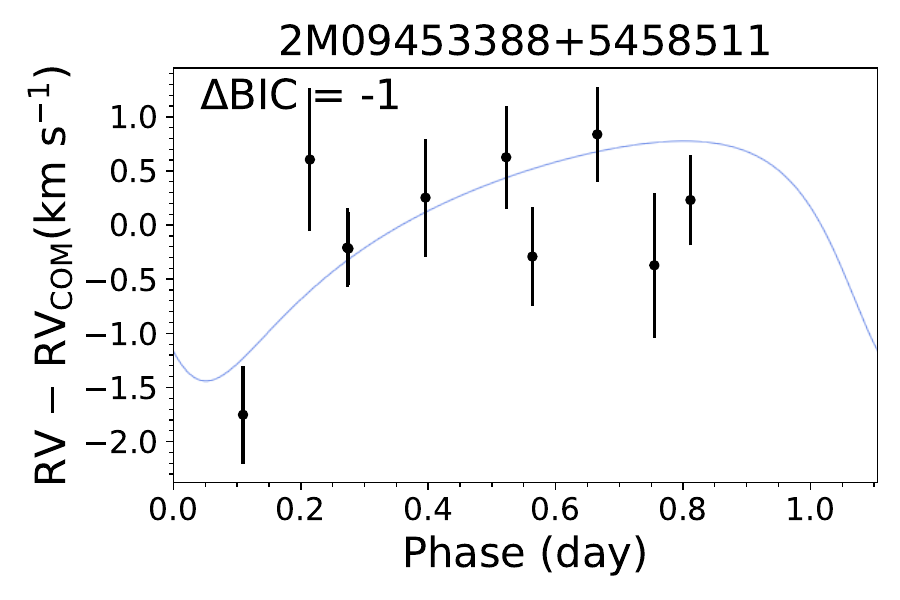} 
\caption{Same as Figure~\ref{fig:binary_orbit_fit} for 09453388$+$5458511. The best-fit orbital parameters are $P_\mathrm{fit}$ = $15^{+120}_{-13}$~day, $K_\mathrm{fit}$ = $1.0^{+0.5}_{-0.5}$~km s$^{-1}$, and $e_\mathrm{fit}$ = $0.31^{+0.3}_{-0.24}$. } 
\end{figure*} 

\begin{figure*} 
\centering 
\plottwo{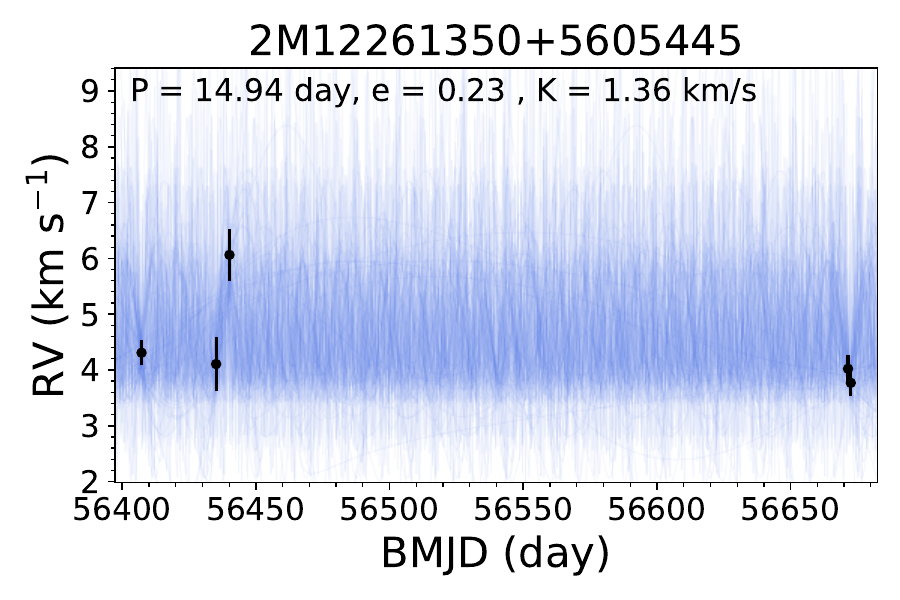}{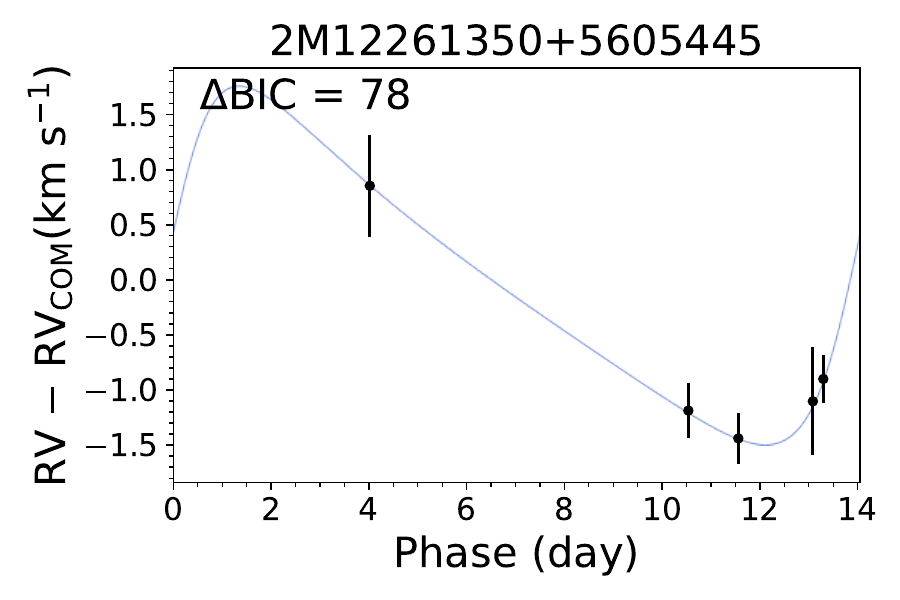} 
\caption{Same as Figure~\ref{fig:binary_orbit_fit} for 12261350$+$5605445. The best-fit orbital parameters are $P_\mathrm{fit}$ = $15^{+48}_{-11}$~day, $K_\mathrm{fit}$ = $1.4^{+0.9}_{-0.5}$~km s$^{-1}$, and $e_\mathrm{fit}$ = $0.23^{+0.3}_{-0.17}$. } 
\end{figure*} 

\begin{figure*} 
\centering 
\plottwo{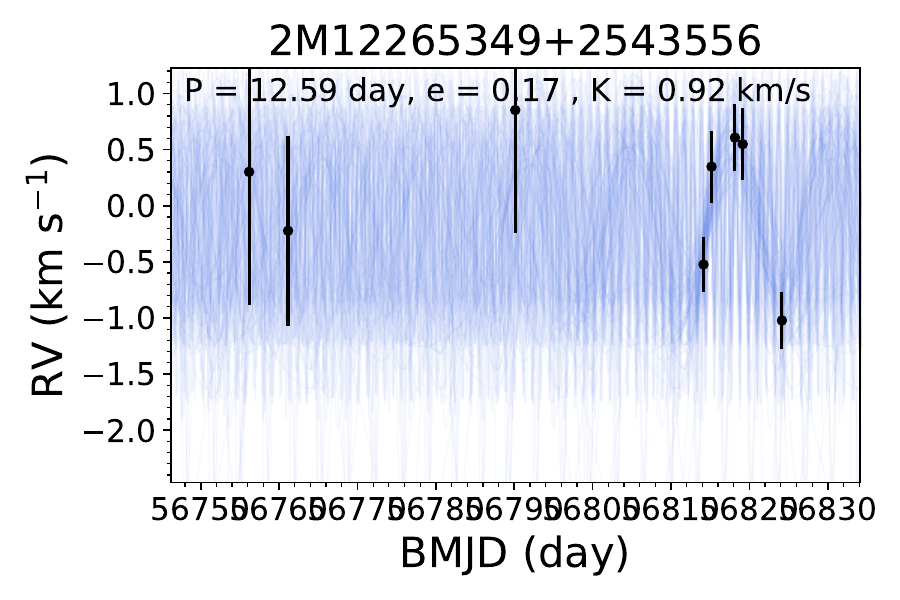}{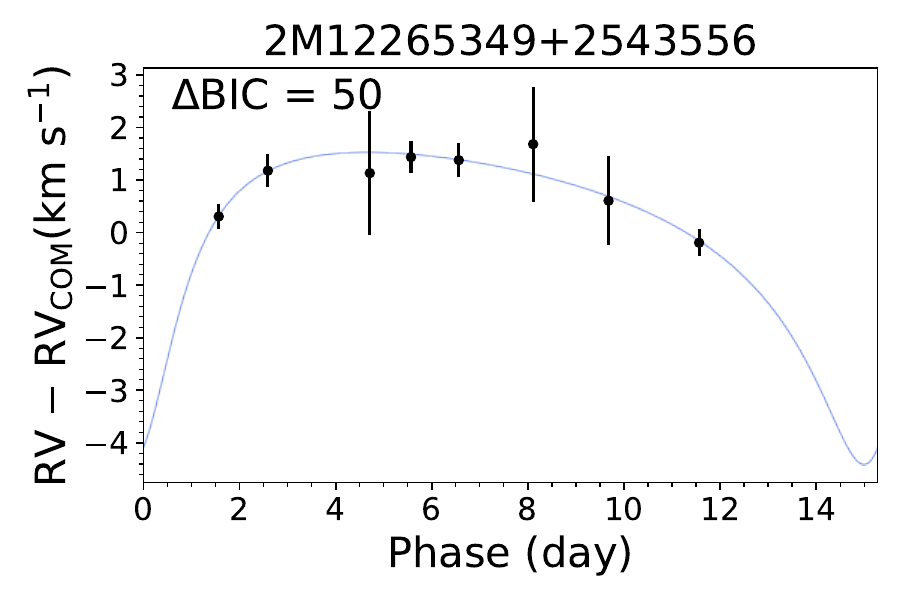} 
\caption{Same as Figure~\ref{fig:binary_orbit_fit} for 12265349$+$2543556. The best-fit orbital parameters are $P_\mathrm{fit}$ = $13^{+3}_{-9}$~day, $K_\mathrm{fit}$ = $0.9^{+0.3}_{-0.2}$~km s$^{-1}$, and $e_\mathrm{fit}$ = $0.17^{+0.24}_{-0.13}$. } 
\end{figure*} 

\begin{figure*} 
\centering 
\plottwo{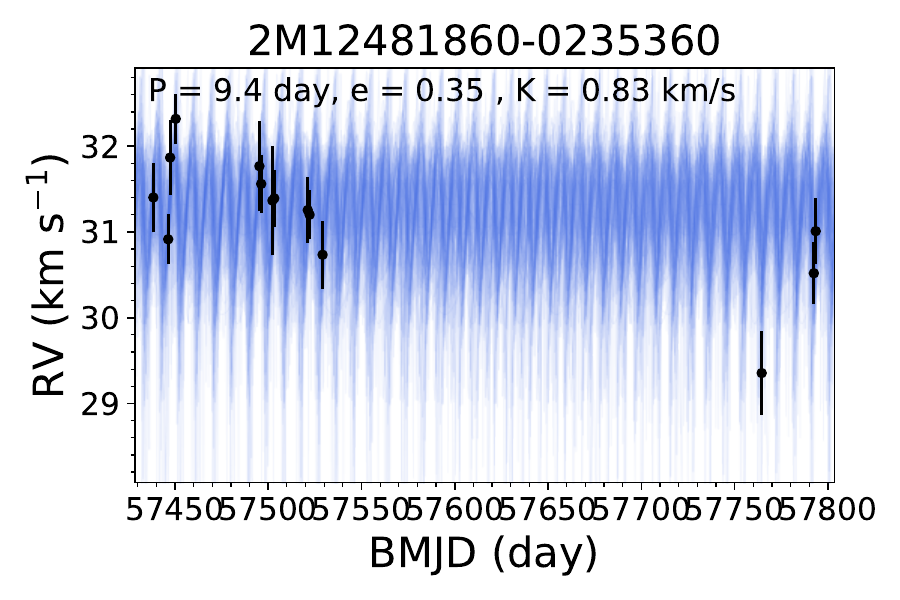}{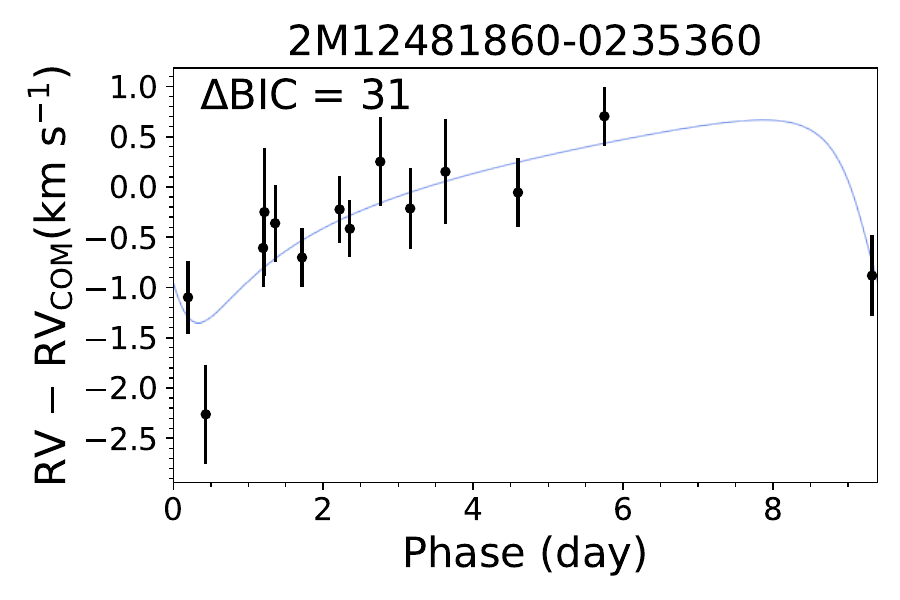} 
\caption{Same as Figure~\ref{fig:binary_orbit_fit} for 12481860$-$0235360. The best-fit orbital parameters are $P_\mathrm{fit}$ = $9^{+5}_{-2}$~day, $K_\mathrm{fit}$ = $0.8^{+0.4}_{-0.3}$~km s$^{-1}$, and $e_\mathrm{fit}$ = $0.35^{+0.28}_{-0.26}$. } 
\end{figure*} 

\begin{figure*} 
\centering 
\plottwo{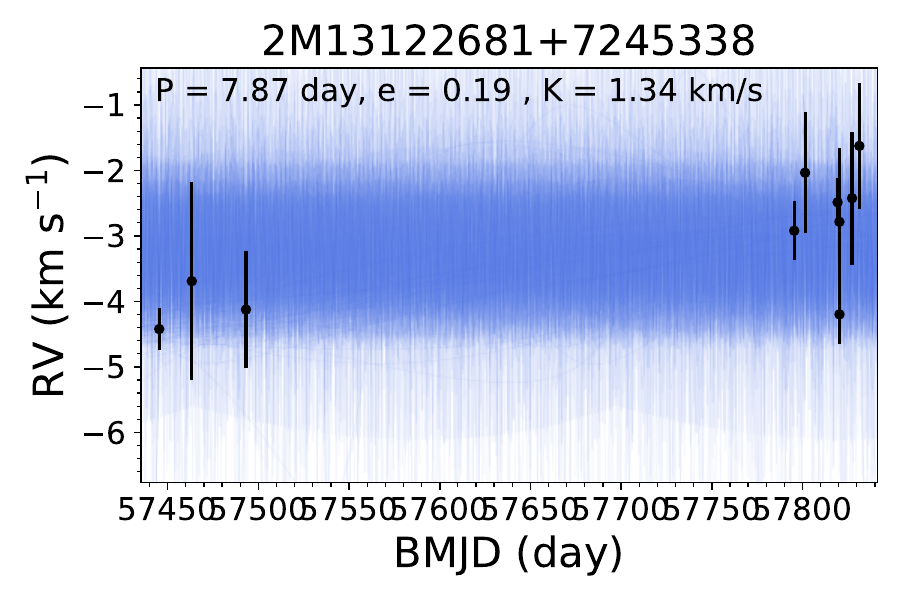}{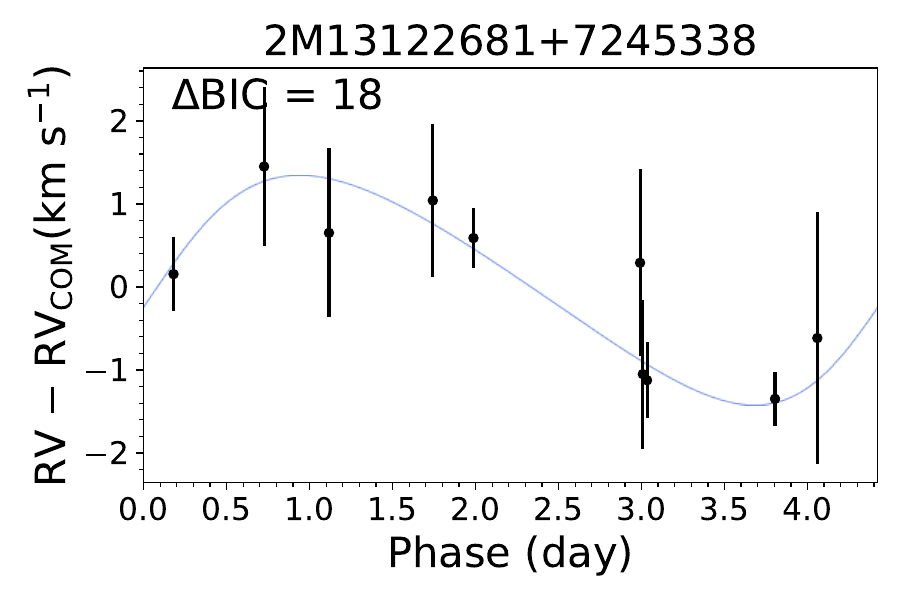} 
\caption{Same as Figure~\ref{fig:binary_orbit_fit} for 13122681$+$7245338. The best-fit orbital parameters are $P_\mathrm{fit}$ = $8^{+1046}_{-6}$~day, $K_\mathrm{fit}$ = $1.3^{+0.8}_{-0.4}$~km s$^{-1}$, and $e_\mathrm{fit}$ = $0.19^{+0.27}_{-0.14}$. } 
\end{figure*} 

\begin{figure*} 
\centering 
\plottwo{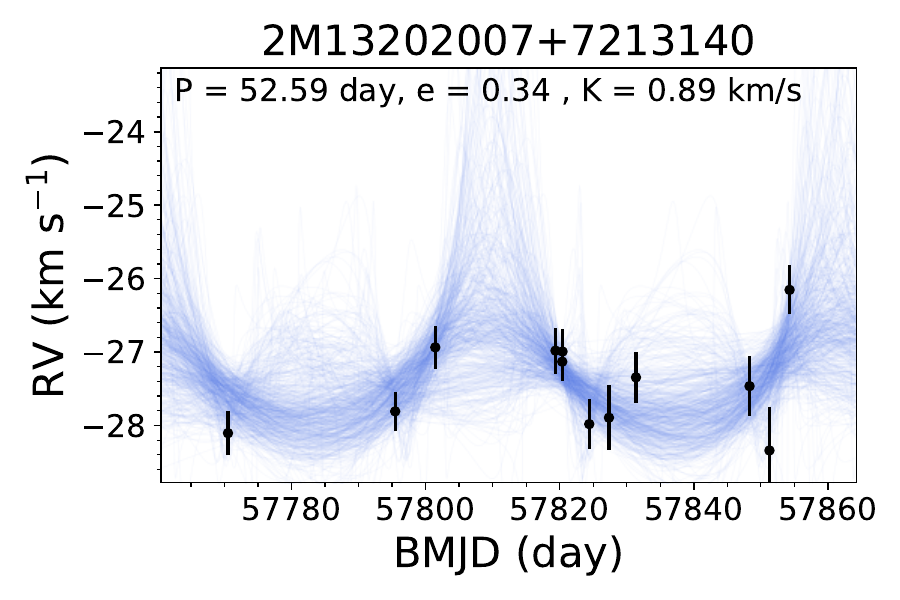}{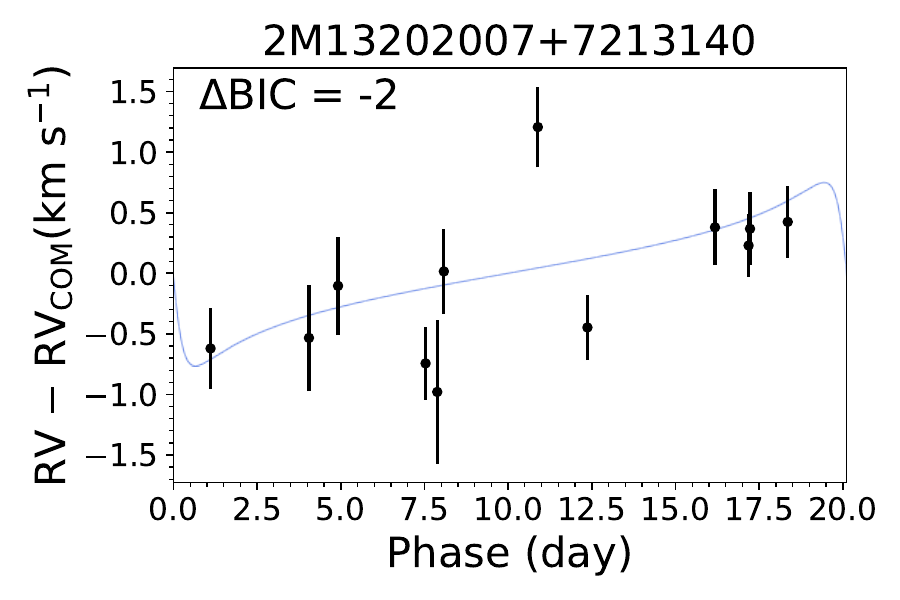} 
\caption{Same as Figure~\ref{fig:binary_orbit_fit} for 13202007$+$7213140. The best-fit orbital parameters are $P_\mathrm{fit}$ = $53^{+4}_{-16}$~day, $K_\mathrm{fit}$ = $0.9^{+0.6}_{-0.5}$~km s$^{-1}$, and $e_\mathrm{fit}$ = $0.34^{+0.27}_{-0.21}$. } 
\end{figure*} 

\begin{figure*} 
\centering 
\plottwo{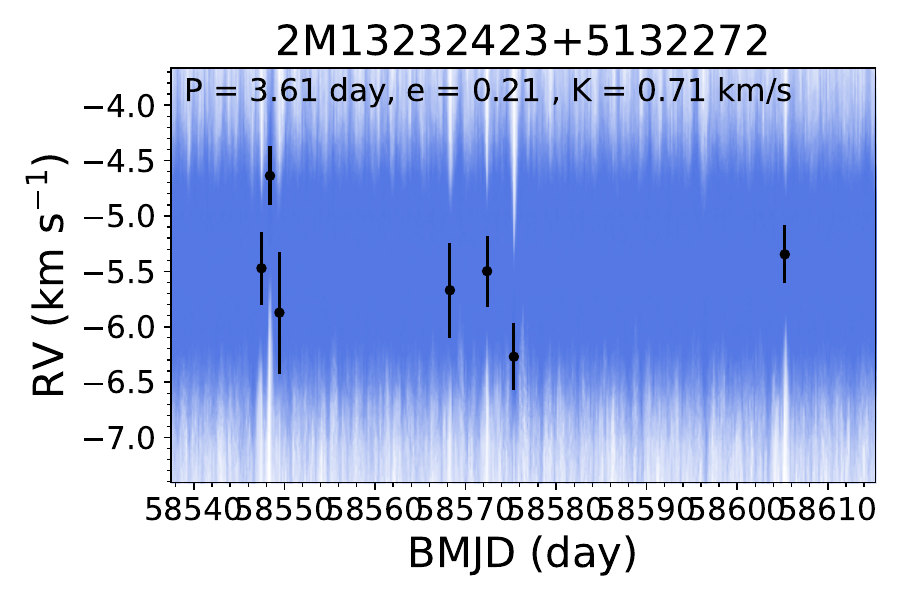}{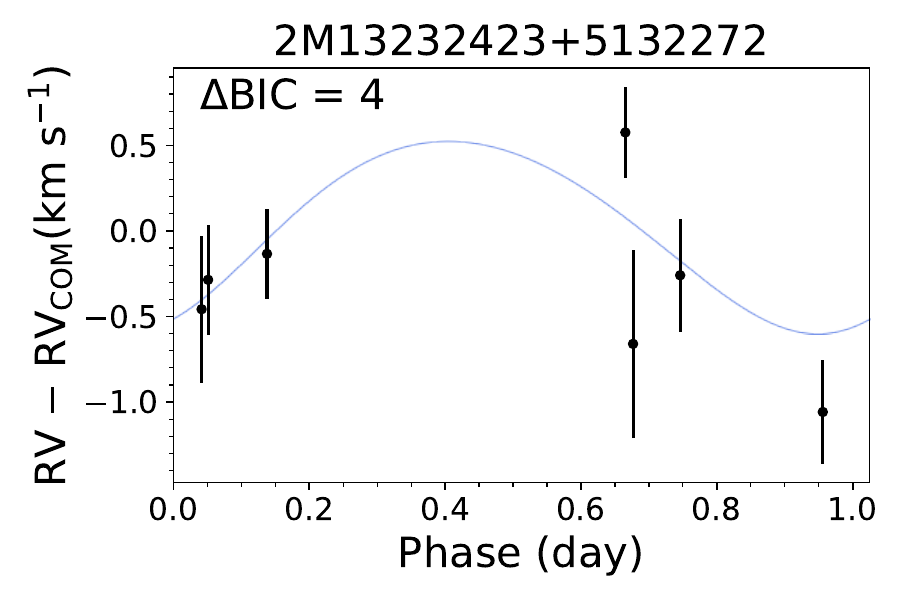} 
\caption{Same as Figure~\ref{fig:binary_orbit_fit} for 13232423$+$5132272. The best-fit orbital parameters are $P_\mathrm{fit}$ = $4^{+8}_{-2}$~day, $K_\mathrm{fit}$ = $0.7^{+0.4}_{-0.3}$~km s$^{-1}$, and $e_\mathrm{fit}$ = $0.21^{+0.33}_{-0.15}$. } 
\end{figure*} 

\begin{figure*} 
\centering 
\plottwo{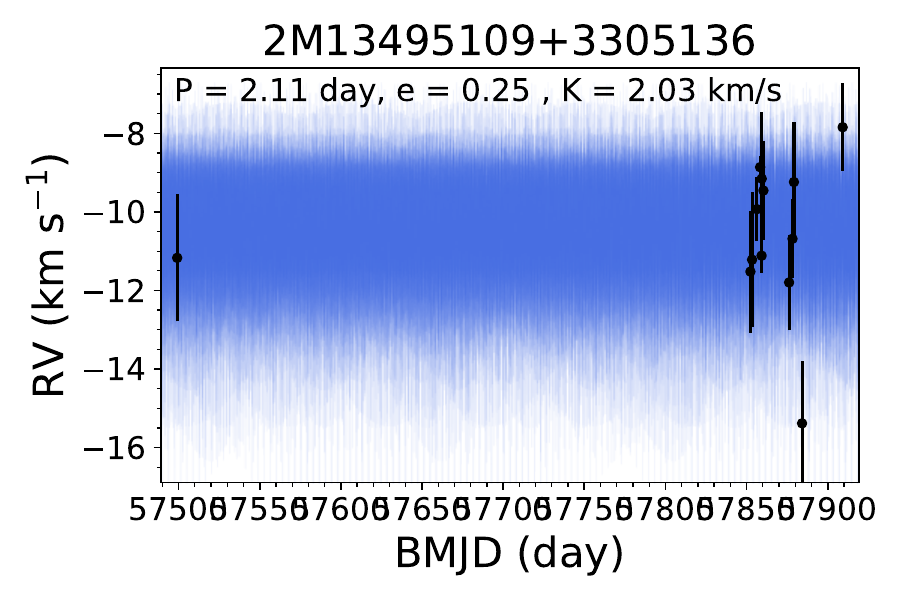}{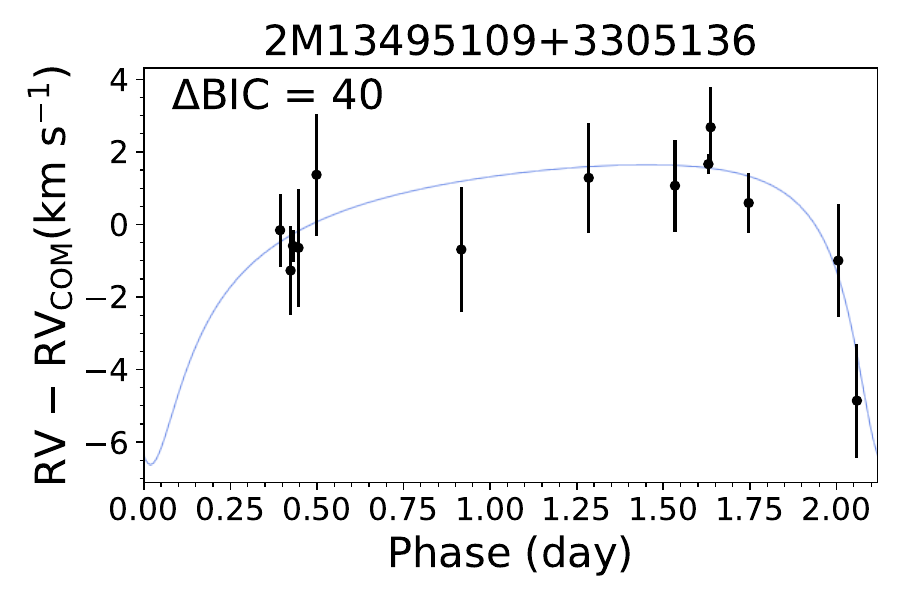} 
\caption{Same as Figure~\ref{fig:binary_orbit_fit} for 13495109$+$3305136. The best-fit orbital parameters are $P_\mathrm{fit}$ = $2^{+2}_{-1}$~day, $K_\mathrm{fit}$ = $2.0^{+1.0}_{-0.7}$~km s$^{-1}$, and $e_\mathrm{fit}$ = $0.25^{+0.32}_{-0.19}$. } 
\end{figure*} 

\begin{figure*} 
\centering 
\plottwo{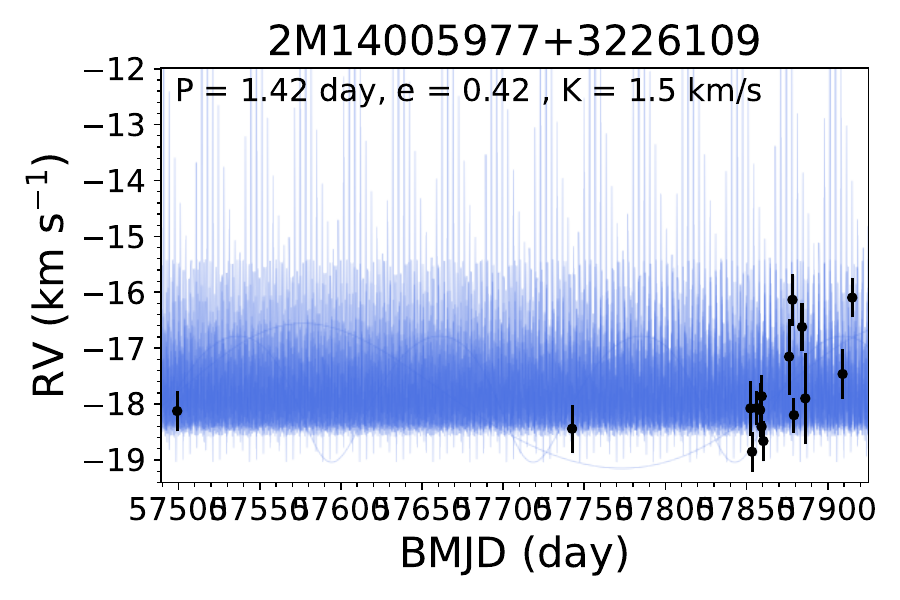}{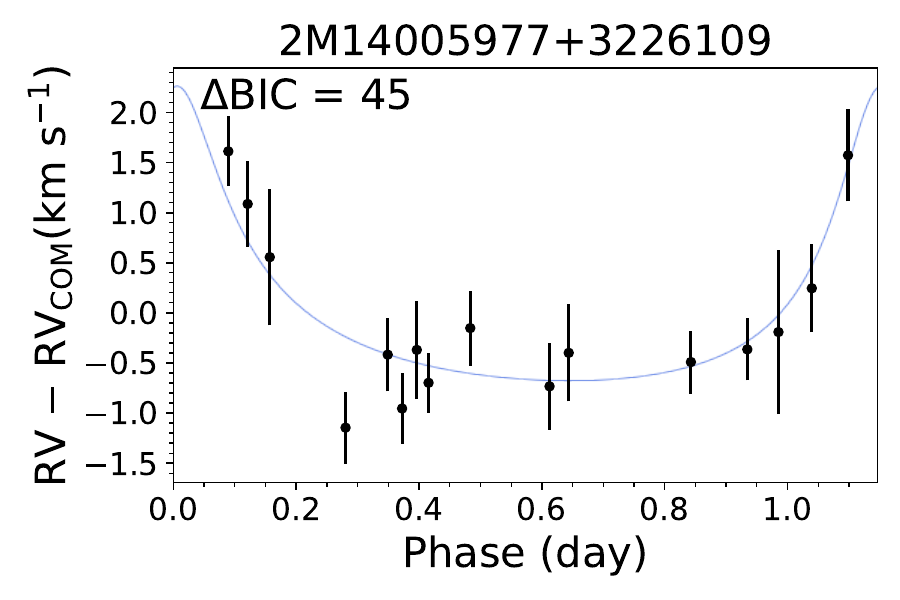} 
\caption{Same as Figure~\ref{fig:binary_orbit_fit} for 14005977$+$3226109. The best-fit orbital parameters are $P_\mathrm{fit}$ = $1^{+321}_{-0}$~day, $K_\mathrm{fit}$ = $1.5^{+1.0}_{-0.4}$~km s$^{-1}$, and $e_\mathrm{fit}$ = $0.42^{+0.26}_{-0.25}$. } 
\end{figure*} 

\begin{figure*} 
\centering 
\plottwo{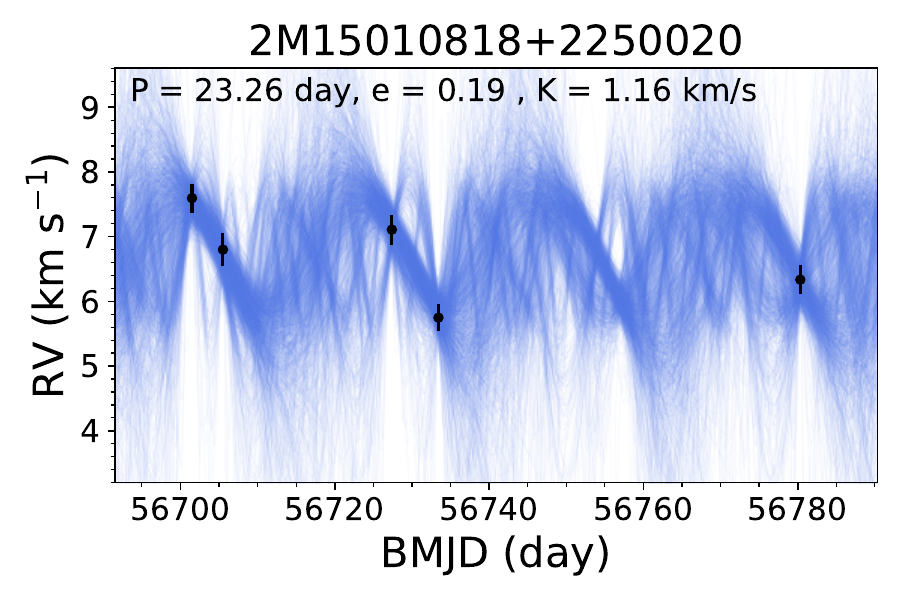}{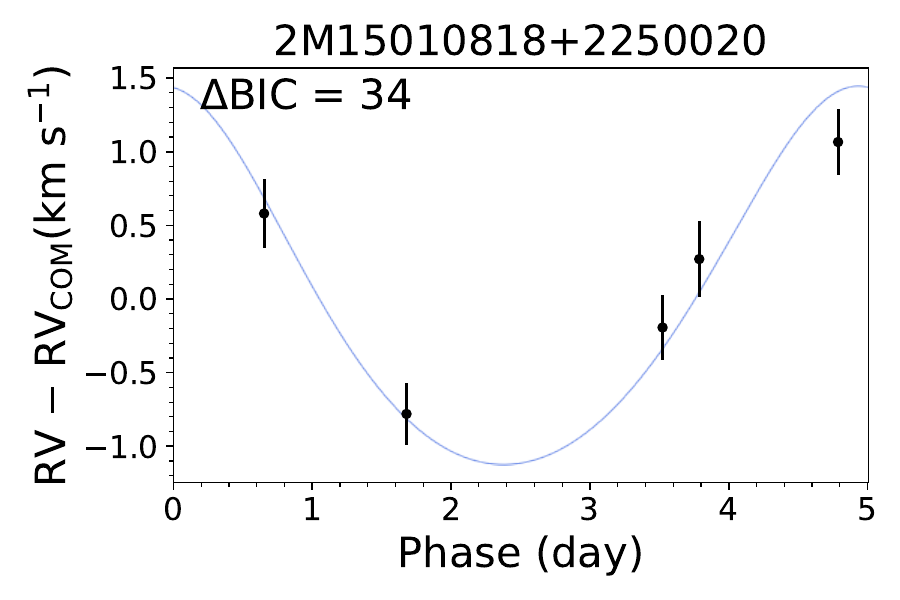} 
\caption{Same as Figure~\ref{fig:binary_orbit_fit} for 15010818$+$2250020. The best-fit orbital parameters are $P_\mathrm{fit}$ = $23^{+1}_{-11}$~day, $K_\mathrm{fit}$ = $1.2^{+0.7}_{-0.3}$~km s$^{-1}$, and $e_\mathrm{fit}$ = $0.19^{+0.28}_{-0.14}$. } 
\end{figure*} 

\begin{figure*} 
\centering 
\plottwo{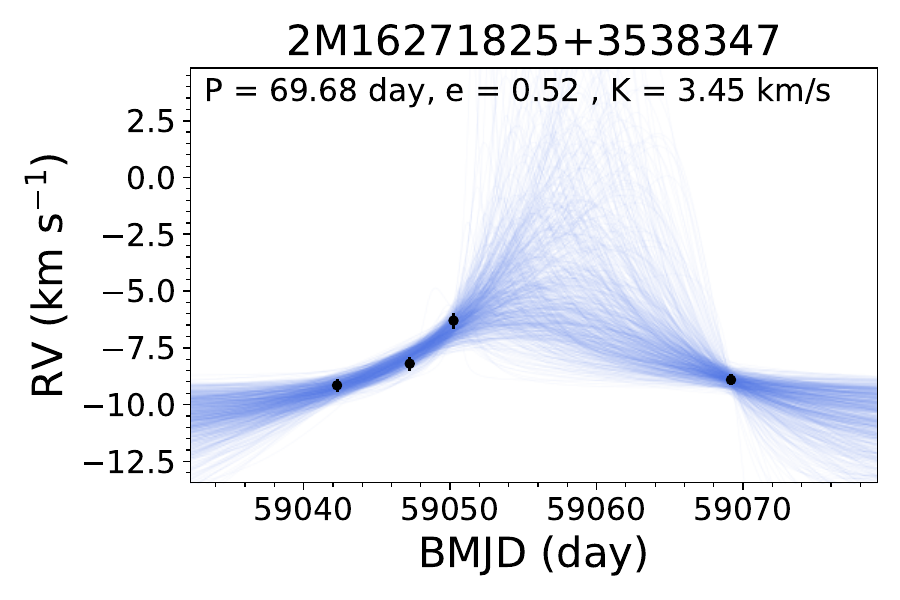}{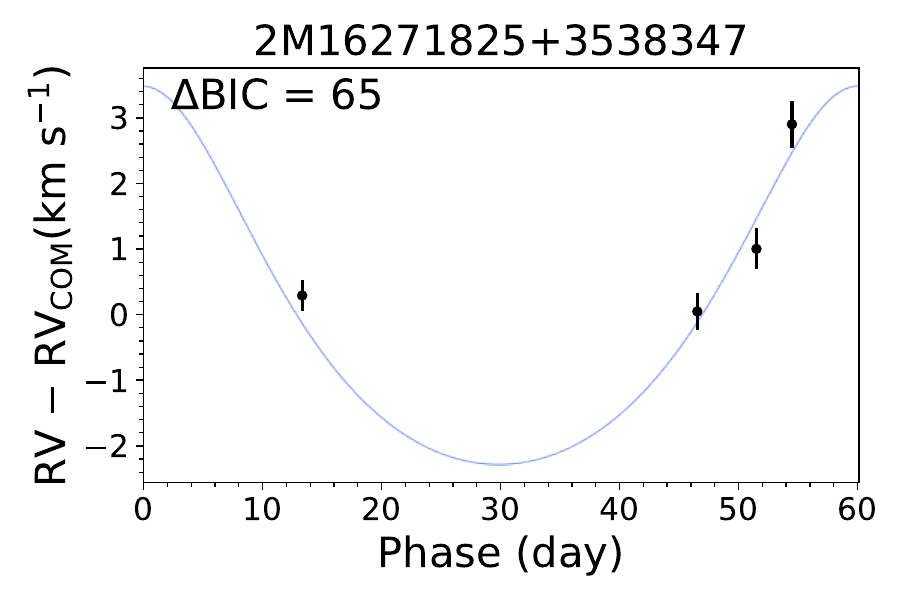} 
\caption{Same as Figure~\ref{fig:binary_orbit_fit} for 16271825$+$3538347. The best-fit orbital parameters are $P_\mathrm{fit}$ = $70^{+18}_{-7}$~day, $K_\mathrm{fit}$ = $3.4^{+2.1}_{-1.3}$~km s$^{-1}$, and $e_\mathrm{fit}$ = $0.52^{+0.17}_{-0.24}$. } 
\end{figure*} 

\begin{figure*} 
\centering 
\plottwo{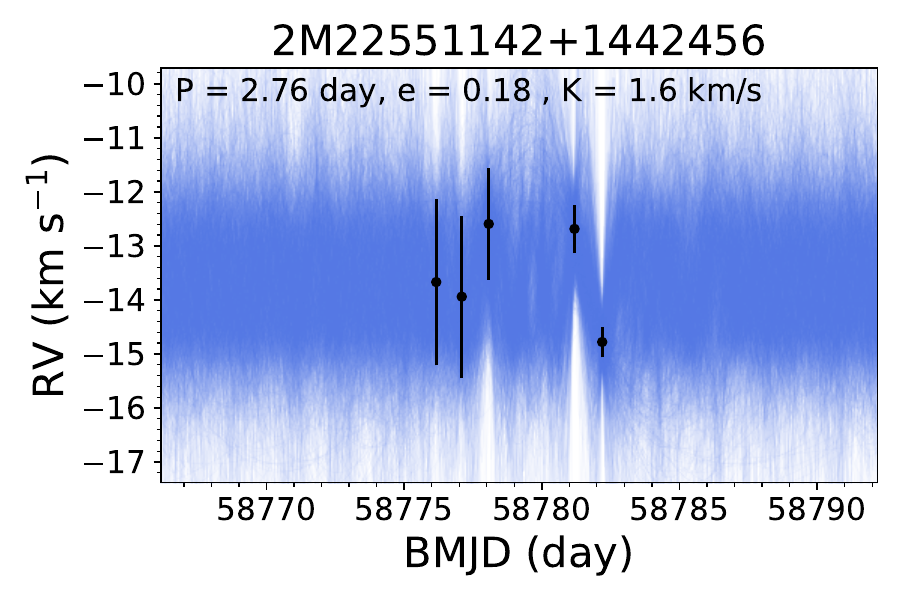}{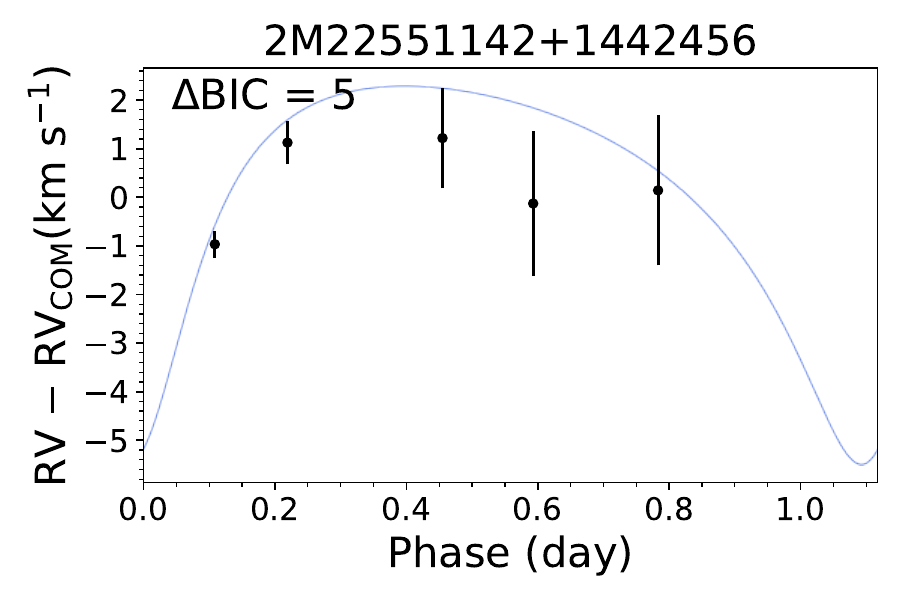} 
\caption{Same as Figure~\ref{fig:binary_orbit_fit} for 22551142$+$1442456. The best-fit orbital parameters are $P_\mathrm{fit}$ = $3^{+5}_{-1}$~day, $K_\mathrm{fit}$ = $1.6^{+0.9}_{-0.5}$~km s$^{-1}$, and $e_\mathrm{fit}$ = $0.18^{+0.26}_{-0.14}$. } 
\end{figure*} 

\clearpage

\bibliography{mylibrary.bib} 
\bibliographystyle{aasjournal}

\end{document}